\documentclass[pra,aps,10pt,twocolumn,amsmath,amssymb,floatfix,footinbib,bibnotes,longbibliography,usenames,dvipsnames]{revtex4-2}
\pdfoutput=1

\usepackage[colorlinks=true,citecolor=blue,linkcolor=magenta]{hyperref}
\usepackage{tikz}
\usepackage{amsmath}
\usepackage{graphicx}
\usepackage{changepage}
\usepackage{amsthm, amssymb}
\usepackage{floatflt}
\usepackage{float}
\usepackage{array}
\usepackage{mathtools}
\usepackage{soul}
\usepackage[normalem]{ulem}

\newcommand{\ci}{\mathrm{i}}

\newcommand{\x}{\tilde{\alpha}}

\def \titlename {Generation of Volume-Law Entanglement by Local-Measurement-Only\\ Quantum Dynamics }
\def \authornames{Surajit Bera$^{1,2}$, Igor V. Gornyi$^{3}$, Sumilan Banerjee$^1$, and Yuval Gefen$^4$}
\def \affiliations{$^1$Centre for Condensed Matter Theory, Department of Physics, Indian Institute of Science, Bangalore 560012, India\\$^{2}$JEIP, USR 3573 CNRS, Collège de France, PSL Research University,\\ 11 Place Marcelin Berthelot, 75321 Paris Cedex 05, France\\$^{3}$Institute for Quantum Materials and Technologies and Institut für Theorie der Kondensierten Materie,\\ Karlsruhe Institute of Technology, 76131 Karlsruhe, Germany\\$^{4}$Department of Condensed Matter Physics, Weizmann Institute of Science, 7610001 Rehovot, Israel
}

\begin{document}
	
	\title{\titlename}
	\author{\authornames}
	\affiliation{\affiliations}
	\email{surajit.bera@college-de-france.fr}
     \email{igor.gornyi@kit.edu}
     \email{sumilan@iisc.ac.in}
     \email{yuval.gefen@weizmann.ac.il}
	\date\today

\begin{abstract}

Repeated local measurements typically have adversarial effects on entangling unitary dynamics, as local measurements usually degrade entanglement. However, recent works on measurement-only dynamics have shown that strongly entangled states can be generated solely through non-commuting random multi-site and multi-spin projective measurements. 
In this work, we explore a generalized measurement setup in a system without intrinsic unitary dynamics and show that volume-law entangled states can be generated through local, non-random, yet non-commuting measurements. Specifically, we construct a one-dimensional model comprising a main fermionic chain and an auxiliary (ancilla) chain, where generalized measurements are performed by locally coupling the system to detector qubits. 
Our results demonstrate that long-time states with volume-law entanglement or mutual information are generated between different parts of the main chain purely through non-unitary measurement dynamics. 
Remarkably, we find that such large-entanglement generation can be achieved using only the measurements of one-body operators. Moreover, we show that measurements of non-local higher-body operators can be used to control and reduce entanglement generation by introducing kinetic constraints to the dynamics. We discuss the statistics of entanglement measures along the quantum trajectories, the approach to stationary distributions of entanglement or long-time steady states, and the associated notions of limited ergodicity in the measurement-only dynamics. 
Our findings highlight the potential of non-random measurement protocols for controlled entanglement generation and the study of non-unitary many-body dynamics.

\end{abstract}
\maketitle 

\section{Introduction}

Measurements and entanglement are fundamental to quantum mechanics. Measurements reveal how an observer influences a quantum system’s evolution, while entanglement plays a crucial role across various fields, from quantum information to condensed matter and high-energy physics
\cite{Calabrese_2004, Casini_2009, RevModPhys.82.277, RevModPhys.91.021001, RyuTakayanagi, Roy2020}. 
Measurements and entanglement are typically perceived as antagonistic, as local measurements tend to collapse quantum states into a local basis, reducing entanglement. This conventional understanding suggests that local quantum measurements produce low-entanglement states, while unitary dynamics exclusively builds entanglement by encoding quantum correlations in non-local operators, making them resilient to local measurements.

On one hand, in recent years, this competition between scrambling \cite{Nahum_PRX2017, Landsman2019} due to entangling unitary dynamics and `unscrambling' due to repeated disentangling local measurements \cite{Misra1977, Wineland_ZenoEffect} has been shown to give rise to unusual dynamical transitions 
\cite{Skinner_PRX2019, Li2019, Chan2019a, Szyniszewski2019a, Bao_Altman_PRB2020, Choi2020, Szyniszewski2020, Jian2020, Gullans2020, Fuji_PRB2020, Tang2020, AlbertonDiehl, Sang2021, Jian2021, Nahum2021, Block2022, Poboiko2024}, namely the measurement-induced phase transitions (MIPT), see Refs.~\cite{Fisher2022, Potter2022} for review. On the other hand, in a parallel line of research, the concept of measurement-induced engineering of quantum states was put forward \cite{Roy2020, Kumar2022, HerasymenkoPRXQ2023, SriramPRB2023, puente2023, Smith2023, morales2023, Chen2023-AKLT, Ravindranath2023, Edd2024, Volya2024, Chen2024efficient, Qian_2024, Langbehn2024}, where non-random measurements can be used to steer an initially trivial state towards a target state characterized by some degree of entanglement. These recent advances challenge the traditional view that measurements are only detrimental to entanglement. This raises a fundamental question: Can local measurements alone, without any intrinsic unitary dynamics, generate highly entangled quantum states?


In the context of MIPTs, an affirmative answer to this question has been given by Ippoliti \textit{et al.}~\cite{Ippoliti_Khemani_PRX} in a class of measurement-only models with repeated projective measurements of finite-range multi-site and multi-spin Pauli-string operators drawn from a random ensemble (see also Ref.~\cite{Klocke2023}). These measurement-only models lead to volume-law-entangled states after many measurements of sufficiently long but finite-range operators, e.g., at least three-site Pauli-string operators. This finding points to the non-commutativity of the measurements as one of the necessary ingredients for producing entangled states
(cf. Ref.~\cite{Bao_Altman_PRB2020, Lunt2020a, VanRegemortel2021a, Lumia2023, Richter2023}).
In the present work, we show that highly entangled states, with volume-law entanglement entropy, as well as volume-law mutual information, can be selectively generated between parts of the system, solely through \emph{local} two-site non-random one-body measurements in the system. 

Non-interacting systems, such as non-interacting fermions, when subjected to repeated local projective or generalized weak measurements along with their entangling unitary dynamics, are
not expected to give rise to steady states with the volume-law scaling of entanglement (which, for monitored fermionic models, is realized only in the presence of interactions~\cite{Poboiko2025, guo2024}).
In particular, in one-dimensional free-fermion chains with U(1) symmetry (charge conservation) \cite{CaoLuca, ChenLucas, Coppola2022, Carollo2022, Rossini2020, Diehl, AlbertonDiehl, Szyniszewski2022, Lumia2023}, the steady states is characterized by area-law entanglement \cite{Igor2, Starchl2024}, where the entanglement entropy $S_\text{EE}(\ell)$ of a subsystem of length $\ell$ varies, in the thermodynamic limit, as $S_\text{EE}(\ell) \propto\ell^{d-1}$ in $d$ spatial dimensions (i.e., is independent of $\ell$ for $d=1$). 
In higher dimensions $d>1$, there is a transition between the area-law phase and the area $\times$ log (``critical''~\cite{ChenLucas}) phase with $S_\text{EE}(\ell)\propto \ell^{d-1}\ln\ell$ \cite{Poboiko2024, Chahine2024}.
Recently, it was demonstrated that specific two-site measurements may induce a subextensive power-law scaling of entanglement in free-fermion models by inducing L\'evy flights of information, even in the measurement-only limit (this limit can be realized by sending the measurement rate to infinity)~\cite{Poboiko2025Levy}.

A related type of non-interacting models---monitored chains of Majorana fermions [lacking the U(1) symmetry]  \cite{Jian2023, Nahum, Graham2023, Merritt2023, Nehra2024}---represents a different symmetry class. In one dimension, these models show a phase transition between an area-law phase and a phase with $\ln^2 \ell$ scaling of the entanglement entropy \cite{Nahum}.
The measurement-only limit of the non-interacting fermionic systems typically hosts only area-law phases; for Majorana fermions, these phases are separated by a critical point with the scaling that is numerically consistent with the critical scaling $S_\text{EE}\propto \ln{\ell}$ \cite{Nehra2024}. 

Similar area-law phases separated by critical point have also been realized for a measurement-only projective transverse-field Ising model \cite{Nicolai_PRB} and other related models \cite{Qian_2024, Cheng_2024, Vu_PRL2024, Kuno_PRB2023}. The volume-law entangled phases and the volume-to-area-law transition have been obtained in higher dimensions \cite{Lavasani2021} or in subsystem error-correcting codes \cite{Benedikt_2024} with multi-site Pauli-operators measurements using the stabilizer formalism \cite{Ippoliti_Khemani_PRX}. In the stabilizer formalism, owing to the classical simulability of the stabilizer codes \cite{gottesman1998}, large systems can be accessed despite strong-entanglement generation.

It is intriguing to investigate whether one can move beyond multi-site and multi-body random projective measurements to a measurement-only dynamics with more local non-random one-body measurements capable of generating entanglement in the system, rather than hindering it. 
Therefore, understanding the potential for engineering local one-body measurements to induce an entangling phase (possibly characterized by the volume-law scaling of entanglement) due to the non-commutativity of such measurements is of particular interest. Additionally, extending beyond projective random Pauli measurements in spin systems or quantum circuit settings to a generalized ancilla-based \cite{Bao_Altman_PRB2020, Koh2022, Doggen2023, doggen2023ancilla} measurement framework in the usual quantum many-body setting could provide valuable insights and a wider range of experimental realizations.

In the present work, we explore an ancilla-based measurement setup in a system without intrinsic unitary dynamics, and show that volume-law entangled states can be generated through local, yet non-commuting measurements. In variance with most studies devoted to MIPTs, here, we use non-random (in space and time) measurements, similar to passive steering protocols \cite{Roy2020}. However, instead of steering towards a predesignated target state, we are concerned with `entanglement steering' that can be realized with a multitude of final states and even without stationary (dark) quantum states.

\begin{figure}[htb]
    \centering    \includegraphics[width=1.\linewidth]{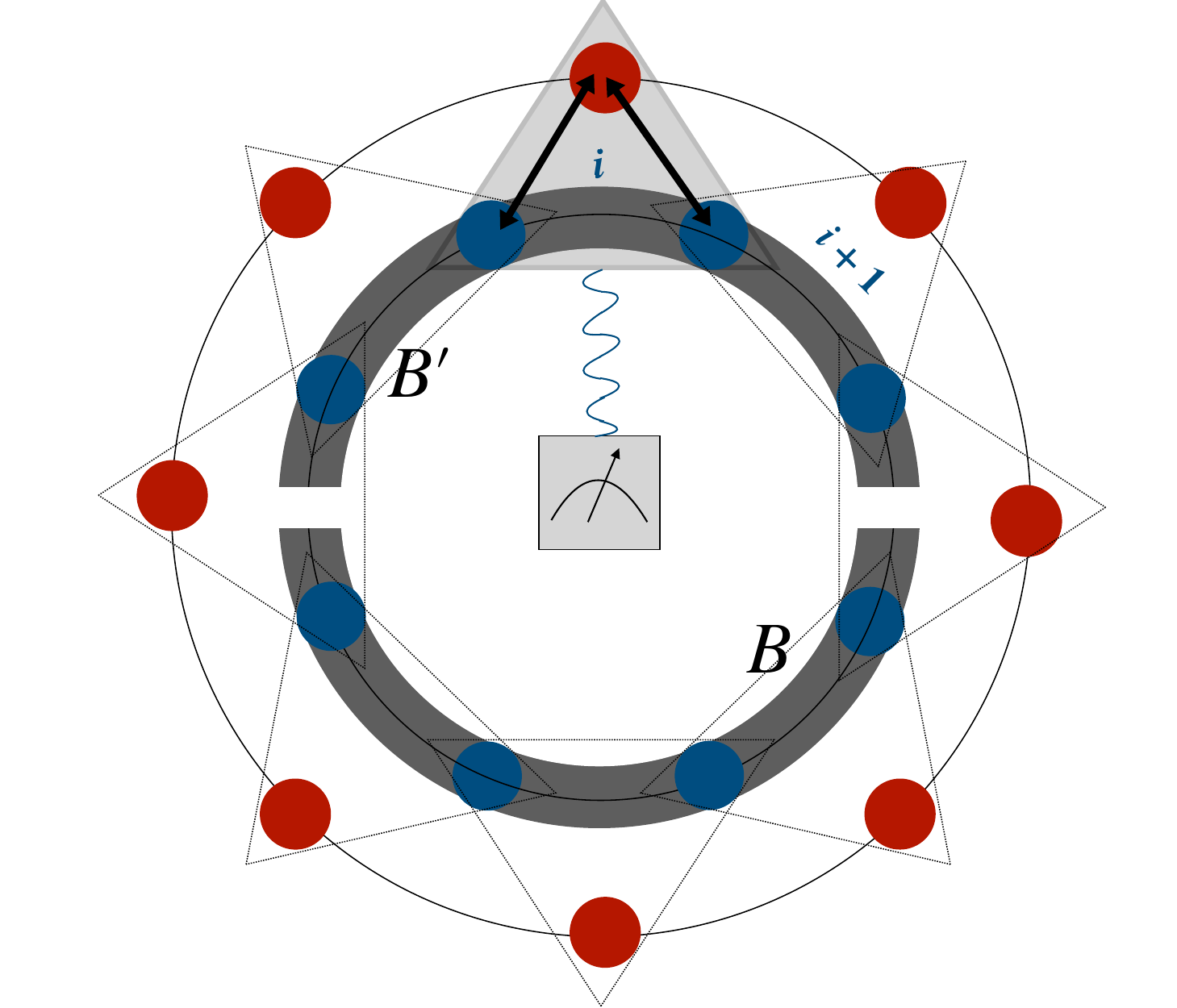}
    \caption{The system-detector setup for measurement only dynamics. Each triangular structure comprises two consecutive sites from the main chain (blue, inner chain) and one from the ancilla or auxiliary chain (red, outer chain), forming a `block'. Because of sharing a fermionic main-chain site, consecutive blocks do not commute. Triggered by the `local' coupling  (wavy lines) between a block (e.g., $i$-th block in shaded triangle) and the detector qubit (square) at the center, particles can hop between the main and ancilla chains (shown as double-headed arrows). There is no direct hopping among the main-chain sites. The subsystems $B$ and $B'$ (shaded half circle comprising half of main-chain sites) and their union $C = B \cup B'$ are used to compute entanglement after each measurement in the detector, which projects the qubit state, initialized in the ${|\!\uparrow\rangle}$ state, to ${|\!\uparrow\rangle}$ (no-click outcome) or  ${|\!\downarrow\rangle}$ (click outcome) state.
}
    \label{fig:mom_cartoon}
\end{figure}

To this end, as shown in Fig.~\ref{fig:mom_cartoon}, we discuss a `local' measurement scheme with fermions, employing a generalized measurement setup, where the system repeatedly interacts with a set of detectors or qubits with local coupling for an instantaneous time. This interaction results in entanglement generation between the system and the detectors. Subsequently, projective measurements are performed on the detectors, yielding either `no-click' (1 or $\uparrow$) or `click' (0 or $\downarrow$) outcomes. Our main results are the demonstration of entanglement generation through local measurement-only non-unitary dynamics and the realization of strongly-entangled long-time steady states and/or phases with volume-law entanglement entropy and volume/logarithmic mutual information among parts of the system. We show that such entanglement generation can be achieved using measurements of only one-body operators via local system-detector coupling. Moreover, we show that by making the model more non-local through higher-body system-detector couplings, one can introduce kinetic constraints to the dynamics and reduce the generation of entanglement.

To introduce `local' system-detector couplings, we consider a \emph{system} composed of a \emph{main} and an \emph{auxiliary (ancilla)} fermionic chain (see Fig.~\ref{fig:mom_cartoon}) coupled with a set of detector quibits. The entire setup is described by a Hamiltonian $$\mathcal{H}=\mathcal{H}_s+\mathcal{H}_d+\mathcal{H}_{sd}.$$ 
There is no intrinsic unitary dynamics of the system ($s$), or of the detectors, i.e., the system and detector Hamiltonians $\mathcal{H}_s=0$ and $\mathcal{H}_d=0$. Thus, the unitary evolution of the setup is governed by the system-detector (measurement) Hamiltonian $\mathcal{H}_{sd}$, which is followed by projective measurements of detector qubits, hence `measurement-only dynamics'. 

In the \emph{one-body measurement model}, the detector, prepared in the ${|\!\uparrow\rangle}$ state, is coupled instantaneously at discrete times with the system through quadratic hopping terms from the $i$-th and $i+1$-th sites of the main chain to the $i$-th site of the auxiliary chain. After the system-detector interaction, the detector qubit is measured projectively in the $\{|\!\uparrow\rangle,|\!\downarrow\rangle\}$ basis. The measurement protocol is sequentially swept clock wise through the entire system (Fig.~\ref{fig:mom_cartoon}) for all the detectors, constituting one \emph{measurement step} for the system. The measurement steps are repeated many times to reach a stationary state, or a stationary distribution of long-time states over many quantum trajectories. A quantum trajectory is obtained by keeping records of all the measurement outcomes across the entire system over all the measurement steps. 

\begin{table}[ht]
    \centering
     {\large \textbf{One-body measurement model}}\\
\setlength{\arrayrulewidth}{0.4mm}
\renewcommand{\arraystretch}{1.5}  
    \resizebox{0.495\textwidth}{!}{%
    \begin{tabular}{|c|c|c|c|c|c|c|c|c|}
        \hline
     \multicolumn{1}{|c|}{} &   \multicolumn{3}{c|}{Initial} &  \multicolumn{3}{c|}{Final} \\
        \hline
        state $\Psi_{\rm in}$\! & $S_B$ &  $S_C$ & $I_{BB'}$ &  $S_B$ &  $S_C$ & $I_{BB'}$ \\
        \hline
        \!$|\Psi_{p}\rangle$, $|\Psi_{\rm rp}\rangle$\!& 0 & 0 & 0 & \!{volume} & 0 & \!{volume} \\
        \hline
        $|\Psi_{s}\rangle$ & $\log L$ & $\log L$ & $\log L$  & $\log L$ & {area} & $\log L$  \\
        \hline
        $|\Psi_{\rm rs}\rangle$ & \!{volume} & \!{volume} & {area} & \!{volume} & {area} & \!{volume} \\
        \hline
    \end{tabular}}
\caption{ Scaling of the trajectory-averaged entanglement entropy and mutual information for the long-time distribution of quantum states for different initial states [see Eqs.~\eqref{eq:Psi-p}, \eqref{eq:Psi-rp}, \eqref{eq:Psi-rs}, and   \ref{eq:Psi-es}] (first column) and entanglement cuts (Fig.~\ref{fig:mom_cartoon}) in the \textit{one-body measurement model} [Eq.~\eqref{eqn:one-body-measurement-model}] as a function of system size $L$. The types of the scaling law are volume ($\propto L$), area ($\sim$ constant), and $\log{L}$. The scaling of the entanglement measures for the initial states is also indicated.   
}
\label{table:noninteracting}
\end{table}

As emphasized above, in this measurement setup, there is no direct hopping between the main chain sites. Fermions can only hop between the main- and ancilla-chain sites. The introduction of the ancilla chain allows us to maintain the desired local nature of the system-detector coupling. 
We show that the above measurement-only dynamics can indeed generate large entanglement between parts of the main chain. For example, mutual information between two halves of the main chain, scaling linearly or logarithmically with the length of the chain, is produced in the long-time limit, starting with initial states with area-law mutual information. The generation of large mutual information in the main chain implies that the dynamics effectively disentangles the auxiliary chain from the main chain, as expected due to the monogamy of entanglement~\cite{Zong_2022}. To illustrate the above points, we consider four different types of initial states for the system: (a) an atypical product state $|\Psi_p\rangle$, (b) a random product state $|\Psi_{rp}\rangle$, (c) an equal superposition of states in the basis of site occupation $|\Psi_s\rangle$, and (d) random superpositions of basis states $|\Psi_{rs}\rangle$. 

In Table~\ref{table:noninteracting}, we summarize our main results for the entanglement entropy between one-half of the main chain and the rest of the system, $S_B$, the entanglement entropy between the main and ancillary chains, $S_C$, and the mutual information, $I_{BB'}$, between two halves of the main chain (Fig.~\ref{fig:mom_cartoon}) for the long-time states obtained starting from the different initial states. 
Furthermore, we analyze the individual quantum trajectory and statistics of an ensemble of trajectories in the one-body measurement model, yielding the following results:
\begin{itemize}
\item[(1)] We find that individual quantum trajectories do not reach a stationary state. The entanglement entropy, e.g., $S_B$, continues to fluctuate substantially as a function of the measurement step even after many steps. 
\item[(2)]
Nevertheless, the distribution of entanglement entropy for an ensemble of
trajectories generated from a particular type of initial states, e.g, $|\Psi_p\rangle$ or $|\Psi_{rs}\rangle$, does converge to a stationary distribution, albeit different ones for different type of initial states.
\item[(3)] Thus, the stationary distributions exhibit a limited form of ergodicity, wherein the same long-time stationary distribution of entanglement measures is attained starting from different initial states within a particular type or set of initial states.
\end{itemize}

We further investigate the effects of measurement-induced interactions by introducing density-dependent hopping between the main and auxiliary chains in the system-detector Hamiltonian. We show that the specific density dependence of the hopping terms and the form of the system-detector coupling introduce kinetic constraints, such that the system can reach stationary states that no longer evolves under measurement-only dynamics.
This system-detector Hamiltonian, of course, introduces a direct density-density coupling between nearest-neighbor main chain sites, and thereby makes the measurement more non-local. 

Typically, direct interaction within the system is expected to make the system more entangled. However, as discussed later, in our measurement scheme, the introduction of interaction reduces the generation of entanglement. This reduction is due to the kinetic constraint on the accessible Hilbert space in the long-time limit. 
The system-size scaling of various trajectory-averaged entanglement measures after many measurement steps, starting with different sets of initial states, is summarized in Table~\ref{table:interacting}. 
The analysis of individual quantum trajectories and statistics of an ensemble of trajectories provides the following insights:


\begin{itemize}
\item[(1)] Each quantum trajectory leads to a single non-equilibrium stationary quantum state at long time. 
\item[(2)] However, different trajectories, even when starting from the same type of initial states, may converge to different stationary quantum states, leading to a distribution of stationary states. The distribution is sharply peaked and consists of a discrete set of stationary states, as opposed to a continuum of final states.
\item[(3)] The Born probability of the quantum trajectory corresponding to the most probable quantum states prevails over the other quantum trajectories. Remarkably, this most probable quantum trajectory is a result of almost all `no-click' measurement outcomes, except for a few at the beginning. Thus, the measurement-only dynamics naturally steers towards a no-click trajectory describable by a purely non-stochastic non-Hermitian Hamiltonian dynamics \cite{XhekMarco, Silva, Wang_2024}. 
\end{itemize}

\begin{table}[ht]
    \centering
    {\large \textbf{Three-body measurement model}}\\
\setlength{\arrayrulewidth}{0.4mm}
\renewcommand{\arraystretch}{1.5}  
    \resizebox{0.495\textwidth}{!}{%
   \begin{tabular}{|c|c|c|c|c|c|c|c|c|c|}
        \hline
     \multicolumn{1}{|c|}{} &   \multicolumn{3}{c|}{Initial} &  \multicolumn{3}{c|}{Final} \\
        \hline
        state $\Psi_{\rm in}$ & $S_B$ &  $S_C$ & $I_{BB'}$ &  $S_B$ & $S_C$ & $I_{BB'}$ \\
        \hline
        \!$|\Psi_{p}\rangle$, $|\Psi_{\rm rp}\rangle$\! & 0 & 0 & 0 & {area} & 0 & {area} \\
        \hline
        $|\Psi_{s}\rangle$ & $\log L$ & $\log L$ & $\log L$ & $\log L$ & $\log L$ & {area} \\
        \hline
        $|\Psi_{\rm rs}\rangle$ & \!{volume} & \!{volume} & {area}  & \!{volume} & \!{volume} & {area} \\
        \hline
    \end{tabular}
    }
    \caption{ Scaling of the trajectory averaged entanglement entropy and mutual information for the long-time distribution of quantum states for different initial states [see Eqs.~\eqref{eq:Psi-p}, \eqref{eq:Psi-rp}, \eqref{eq:Psi-rs}, and \eqref{eq:Psi-es}] (first column) and entanglement cuts (Fig.~\ref{fig:mom_cartoon}) in the \textit{three-body measurement model} [Eq.~\eqref{eqn:HsdInt}]  as a function of system size $L$. The types of the scaling law are volume ($\propto L$), area ($\sim$ constant), and $\log{L}$. The scaling of the entanglement measures for the initial states is also indicated. 
     }
    \label{table:interacting}
\end{table}

Our paper is organized as follows. We first introduce the one-body measurement and three-body measurement models, measurement protocols, initial states, Kraus operators, the role of auxiliary chain in measurement dynamics and various entanglement measures utilized in this work in Section \ref{sec:Models_meas_protocol}. 
We then discuss evolution of entanglement along the quantum trajectories, associated statistical properties over an ensemble of trajectories, and dependence on initial states under one-body measurement and three-body measurement models in Section \ref{sec:StatOnebodyMeasurement} and Section \ref{sec:StatThreebodyMeasurement} respectively. Our main results on the generation of entanglement in the long-time states and the system-size scaling of various entanglement measures are discussed in Section \ref{sec:EntGen}. Finally, we conclude with a discussion and outlook in Sections \ref{sec:Discussion}. We provide additional details in the Appendices.

\section{Measurement-only models}\label{sec:Models_meas_protocol}

In this Section, we introduce the system-detector Hamiltonian and the measurement protocol, and illustrate the elementary steps of the measurement-only dynamics for the generalized measurements of one-body and three-body operators. We also discuss various initial states, the role of the auxiliary chain in measurement dynamics, and entanglement measures used to characterize the ensuing quantum states.

\subsection{One-body measurement model} \label{sec:NonIntModel}

\subsubsection{Model: system-detector coupling}

We begin with the one-body measurement model.
As shown in Fig.~\ref{fig:mom_cartoon}, we consider a fermionic system consisting of the main chain with fermionic operators $\{c_i,c_i^\dagger\}$  on its sites ($i=1,\ldots, L/2$; $L$ even) and the ancilla chain with operators $\{a_i,a_i^\dagger\}$. Generalized measurements are performed on the system by coupling the system at discrete times, sequentially from $i=1$ to $i=L/2$ with the detector qubit (spin $1/2$) described by Pauli operators $\{\boldsymbol{\sigma}=(\sigma^x,\sigma^y,\sigma^z)\}$ through the following system-detector Hamiltonian:
\begin{subequations}\label{eqn:one-body-measurement-model}
\begin{align}
\mathcal{H}_{sd}(t)&=\sum_{i,m}H_{sd}^{(i,i+1)}\delta(t-t_{i,m}), \label{eq:Hsdt}\\
  H^{(i, i+1)}_{sd}&= \alpha\,(c^{\dagger}_i+c^{\dagger}_{i+1})\,a_i \,\sigma^{x} + \text{H.c.} \label{eq:HsdNonInt}
\end{align}
\end{subequations}
Here, $t_{i,m}$'s are (arbitrary) discrete times satisfying $0<t_{i,m}<t_{i+1,m}$ and $t_{i,m}<t_{j,m+1}$ for all $i,j=1,\ldots, L/2$, and the integer $m$ denotes one measurement step corresponding to a full sequential clock-wise sweep of measurements across the entire system, as shown in Fig.~\ref{fig:mom_cartoon}. All the measurements in the $m$-th step are done between times $t_{m-1}$ and $t_{m}$, i.e., $t_{m-1}<t_{i,m}<t_m$ for all $i$.

The time dependence of Hamiltonian \eqref{eq:Hsdt} corresponds to the instantaneous coupling between the system chain and detectors (as described by the delta-functions of time). Realistically, the system-detector coupling takes a finite time $\tau$. However, since we are interested in the state of the system between the coupling events (when the total Hamiltonian is zero), and the system-detector coupling is time-independent during the interaction time $\tau$, we reduce the coupling in Eq.~\eqref{eq:Hsdt} to the delta-function form. The strength of the system-detector coupling is represented by the dimensionless constant $\alpha$
(which, for finite $\tau$, would be proportional to $\tau$). 

We apply a periodic boundary condition, where the $L/2$-th detector couples the $L/2$-th and the first ($i=1\equiv L/2+1$) sites. Before each measurement of the detector, it is prepared in the state ${|\!\uparrow\rangle}$ at times $t_{i,m}^-$, infinitesimally before the system-detector interaction times. The $z$ component $S^z$ of the detector qubit is measured projectively at times $t_{i,m}^+$. The entire process of detector state preparation, system-detector interaction, and the subsequent projective measurement on the detector constitutes a single generalized measurement. 
As evident from the form of $H^{(i,i+1)}_{sd}$ in Eq.~\eqref{eq:HsdNonInt}, measurements in two neighboring blocks, in general, do not commute since the blocks share a common fermionic site.

Essentially, the operator $$\hat{\mathcal{M}}_{i,i+1} = (c_i^\dagger a_i + \mathrm{H.c.}) + (c_{i+1}^\dagger a_i + \mathrm{H.c.}),$$ which is a sum of two one-body hopping terms between the main and ancilla chains of the system, is weakly measured in this process. It is important to note that, unlike local occupation or parity measurements~\cite{AlbertonDiehl,CaoLuca,ChenLucas,Poboiko2024,Igor2,Dganit2024,Merritt2023,Nehra2024}, this type of one-body measurement dynamics does not preserve the Gaussianity of the pure states. Indeed, in our case, the measurement operator is such that $\mathcal{M}^2 \neq \mathcal{M}$ and $\mathcal{M}^2 \neq \mathbf{I}$, which breaks down the Gaussianity~ \cite{Nehra2024,Lumia2023}.

In the above measurement protocol, we do not have any intrinsic Hamiltonian for either the main or the auxiliary chain. Thus, the system does not possess any unitary dynamics on its own. To make the system-detector coupling local in the main chain, we do not have any term that couples two main-chain sites. The coupling of the form $\sim (n_i+n_{i+1}) \sigma^x_i$ is trivial and cannot entangle two main chain sites. Thus, the auxiliary chain is introduced to incorporate a minimal quadratic-in-fermion term that can generate entanglement while preserving fermion parity and the total fermion number in the combined system of the main and auxiliary chains.

\subsubsection{Blockwise unitary evolution} \label{sec:Blockwise-evol_non-int}

Each measurement acts on the system in a blockwise fashion, as shown in Fig.~\ref{fig:mom_cartoon}. Here, we discuss how the instantaneous switching on of the system-detector Hamiltonian acts in the computational basis states in the $i$-th block. The basis states are spanned by the occupation numbers $n^c$ of the main chain sites, $n^a$ of the ancilla site, and $S^z$ of the detector spin in the $i$-th block, which consists of two main chain sites, $i$ and $i+1$, one auxiliary site $i$, and the detector. The computational basis state in the $i$-th block is given by 
$$|n\rangle_i \otimes |S^{z}_i\rangle = |n^{c}_i n^{c}_{i+1}\rangle \otimes |n^{a}_i\rangle \otimes |S^{z}\rangle.$$ 
Here, $|n\rangle_i=|n^{c}_i n^{c}_{i+1}\rangle \otimes |n^{a}_i\rangle$ represents the basis states for the system in the block, where there are $n = 1,\ldots, 2^3$ states in the local Hilbert space of the block.


The unitary operator $\mathcal{U}_i$ corresponding to the instantaneous switching on  $H^{(i,i+1)}_{sd}$ in the $m$-th measurement step is given by
\begin{align}
    \mathcal{U}_i &= \mathcal{T}_t\exp\left[-\ci \int_{t_{i,m}^-}^{t_{i,m}^+}\! dt \, H^{(i,i+1)}_{sd} \delta(t-t_{i,m})\right]\! = e^{-\ci H^{(i,i+1)}_{sd}}\!.
    \label{eq:UnitaryOp}
\end{align}
Here, $\mathcal{T}_t$ denotes time ordering. The action of the unitary operator $\mathcal{U}_i$ on the basis state in the $i$-th block is obtained by expanding $\exp[{-\ci H^{(i,i+1)_{sd}}}]$ and regrouping the terms, as discussed in detail in {Appendix~\ref{sup_sec:nonintUact}}. The particle number $N_p$ in the block is conserved under the unitary $\mathcal{U}_i$, and hence we can determine the action of the unitary operator $\mathcal{U}_i$ separately on the basis states in each particle sector. As already mentioned, the detector is initially prepared in the up-spin state $|S^{z}\rangle = |\!\uparrow\rangle$. We illustrate below the action of $\mathcal{U}_i$ on each particle number sector.

{\it $N_p=1$ sector:}
In this sector, we get
\begin{subequations}\label{eq:UNp1}
\begin{align}
\mathcal{U}_i|00\rangle |1\rangle  |\!\uparrow\rangle &= \cos{\x}|00\rangle|1\rangle |\!\uparrow \rangle  \notag\\
&-\frac{i\sin \x}{\sqrt{2}} \Big(|01\rangle  + |10\rangle \Big) |0\rangle |\!\downarrow\rangle,  \label{eqn:nonint_Ua}\\
\mathcal{U}_i |01\rangle|0\rangle|\!\uparrow\rangle &= \bigg({\frac{\cos \x+1}{2}|01\rangle} + \frac{\cos \x-1}{2}|10\rangle\bigg)|0 \rangle |\!\uparrow \rangle \label{eqn:nonint_Ub}
\notag \\ &
- \frac{i\sin{\x}}{\sqrt{2}} |00\rangle|1\rangle |\!\downarrow\rangle, 
\\
\mathcal{U}_i |10\rangle|0\rangle|\!\uparrow\rangle &= \bigg({\frac{\cos \x+1}{2}|10\rangle} + \frac{\cos \x-1}{2}|01\rangle\bigg)|0 \rangle |\!\uparrow \rangle  \label{eqn:nonint_Uc}
\notag \\ &
- \frac{i\sin{\x}}{\sqrt{2}} |00\rangle|1\rangle |\!\downarrow\rangle. 
\end{align}
\end{subequations}
Here, $\tilde{\alpha}=\sqrt{2}\alpha$ and the action of the unitary on all the basis states is periodic in $\tilde{\alpha}$ with periodicity $2\pi$. 


{\it $N_p=2$ sector:}
We can get the action of the $\mathcal{U}_i$ in $N_p=2$ sector by replacing 1 by 0 or 0 by 1 in Eqs.~\eqref{eq:UNp1}. This is because of the  particle-hole symmetry of the Hamiltonian. We can also explicitly find it by expanding  $\exp\big[{-\ci H^{(i,i+1)}_{sd}}\big]$, as discussed in {Appendix~\ref{sup_sec:nonintUact}}. 


{\it $N_p=0$ and $N_p=3$ sectors: } In these sectors, the action of the unitary operator is trivially an identity operation.

We find that the action of instantaneous switching on and off the system-detector interaction yields entanglement between the system and the detector. When we perform a measurement on the detector, we either obtain a `no-click' ($\uparrow$) or `click' ($\downarrow$) outcome according to the Born probability. An important observation is that an entangled state between two sites of the main chain is generated in both the click and no-click outcomes, depending on the basis states [Eqs.~\eqref{eq:UNp1}]. Thus, we are able to engineer a system-detector coupling whose microscopic action on basis states within a block can generate either an entangled or a disentangled state between two main chain sites. 

The key question is whether an entangled many-body state can be obtained in a large system by repeatedly switching on the system-detector interaction on different blocks and subsequent measurements. We find that blockwise action and repetitive measurement protocol can lead to strongly entangled long-time states even for measurements of two-site one-body operators, with mutual information within the main chain scaling linearly with $L$ (volume law) or as $\log{L}$, as summarized in Table~\ref{table:noninteracting} and discussed in detail in Sec.~\ref{sec:StatOnebodyMeasurement}.

\subsection{Three-body measurement model}\label{sec:IntModel}

\subsubsection{Model: system-detector coupling} 
For three-body measurements, we have the same setup of Fig.~\ref{fig:mom_cartoon} and an instantaneous system-detector Hamiltonian as in Eq.~\eqref{eq:Hsdt},  with direct density-density interaction in the detector-induced hopping between the main and auxiliary chains. Namely, $H^{(i,i+1)}_{sd}$ of Eq.~\eqref{eq:HsdNonInt} is replaced with
\begin{align}\label{eqn:HsdInt}
  \tilde{H}^{(i, i+1)}_{sd}\! &=  \!\Big( \alpha\big[(c^{\dagger}_i+c^{\dagger}_{i+1})\,a_i\,(1-n^c_i) (1-n^c_{i+1}) \big]\sigma^{-}\! +\!\text{H.c.}\! \Big) \notag\\& + 
  \Big(\alpha\big[a^{\dagger}_i\,(c_i+c_{i+1})\,n^c_i n^c_{i+1} \big]\sigma^{-}\! +\! \text{H.c.} \Big). 
\end{align}
In the above expression, $n^c_i=c^{\dagger}_i c_i$ is the number operator for site $i$ of the main chain and $\sigma^{\pm}$ is the spin-raising or lowering operator for the detector qubit. As we show below, the system-detector coupling \eqref{eqn:HsdInt} produces dynamics only when the main-chain sites $i$ and $i+1$ are empty (occupied) and the ancilla site $i$ is occupied (empty). This corresponds to severe kinetic constraints on the measurement-only dynamics.

\subsubsection{Blockwise unitary evolution}\label{sec:Blockwise-evol_int}

As in the case of one-body measurements, we obtain the action of the unitary operator $\mathcal{U}_i = \exp\big[{-\ci \tilde{H}^{(i,i+1)}_{sd}}\big]$ in the computational basis state, i.e., on $|n\rangle_i \otimes |S^z\rangle$ for $i$-th block.
We demonstrate the action of $\mathcal{U}_i$ in all the particle sectors below and, with more details, in {Appendix~\ref{sup_sec:intUact}}.

{\it $N_p=1$ sector:} In this case,
\begin{subequations}\label{eq:UNp1_Int}
\begin{align}
\mathcal{U}_i|00\rangle |1\rangle  |\!\uparrow\rangle &= \cos{\x}|00\rangle|1\rangle |\!\uparrow \rangle  \notag\\
&-\frac{i\sin \x}{\sqrt{2}} \Big(|01\rangle  + |10\rangle  \Big)|0\rangle |\!\downarrow\rangle, 
\label{eqn:int_Ua}
\\
\mathcal{U}_i |01\rangle|0\rangle|\!\uparrow\rangle &= |01\rangle |0\rangle|\!\uparrow \rangle, \label{eqn:int_Ub}
\\
 \mathcal{U}_i |10\rangle |0\rangle  |\!\uparrow\rangle &= |10\rangle |0\rangle  |\!\uparrow \rangle. 
 \label{eqn:int_Uc}
\end{align}
\end{subequations}
As earlier, the action of the unitary is $2\pi$-periodic in $\x = \sqrt{2}\alpha$. 
Since the probabilities of no-click ($\uparrow$) and click ($\downarrow$) outcomes are $P_{\uparrow} = \cos^2 \x$ and $P_{\downarrow} = \sin^2 \x$, respectively, for the evolution of the $|00\rangle|1\rangle$ state in Eq.~\eqref{eqn:int_Ua}, we further observe periodicity in $\x$ between $0$ and  $\pi$. For $\x=\pi/4$, the spin-up and spin-down read-outs are equally probable. 

{\it $N_p=2$ sector:}
Since the system-detector Hamiltonian \eqref{eqn:HsdInt} exhibits the particle-hole symmetry for each block, the action of the unitary on basis states in the $N_p=2$ sector is obtained by interchanging 1 and 0 in Eqs.~\eqref{eq:UNp1_Int} for the $N_p=1$ sector. 

{\it $N_p=0$ and $N_p=3$ sectors:} In these sectors, the action of the unitary is trivially identity operation, i.e.,  $\mathcal{U}_i=I$. 

We find that the unitary operator acts on the basis state in the $i$-th block non-trivially only when the two main chain sites $(i, i+1)$ are either both filled or both empty, i.e., only the basis states $|00\rangle|1\rangle$ or $|11\rangle|0\rangle$ evolve under the measurement dynamics. During the measurement process, a click ($\downarrow$) outcome in the $i$-th block results in the entangled pair $(|01\rangle + |10\rangle)$ between the main chain sites $i$ and $i+1$. Once we obtain this entangled pair, further action of the unitary $\mathcal{U}_i$ on this block amounts to a trivial identity operation, as shown in Eqs.~\eqref{eqn:int_Ub} and \eqref{eqn:int_Uc} for the $N_p=1$ sector, and equivalently in the $N_p=2$ sector. This is in sharp contrast to the non-interacting case, where all the basis states in the $N_p=1$ or $N_p=2$ sectors non-trivially evolve [Eqs.~\eqref{eq:UNp1}] under measurements.


\subsection{Initial states and measurement dynamics}\label{sec:measurementprotocol}

We consider a system of size $L$ (even) with $L/2$ sites in each of the main and auxiliary chains.
We represent the (pure) quantum state of the system as $|\Psi(t_{i,m}^-)\rangle$ and the combined state of the system and the detector as $|\Phi_i(t_{i,m}^-) \rangle$ just before the the $m$-th measurement in the $i$-th block at discrete times $t_{i,m}$, as described in Sec.~\ref{sec:NonIntModel}. The detector is prepared in the state $|\!\uparrow\rangle_i$ before the measurement. Thus, 
\begin{align}\label{eq:StateBeforeMeas}
   |\Phi_i(t_{i,m}^-) \rangle = |\Psi(t_{i,m}^-)\rangle \otimes |\!\uparrow\rangle.
\end{align}
In particular, initially at time $t=0$, before any measurements are done, 
we prepare the system in either a generic entangled state or an unentangled product state $|\Psi(0)\rangle $ in the computational basis 
$$|n\rangle = |n^c_1\cdots n^c_in^c_{i+1}\cdots n^c_{L/2}\rangle \otimes |n^a_1\cdots n^a_i\cdots n^a_{L/2}\rangle$$ 
for {total $L/2$ particles, or the half-filled sector of the combined system of the main and auxiliary chains}. Specifically, we consider four types of initial states: 
\begin{enumerate}
\item An atypical product state where all the main chain sites are filled, i.e., 
\begin{equation}
|\Psi(0)\rangle=|\Psi_p\rangle = |111...\rangle \otimes |000...\rangle;
\label{eq:Psi-p}
\end{equation} 
\item a generic random product state \begin{equation} 
|\Psi(0)\rangle=|\Psi_{rp}\rangle= |n^c_0 n^c_1\cdots\rangle \otimes |n^a_0 n^a_1\cdots\rangle,
\label{eq:Psi-rp}
\end{equation}
where $n^{c/a}$ are randomly chosen between 1 or 0 with the constraint of total $L/2$ fermions; 
\item a random superposition state 
\begin{equation}
|\Psi(0)\rangle=|\Psi_{rs}\rangle= \sum^D_{n=1} \beta_n |n\rangle, 
\label{eq:Psi-rs}
\end{equation}
where real coefficients $\beta_n$'s are drawn from a uniform random distribution with the normalization constraint $\sum_n\beta_n^2=1$; 
\item an `equal superposition' 
\begin{equation}
 |\Psi(0)\rangle=|\Psi_{s}\rangle= \frac{1}{\sqrt{D}} \sum_n |n\rangle;
 \label{eq:Psi-es}
\end{equation} 
\end{enumerate}
Here, $D=2^L$ is the Hilbert-space dimension of the system.

After switching on the instantaneous interaction in the $i$-th block, the state in Eq.~\eqref{eq:StateBeforeMeas} evolves to
\begin{align}
|\Phi (t_{i,m}) \rangle &= \mathcal{U}_i |\Phi(t_{i,m}^-)\rangle \notag\\ 
&=|\Psi_\uparrow(t_{i,m})\rangle \otimes |\!\uparrow\rangle + |\Psi_\downarrow(t_{i,m})\rangle \otimes |\!\downarrow\rangle.
\end{align}
The system states $|\Psi_\sigma(t_{i,m})\rangle$ (here, $\sigma=\uparrow,\downarrow$) can be obtained from the action of the unitary operator $\mathcal{U}_i$ on the basis states, $|n\rangle_i \otimes |S^z_i\rangle$, as discussed in Secs.~\ref{sec:NonIntModel} and \ref{sec:IntModel}. 
Subsequently, the projective measurement on the $i$-th detector produces the $\sigma=\uparrow,\downarrow$ outcome for the detector with probability $P_\sigma(t_{i,m})=\langle\Psi_\sigma(t_{i,m})|\Psi_\sigma(t_{i,m})\rangle$, and collapses the state of the system just after measurement at $t_{i,m}^+$ to 
\begin{align}
    |\Psi(t_{i,m}^+)\rangle &= \frac{|\Psi_\sigma(t_{i,m})\rangle}{\sqrt{P_{\sigma}(t_{i,m})}}.
\end{align}

The above generalized measurement protocol is repeated in a blockwise fashion sequentially across the system (Fig.~\ref{fig:mom_cartoon}) for the $m$-th measurement step on the entire system between times $t_{m-1}$ and $t_m$. Many such measurement steps are performed starting at time $t_0=0$ from the four types of initial states specified above. The full record of the measurement outcomes 
$\{\sigma_{i,m}\}$ constitutes a quantum trajectory, where $\sigma_{i,m}=\uparrow,\downarrow$ is the read-out registered by the detector qubit for the $i$-th block in the $m$-th measurement step.


\subsection{Kraus operators}

The time evolution of the state of the system due to the generalized measurement in the $i$-th block for the $m$-th measurement step with the detector outcome $\sigma$ can be described compactly using the Kraus operators, namely,
\begin{align} |\Psi(t_{i,m}^+)\rangle=\frac{K_{i\sigma}|\Psi(t_{i,m}^-)\rangle}{\|K_{i\sigma}|\Psi(t_{i,m}^-)\rangle\|},  
\end{align}
where the Kraus operator 
$$K_{i\sigma}=  \oplus_{\mathcal{N}=0,1,2,3} K^{(N_p=\mathcal{N})}_{\sigma} $$ 
can be decomposed into a block-diagonal form in the different particle number sectors of the $i$-th block, like the unitary $\mathcal{U}_i$.

\subsubsection{Kraus operators for one-body measurement model}

For the one-body measurement model of Sec.~\ref{sec:NonIntModel}, in the $N_p=1$ particle sector, the Kraus operators take the form:
\begin{subequations}\label{eq:KrausOneBody_1}
\begin{align}
    K^{(N_p=1)}_{i\uparrow} &= \begin{pmatrix}
        \cos \tilde{\alpha}& 0 & 0\\
        0 & \dfrac{\cos \tilde{\alpha} + 1}{2} & \dfrac{\cos \tilde{\alpha}-1}{2} \\ 
        0 &\dfrac{\cos \tilde{\alpha} - 1}{2}& \dfrac{\cos \tilde{\alpha} + 1}{2}
    \end{pmatrix}, \\[0.2cm]
    K^{(N_p=1)}_{i\downarrow} &= \begin{pmatrix}
       0& -\dfrac{ \ci\sin \tilde{\alpha}}{\sqrt{2}} &  -\dfrac{ \ci\sin \tilde{\alpha}}{\sqrt{2}}\\
         -\dfrac{\ci\sin \tilde{\alpha}}{\sqrt{2}} & 0 &0\\ 
         -\dfrac{\ci\sin \tilde{\alpha}}{\sqrt{2}} & 0 & 0
    \end{pmatrix} .
\end{align}
\end{subequations}
These Kraus operators obey the completeness relation: $$\sum_{\sigma}K^{\dagger}_{i\sigma}K_{i\sigma}=I_{3\times 3}.$$ 
Similarly, the Kraus operators for the $N_p=2$ sector can be written as
\begin{subequations}\label{eq:KrausOneBody_2}
\begin{align}
    K^{(N_p=2)}_{i\uparrow} &= \begin{pmatrix}
        \dfrac{\cos \tilde{\alpha} + 1}{2} & \dfrac{\cos \tilde{\alpha}-1}{2} & 0\\ 
        \dfrac{\cos \tilde{\alpha} - 1}{2}& \dfrac{\cos \tilde{\alpha} + 1}{2} & 0 \\
        0 & 0 & \cos{\tilde{\alpha}}
    \end{pmatrix}, \\[0.2cm]
    K^{(N_p=2)}_{i\downarrow} &= \begin{pmatrix}
       0& 0 &  -\dfrac{\ci\sin \tilde{\alpha}}{\sqrt{2}}\\
        0 & 0 & -\dfrac{\ci\sin \tilde{\alpha}}{\sqrt{2}} \\ 
         -\dfrac{\ci\sin \tilde{\alpha}}{\sqrt{2}} & -\dfrac{\ci\sin \tilde{\alpha}}{\sqrt{2}} & 0
    \end{pmatrix}.
\end{align}
\end{subequations}

In the $N_p=0$ and $N_p=3$ sectors, the Kraus operator is trivially the identity operator $K_{i\uparrow}^{(N_p=0,3)}=I_{1\times 1}$ with definite outcome $\sigma=\uparrow$, since the detector is always prepared in the $|\uparrow\rangle$ state before the measurement and the system-detector unitary $\mathcal{U}_i=I_{1\times 1}$. It is important to note that the Kraus operators in our case are not projectors, i.e., they do not satisfy $K^2_{i\sigma}= K_{i\sigma}$ for generic value of $\alpha\neq 0$.

\subsubsection{Kraus operators for three-body measurement model}

For the three-body measurement model of Sec.~\ref{sec:IntModel}, in the $N_p=1$ particle sector, we have
\begin{align}
    K^{(N_p=1)}_{i\uparrow} &= \begin{pmatrix}
        \cos \tilde{\alpha}& 0 & 0\\
        0 & 1 & 0 \\ 
        0 & 0 & 1 
    \end{pmatrix}, \\[0.2cm]
    K^{(N_p=1)}_{i\downarrow} &= \begin{pmatrix}
       0 & 0 & 0\ \\
         -\dfrac{\ci\sin \tilde{\alpha}}{\sqrt{2}} & 0 & 0\ \\ 
         -\dfrac{\ci\sin \tilde{\alpha}}{\sqrt{2}} & 0 & 0\
    \end{pmatrix}. 
\end{align}
Similarly, the Kraus operators for the $N_p=2$ sector can be written as
\begin{align}
    K^{(N_p=2)}_{i\uparrow} &= \begin{pmatrix}
        1 & 0 & 0 \\ 
        0 & 1 & 0 \\
        0 & 0 & \cos{\tilde{\alpha}}
    \end{pmatrix}, \\[0.2cm]
    K^{(N_p=2)}_{i\downarrow} &= \begin{pmatrix}
       0 & 0 &  -\dfrac{ \ci\sin \tilde{\alpha}}{\sqrt{2}}\\
        0 & 0 & -\dfrac{ \ci\sin \tilde{\alpha}}{\sqrt{2}} \\ 
         0 & 0 & 0
    \end{pmatrix}.
\end{align}
In the $N_p=0$ and $N_p=3$ sectors, the Kraus operator is trivially the identity operator $K_{i\uparrow}^{(N_p=0,3)}=I_{1\times 1}$ with the only outcome $\sigma=\uparrow$.

\subsubsection{Comparison with projective measurement}
We can compare the above generalized measurements with the corresponding projective measurements. In the one-body measurement model of Sec.~\ref{sec:NonIntModel}, the operator $\hat{\mathcal{M}}_{i,i+1}=(c_i^\dagger a_i+\mathrm{H.c.})+(c_{i+1}^\dagger a_i+\mathrm{H.c})$ is weakly measured leading to the Kraus operators of Eqs.~\eqref{eq:KrausOneBody_1} and \eqref{eq:KrausOneBody_2} depending on the detector outcome. Instead, in a projective measurement of $\hat{\mathcal{M}}_{i,i+1}$, the system operator in the $i$-th block yields one of its eigenvalues $\mathcal{M}=0,\pm \sqrt{2}$ as the measurement outcome. The evolution of the system can again be represented by the Kraus operators for the measurement on the $i$-th block in the $m$-th step.
The Kraus operators, e.g., for $N_p=1$ particle sector, are then given by the projectors $|\mathcal{M}\rangle \langle \mathcal{M}|$ into the eigenstates of $\hat{\mathcal{M}}_i$, i.e.,
\begin{subequations}
\begin{align}
    K^{(N_p=1)}_{i,0} &= \begin{pmatrix}
        0 & 0 & 0\\ 
        0& \frac{1}{2} & -\frac{1}{2} \\
        0 & -\frac{1}{2} & \frac{1}{2}
    \end{pmatrix}, 
    \label{proj-1}\\
   K^{(N_p=1)}_{i,\pm \sqrt{2}} &= \begin{pmatrix}
        \frac{1}{2} & \pm \frac{1}{2\sqrt{2}} & \pm \frac{1}{2\sqrt{2}}\\ 
        \pm \frac{1}{2\sqrt{2}}& 1 & 1 \\
        \pm \frac{1}{2\sqrt{2}} & 1 & 1
    \end{pmatrix}. 
    \label{proj-2}
\end{align}
\end{subequations}
These Kraus operators again satisfy $\sum_{\mathcal{M}} K_{\mathcal{M}}^\dagger K_{\mathcal{M}}=I_{3\times 3}$. By comparing Eqs.~\eqref{proj-1} and \eqref{proj-2} with the corresponding $N_p=1$ Kraus operators in Eqs.~\eqref{eq:KrausOneBody_1} for the generalized measurement model, we see that generalized measurements do not reduce to the projective measurement for any value of $\tilde{\alpha}$. However, the no-click limit of the quantum trajectories for $\tilde{\alpha}=\pi/2$, where all the generalized measurement outcomes for the detector qubits are $\sigma=\uparrow$, is the same as the no-click limit of the projective measurement trajectories. In the latter case, the no-click limit corresponds to $\mathcal{M}=0$ for all the outcomes of measurements of the system operators $\hat{\mathcal{M}}_{i,i+1}$.

\subsection{Role of the auxiliary chain in measurement dynamics}
\label{sec:Role-aux-chain}

The blockwise unitary evolution generated by the system-detector Hamiltonian, as described in Secs.~\ref{sec:Blockwise-evol_non-int} and \ref{sec:Blockwise-evol_int}, yields an entangled pair between two consecutive main-chain sites, depending on the measurement outcome (click or no-click) and the particle configuration $(|n^{c}_i n^{c}_{i+1}\rangle \otimes |n^{a}_i\rangle)$ of the basis state in the block. Crucially, the ancilla site (Fig.~\ref{fig:mom_cartoon}) never entangles with any main-chain site when the measurements are performed on the occupation basis states $|n\rangle$. 
Moreover, as we demonstrate below, the ancilla chain also remains unentangled under the measurement dynamics starting from $|\Psi\rangle=|\psi_{\rm main}\rangle\otimes |\psi_{\rm aux}\rangle$, where the initial state of the main chain $|\psi_{\rm main}\rangle$ is arbitrary, but the  the ancilla chain is in a product state $|\psi_{\rm aux}\rangle=|n^a_1\dots n^a_i \dots n^a_{L/2}\rangle$. On the contrary, the ancilla chain remains entangled with the main chain via the measurement dynamics starting from a generic superposition state $|\Psi\rangle\neq |\psi_{\rm main}\rangle\otimes |\psi_{\rm aux}\rangle$. 

First, we consider the evolution of a generic initial main-auxiliary product state ${|\Psi\rangle=|\psi_{\rm main}\rangle\otimes |\psi_{\rm aux}\rangle}$. Written explicitly in the occupation basis, this state takes the form:
\begin{equation}\label{eq:generic-psi-prod}
|\Psi\rangle=\Big(\sum_{k} \beta_k|b_k\rangle\Big) \otimes |n^a_1\cdots n^a_i \cdots n^a_{L/2} \rangle,
\end{equation}
where $|b_k\rangle=|n^c_1\cdots n^c_i n^c_{i+1}\cdots n^c_{L/2}\rangle$ for $k=1,\ldots, 2^{L/2}$ represents different particle configurations for the main chain, and $\beta_k$ are real or complex coefficients. The quantum states in Eqs.~\eqref{eq:Psi-p} and \eqref{eq:Psi-rp} are special cases of Eq.~\eqref{eq:generic-psi-prod}, with only a single nonzero coefficient $\beta_n=1$. 

Below, we discuss the action of the $i$-th block in the $m$-th measurement step on the initial system-detector state:
\begin{equation}
|\Phi_i(t^{-}_{i,m})\rangle = \Big(\sum_{k} \beta_k|b_k\rangle\Big) \otimes |n^a_1\cdots n^a_i \cdots n^a_{L/2} \rangle\otimes |\uparrow\rangle_i.
\end{equation}
The measurement outcome in the $i$-th block depends on the particle configuration of the $(i, i+1)$-th main chain sites and the $i$-th auxiliary site:
\begin{equation}
|n^c_1\cdots \boldsymbol{n^c_i n^c_{i+1}}\cdots n^c_{L/2}\rangle\otimes |n^a_1\cdots \boldsymbol{n^a_i}\cdots n^a_{L/2} \rangle.
\end{equation}
A key observation is that all the above basis states constituting $|\Phi_i(t^{-}_{i,m})\rangle$ have the same occupation number $n^a_i$ (0 or 1) for the $i$-th auxiliary site. In the blockwise evolution described in Secs. \ref{sec:Blockwise-evol_non-int} and \ref{sec:Blockwise-evol_int}, the auxiliary site occupation remains unchanged at $n^a_i$ for no-click ($\uparrow$), whereas the occupation changes to $(1 - n^a_i)$ for click outcomes ($\downarrow$).

In the spin-up sector (no-click), for each configuration $|b_k\rangle\otimes |...n^a_i...\rangle$, we obtain:
\begin{equation}
|b_k\rangle\otimes |\cdots n^a_i\cdots \rangle \to \Big(\sum_{l}\gamma^{\uparrow}_{kl} |b_{l}\rangle\Big)\otimes |\cdots n^a_i\cdots \rangle.
\end{equation}
Similarly, in the spin-down sector (click), for each configuration $|b_k\rangle\otimes |\cdots n^a_i\cdots \rangle$, we obtain:
\begin{equation}
|b_k\rangle\otimes |\cdots n^a_i\cdots\rangle \to \Big(\sum_{l}\gamma^{\downarrow}_{kl} |b_{l}\rangle\Big)\otimes |\cdots(1-n^a_i)\cdots\rangle.
\end{equation}
Here, the coefficients $\gamma^{\uparrow}_{kl}$ and $\gamma^{\downarrow}_{kl}$ correspond to the amplitudes for $|b_l\rangle$ after the action of the system-detector unitary operator, and the subscript $k$ emphasizes that these coefficients depend on the particle configuration $|b_k\rangle$ before the unitary operation. When an entangled pair is generated, there are \textit{two} nonzero coefficients $\gamma^{\uparrow}_{kl}$ (or $\gamma^{\downarrow}_{kl}$); otherwise, there is only \textit{one} nonzero coefficient. The exact values of $\gamma^{\uparrow}_{kl}$ (or $\gamma^{\downarrow}_{kl}$) depend on the system-detector coupling strength, as described in Secs. \ref{sec:Blockwise-evol_non-int} and \ref{sec:Blockwise-evol_int}.

Finally, the action of the unitary and the projective measurement of the detector qubit in the $i$-th block results in the following no-click and click states,
\begin{align}
   &|\Psi_{\uparrow}(t_{i,m})\rangle = \Big(\sum_{k,l} \beta_k \gamma^{\uparrow}_{kl}|b_{l}\rangle \Big) \otimes |n^a_1\dots n^a_i \dots n^a_{L/2} \rangle, \notag \\
   &|\Psi_{\downarrow}(t_{i,m})\rangle = \Big(\sum_{k,l}\beta_k \gamma^{\downarrow}_{kl}|b_l\rangle\Big) \otimes |n^a_1\dots (1-n^a_i).. n^a_{L/2} \rangle,
\end{align}
respectively with the Born probability $$P_\sigma(t_{i,m}) = \langle\Psi_\sigma(t_{i,m})|\Psi_\sigma(t_{i,m})\rangle.$$

Thus, starting from a product state configuration of the auxiliary chain, the measurement dynamics preserves this product state structure while progressively generating entanglement within the main chain. This implies that auxiliary chain sites do not entangle with the main chain or among themselves. The auxiliary chain, therefore, plays a crucial role in facilitating the generation of entanglement in the main chain without acting as a bath, provided the initial state is of the form given in Eq.~\eqref{eq:generic-psi-prod}. 

For a generic superposition state, where the quantum state of the system is of the following form:
\begin{equation}\label{eq:generic-superposition}
    |\Psi\rangle = \sum_{l} \bigg(\sum_{k} \beta_{k,l} |b_k\rangle\bigg)\otimes |a_l\rangle,
\end{equation}
where $|b_k\rangle=|n^c_1...n^c_i ...n^c_{L/2}\rangle$ and $|a_l\rangle=|n^a_1...n^a_i...n^a_{L/2}\rangle$ ($k,l=1, \ldots, 2^{L/2}$), each product state of the form
$$
\bigg(\sum_{k} \beta_{k,l} |b_k\rangle\bigg)\otimes |a_l\rangle
$$ 
evolves like the generic product state of Eq.~\eqref{eq:generic-psi-prod}, as described above. In the superposition of all these generic product states in Eq.~\eqref{eq:generic-superposition}, the auxiliary chain remains entangled with the main chain. In this case, the auxiliary chain is expected to behave as a bath to the main chain. In the following sections, we illustrate with our numerical findings how the auxiliary chain evolves under the one- and three-body measurement dynamics starting from superposition states, Eq.~\eqref{eq:Psi-rs} and Eq.~\eqref{eq:Psi-es}.

\subsection{Entanglement measures}
\label{sec:EntanglementCharacterization}

We characterize a quantum state $|\Psi(t_m)\rangle$ at time $t_m$, i.e., after the $m$-th measurement step is completed, by its entanglement properties, utilizing von Neumann's entanglement entropy and the associated mutual information. For computing entanglement in the system, we divide the system into two parts, namely subsystem $A$ and its complement $A^\mathrm{c}$. The von Neumann entanglement entropy is obtained from
\begin{align}
    S_{A}(t_m) &= -\text{Tr}_A \big[\rho_A(t_m) \ln \rho_A(t_m)\big],
\end{align}
where the reduced density matrix $\rho_A(t_m)$ is given by $$\rho_A(t_m)=\mathrm{Tr}_{A^\mathrm{c}}\big[\,|\Psi(t_m)\rangle \langle\Psi(t_m)|\,\big].$$




Since our system consists of the main chain plus the auxiliary chain, we can choose different subsystems to extract various types of quantum correlations between a subsystem $A$ and the rest of the system, denoted as $A^\mathrm{c}$. We focus on three different types of subsystems $(B,\, B',\, C=B\cup B')$ within the main chain, as illustrated in Fig.~\ref{fig:mom_cartoon}, to extract the quantum correlations between different parts of the system. 
The subsystem cut $C$ extracts the entanglement between the auxiliary chain and main chain. For a generic initial main-auxiliary product state [Eq.~\eqref{eq:generic-psi-prod}], the entanglement entropy $S_C$, quantifying entanglement between the main and auxiliary chains, remains exactly \textit{zero} in each measurement step.

For a generic superposition state, the main chain is expected to get entangled with the ancilla chain under the measurement dynamics. Thus, the main chain should be described by a mixed-state reduced density matrix. Since our main focus is the entanglement generation within the main chain, we characterize it via mutual information between two halves $B$ and $B'$ of the main chain, i.e.,
\begin{align}
    I_{BB'} &= S_{B} + S_{B'}- S_{B\cup B'}
\end{align}
Unlike the von Neumann entropy, the mutual information is typically a good measure of entanglement for mixed states, except for a few special cases~\cite{Tarun}, like a system close to a finite-temperature continuous phase transition with diverging correlation length. For a thermal mixed state, the mutual information typically follows an area law \cite{Wolf,MatthewB,Roger}. At the same time, the von Neumann entropy calculated for the subsystem in the thermal state would generically yield a volume-law scaling characteristic of the thermal fluctuations rather than of quantum entanglement. 

In our setup, the mutual information within the main chain will follow an area law similar to that in the thermal state, if the auxiliary chain gets strongly entangled with the main chain and acts like a bath. We show below that this is not the case for the long-time states in the one-body measurement model of Sec.~\ref{sec:NonIntModel}, see Table~\ref{table:noninteracting}. In contrast, the long-time mutual information within the main chain does indeed follow the area law in the higher-body measurement model of Sec.~\ref{sec:IntModel}, see  Table~\ref{table:interacting}.  

\subsection{Quantum trajectories and long-time states} \label{sec:QuantumTrajectory}


As discussed in Sec.~\ref{sec:measurementprotocol}, a quantum trajectory in our measurement-only dynamics is generated by sequentially switching on the instantaneous system-detector interaction $ H^{(i,i+1)}_{sd} $ within each block, followed by a projective measurement on the detector in a clockwise manner, as illustrated in Fig.~\ref{fig:mom_cartoon}. This process is repeated $t_m \times (L/2)$ times, where the factor $L/2$ accounts for the number of blocks in a single measurement sweep, to reach the measurement step $t_m$ for a system of size $L$. All quantum trajectories are sampled using Born's rule measurement probabilities dictated by quantum mechanics. 

To implement the sampling of quantum trajectories with Born's probabilities---determined by the quantum state prior to measurement in the detector---we employ standard Monte Carlo sampling techniques. Before switching on the system-detector interaction, the detector is initialized in the spin-up state. Consequently, the quantum state of the detector collapses to the spin-up state with an acceptance ratio is determined by 
$$P_{\uparrow}(t_{i,m}) = \langle \Psi_{\uparrow}(t_{i,m}) | \Psi_{\uparrow}(t_{i,m}) \rangle.$$ Numerically, a random number $r$ is drawn uniformly from the interval $[0,1]$, and if $r < P_{\uparrow}(t_{i,m})$, the quantum state collapses to the spin-up state; otherwise, it collapses to the spin-down state (`click'). 

It is important to emphasize that quantum trajectories generated with the Born-weighted sampling and those sampled with uniform weight (where measurement outcomes are independent of the quantum state prior to measurement) correspond to different replica limits in the analytical formulations of measurement-induced transitions~\cite{Jian2023, Nahum, Igor2, Poboiko2025, guo2024, xiao2025}. In the forced-measurement scenario, the replica limit corresponds to the conventional replica limit, where the number of replicas $R$ is sent to zero, whereas Born-rule sampling corresponds to the replica limit $R \to 1$. 

After generating quantum trajectories, we analyze the evolution of quantum states by characterizing their entanglement properties along individual trajectories for both the one-body and three-body measurement models. We then examine the statistical properties of the quantum trajectories, focusing on the distribution of entanglement entropy under measurement-only dynamics. In particular, we characterize the long-time distribution of entanglement entropy, first for the one-body measurement model and subsequently for the three-body measurement model.

\section{Evolution of entanglement in the one-body measurement model }\label{sec:StatOnebodyMeasurement}

We first discuss the evolution of entanglement along individual quantum trajectories, focusing mainly on the entanglement entropy $S_B$ for the $B$ entanglement cut shown in Fig.~\ref{fig:mom_cartoon}. We find that the distribution of $S_B$ at long times $t_m$ over an ensemble of trajectories converges to a stationary distribution, although individual quantum trajectories do not reach a stationary state, or even a stationary value of $S_B(t_m)$. This stationary distribution exhibits a \emph{limited form} of ergodicity, wherein the long-time stationary distributions for the same type of initial states, e.g., the random product states $|\Psi_{rp}\rangle$, Eq.~(\ref{eq:Psi-rp}), converge to identical distributions.

\begin{figure}[ht]
    \centering
\includegraphics[width=1.0\linewidth]{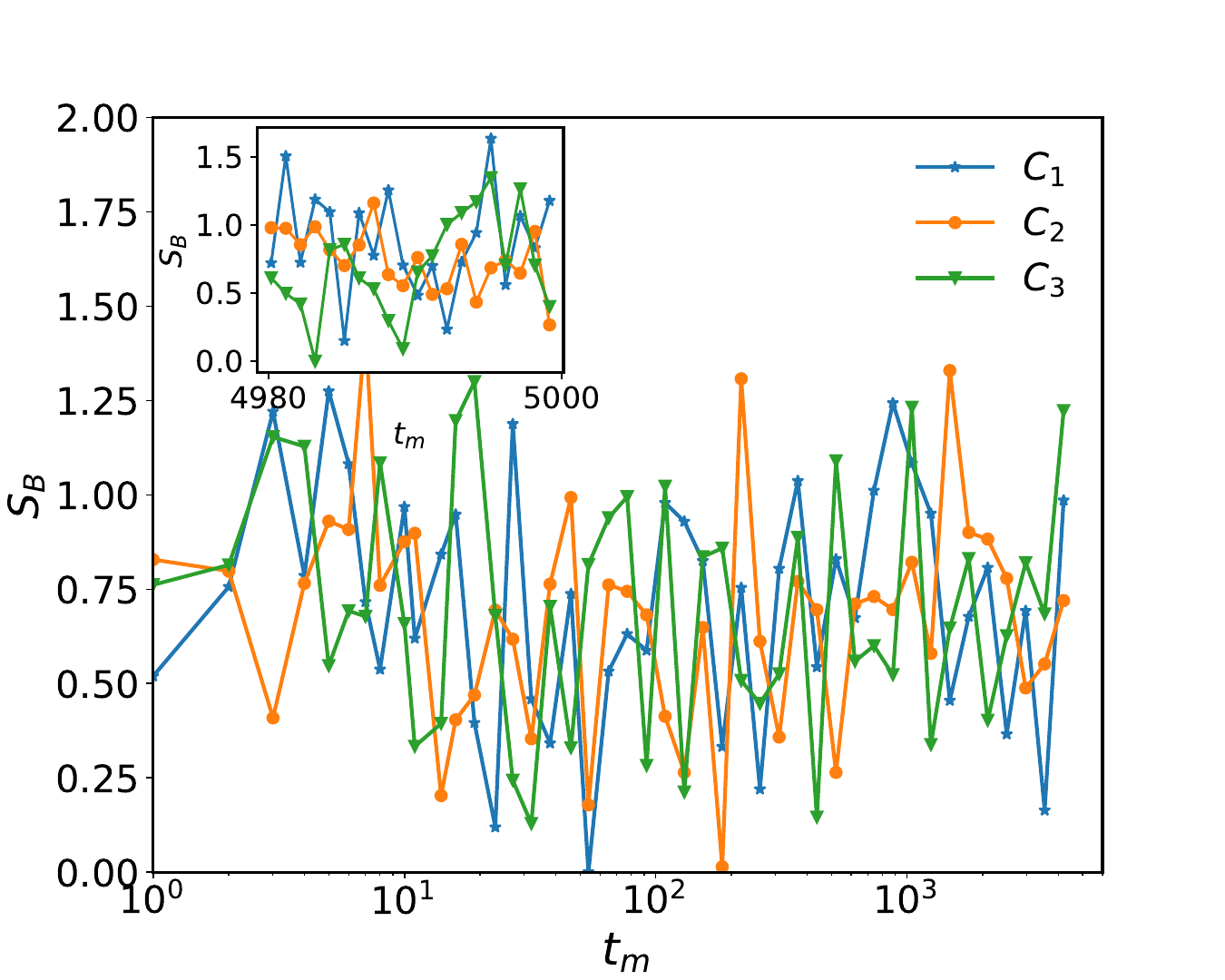}
\caption{Entanglement entropy $S_B$ for the half of the main chain as a function of discrete measurement times {$t_m$} for three different quantum trajectories, $\mathcal{C}_k$ $(k=1, 2, 3)$, shown on a logarithmic scale along the $ x $-axis. These trajectories are sampled for $\x=\pi/4$ and a system of size $ L=12 $, initialized in a product state, Eq.~(\ref{eq:Psi-p}). The inset provides a closer view of $S_B $ over the last 20 measurement sweeps on linear scale, illustrating the persistent oscillations around a finite value of $ S_B $. This indicates the absence of convergence to a stationary state.}
\label{fig:Nonint_cfill_traj}
\end{figure}

\subsection{Entanglement along quantum trajectories}
\label{subsub:entanglement-q-traj}

In Fig.~\ref{fig:Nonint_cfill_traj}, we show $S_B$ as a function of the discrete measurement time $t_m$ for three representative quantum trajectories $\mathcal{C}_k$ $(k=1, 2, 3)$, generated for $L=12$ starting from a specific product state $ |\Psi_p\rangle$ defined in Eq.~\eqref{eq:Psi-p}. 
We clearly see that entanglement is generated under the measurement-only dynamics. However, $S_B$ in each individual quantum trajectory, instead of saturating to a specific value, fluctuates strongly around a finite mean up to arbitrarily long times. 
As shown in Appendix~\ref{sup_sec:QuantumTraj_PsiRs}, the fluctuations around a finite mean do not appear to decrease with increasing $L$, indicating that the entanglement entropy of individual trajectories does not reach a stationary value in the large-system limit.   
This implies that these trajectories do not reach a stationary state. The fraction of click outcomes in these quantum trajectories has a finite value, as discussed in Appendix~\ref{sup_sec:BornProbClick_NI}. Similar results are found for entangled initial states of the random-superposition ($|\Psi_{rs}\rangle$)  and equal-superposition ($|\Psi_s\rangle$) types, Eqs.~\eqref{eq:Psi-rs} and \eqref{eq:Psi-es}, see {Appendix~\ref{sup_sec:QuantumTraj_PsiRs}}.


\begin{figure}[ht]
    \centering
\includegraphics[width=1.0\linewidth]{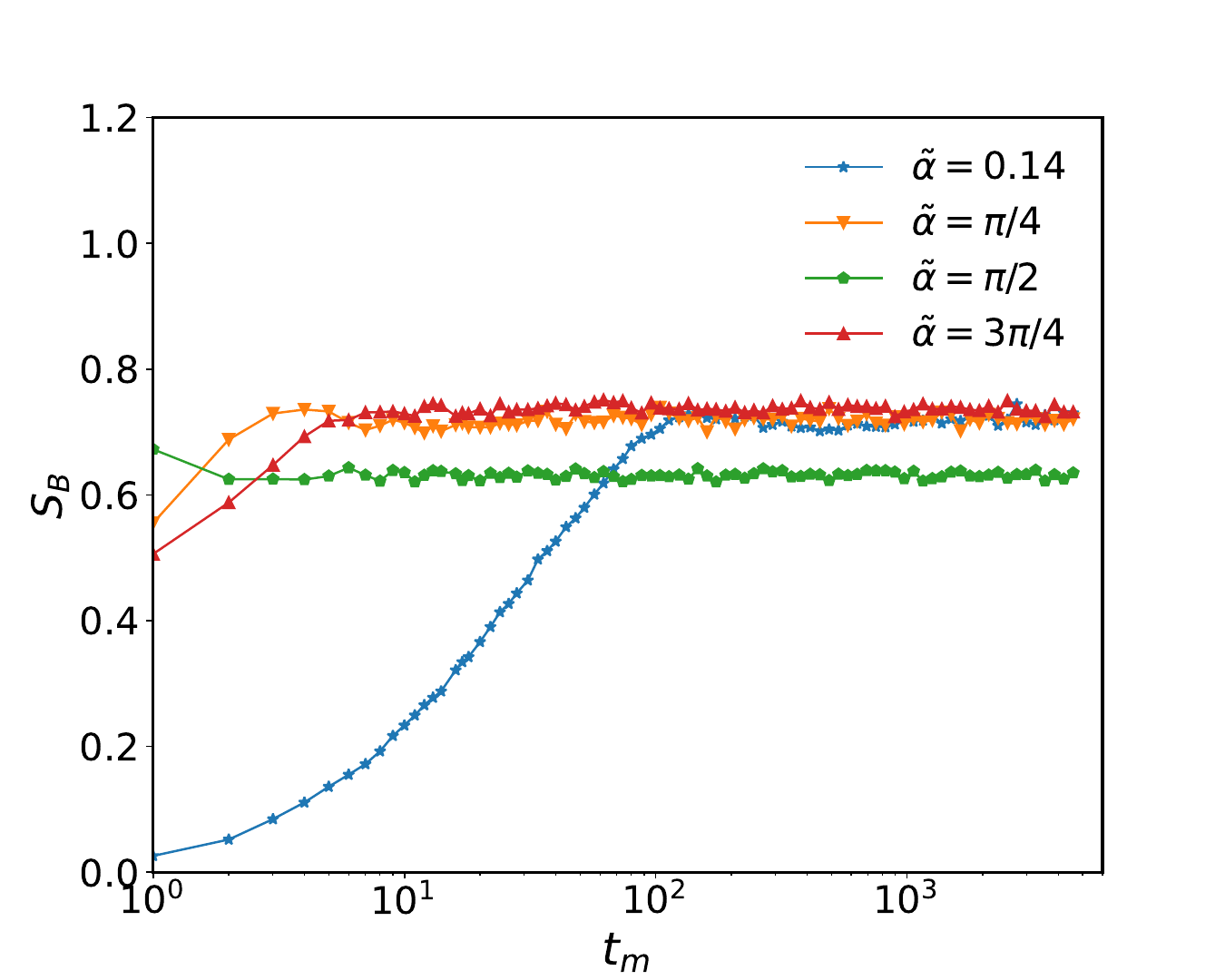}
\caption{The trajectory-averaged entanglement entropy $S_B$ as a function of discrete measurement times {$t_m$}, shown for $\tilde{\alpha}=0.14,\, \pi/4,\, \pi/2, 3\pi/4$ on logarithmic scale along the $ x $-axis. The data points are computed for the system size $ L=12 $, when the system is initialized in a product state [Eq.~\eqref{eq:Psi-p}]. The average $S_B$ shows convergence to stationary values in the long-time limit.}
\label{fig:Nonint_cfill_traj_avg}
\end{figure}

\subsection{Long-time states and stationary distribution for ensemble of trajectories}

\subsubsection{Product initial state}

{We plot the entanglement entropy $S_B(t_m)$ averaged over 4000 trajectories starting from the same initial product state $|\Psi_p\rangle$, Eq.~\eqref{eq:Psi-p}, in Fig.~\ref{fig:Nonint_cfill_traj_avg}. As is evident, the average entanglement entropy reaches a stationary value in the long-time limit, implying the approach of entanglement to a steady state in the trajectory-averaged sense.

\begin{figure}[ht]
    \centering
\includegraphics[width=1.0\linewidth]{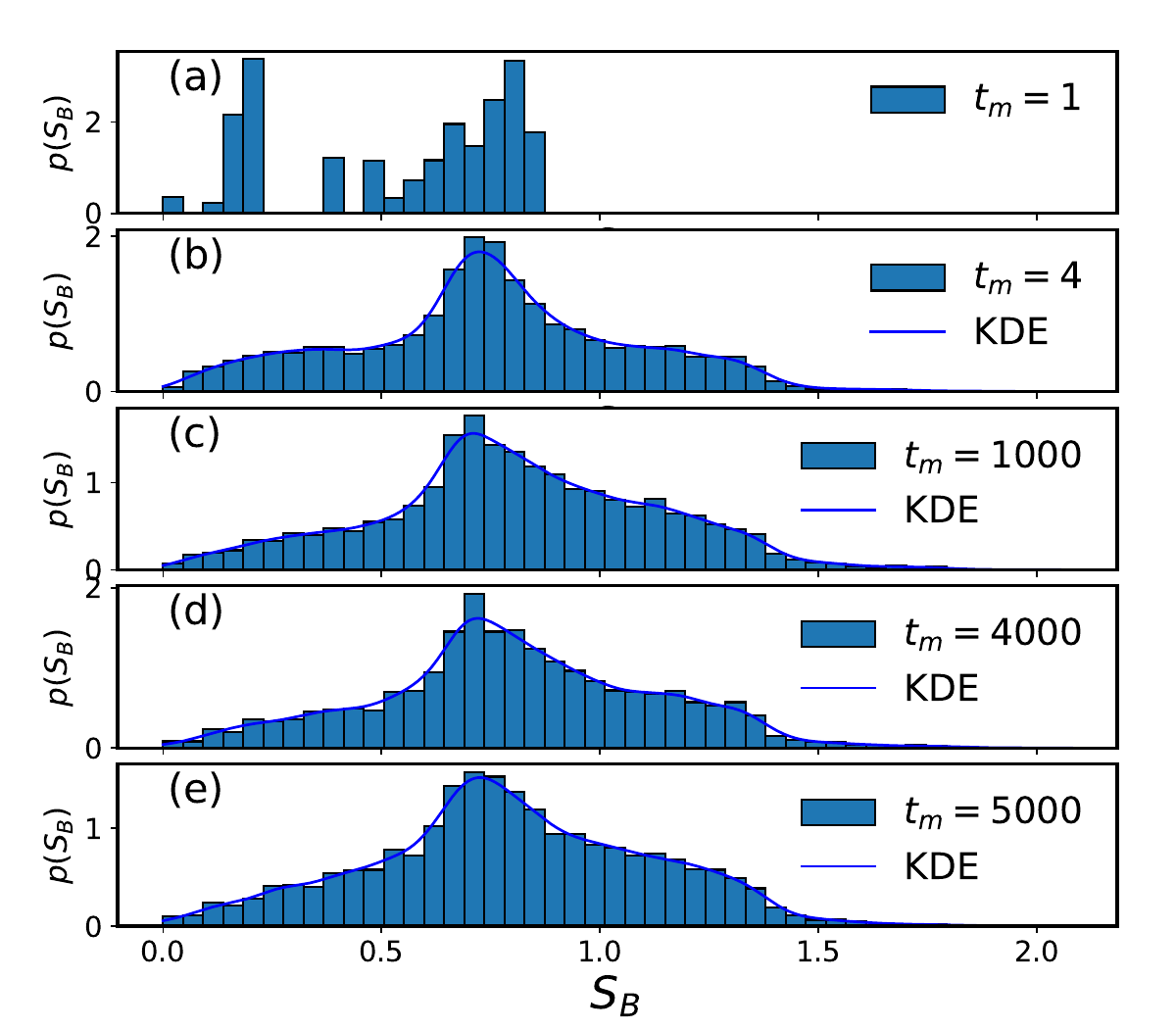}
\caption{The probability density function {$p(S_B)$} of \( S_B \) at different measurement steps \( t_m \) for the product initial state \( |\Psi_p\rangle \) with \( \tilde{\alpha} = 3\pi/4 \) and $L=12$. The histograms show the numerically obtained distributions, whereas the smooth curves are obtained with the kernel density estimation (KDE) for distributions. As \( t_m \) increases, $p(S_B)$ broadens and converges to a stationary distribution.}
 \label{fig:NI_distvstime_cfill}
\end{figure}

To further understand and characterize the approach of ensemble-averaged entanglement to the stationary state, we study the time evolution of the probability density function {$p(S_B)$} for the distribution of $S_B$ for the same ensemble of 4000 quantum trajectories, as shown in Fig.~\ref{fig:NI_distvstime_cfill}. 
The initial distribution at $t_0=0$ is a delta function peaked at $S_B=0$. At very early times, as depicted in Fig.~\ref{fig:NI_distvstime_cfill}(a), the distribution begins to spread but remains relatively narrow. As time progresses, the distribution broadens significantly, with $S_B$ values forming a continuum between 0 and the maximum possible value $(L/4) \log 2$ of $S_B$. Notably, the successive distributions at larger times, shown in Figs.~\ref{fig:NI_distvstime_cfill}(d)-(e), become nearly identical, suggesting that the system has attained a stationary distribution $p(S_B)$. The results are shown for parameter $ \x = 3\pi/4 $ with system size $L=12$. Similar results are obtained for other values of $\tilde{\alpha}$ and system sizes. 

To quantify stationarity and compare the distributions, we smoothen the numerically obtained distributions using kernel density estimation (KDE) \cite{Kde_Węglarczyk} to obtain the smooth probability distribution $p(S_B)$. We then compute the total variation distance (TVD) or the metric distance between these distributions. The TVD between two distributions $p(x)$ and $q(x)$ for $0\leq x\leq \infty$ is defined as:
\begin{align}\label{Eq:tvd}
   \text{TVD} = \frac{1}{2} \int_0^{\infty} dx \, |p(x) - q(x)|. 
\end{align}
For our analysis, we take $q(x)$ as the long-time distribution. By definition, the TVD is zero when $p(x)$ and $q(x)$ are identical. Thus, the metric distance should approach zero when stationarity of the distribution is attained, as we find below.

\begin{figure}[htb]
    \centering
\includegraphics[width=1.0\linewidth]{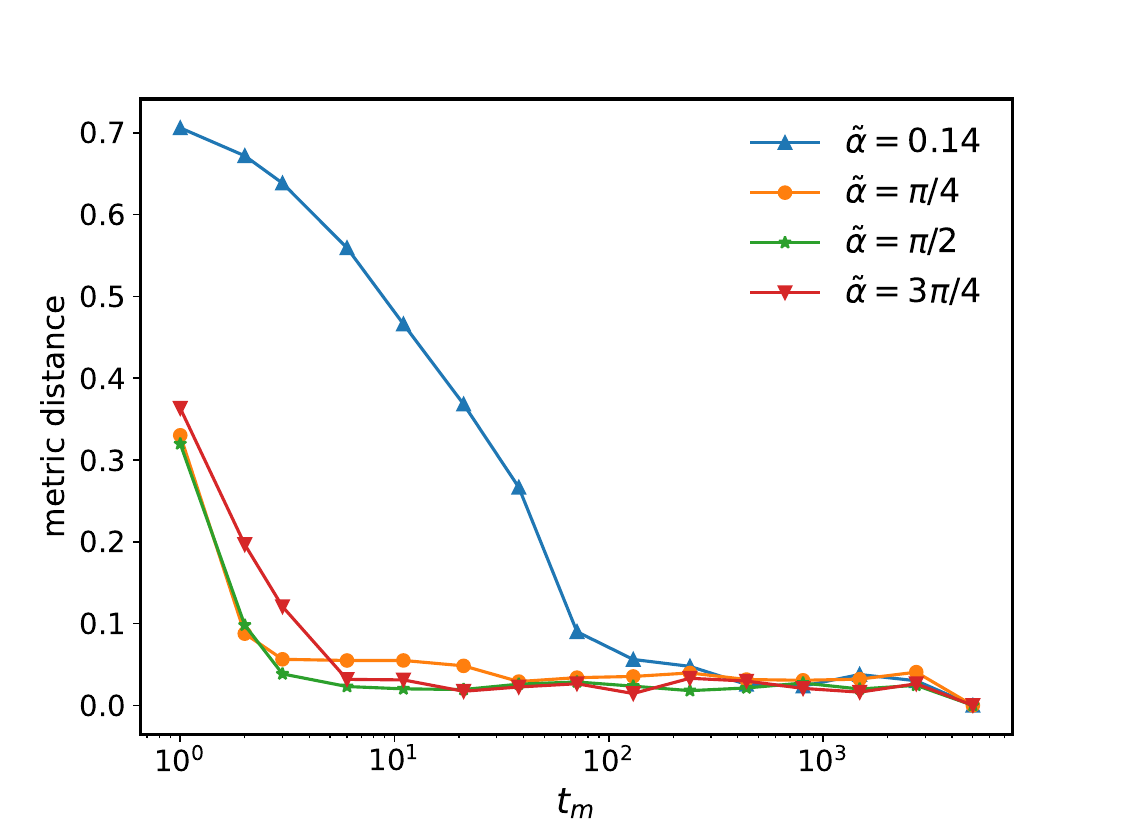}
\caption{The \emph{metric distance} or total variation distance (TVD) between the distribution at time \( t_m \) and the long-time distribution at {$t_m=5000$}, plotted as a function of \( t_m \) for different values of \( \tilde{\alpha} \) and $L=12$. The results show a rapid decrease in TVD, indicating convergence to a stationary distribution. A much slower convergence is observed for small \( \tilde{\alpha}=0.14 \). }
\label{fig:NI_TVD_vs_time}
\end{figure}

We calculate the TVD at various measurement times with respect to the long-time distribution at {time $t_m=5000$}. The evolution of TVD as a function of $t_m$ is shown in Fig.~\ref{fig:NI_TVD_vs_time}. We observe that after an initial transient period, the TVD becomes very small and remains approximately constant, indicating that the successive distributions have become very similar and are approaching a stationary distribution. Additionally, we find that, as $ \tilde{\alpha} $ approaches 0, the time scale required to reach a stationary distribution increases significantly compared to other values of $ \tilde{\alpha} $.

\subsubsection{Random superposition and other initial states} 
We perform a similar analysis for the initial random superposition state $|\Psi_{rs}\rangle$, Eq.~\eqref{eq:Psi-rs}. The evolution of the resulting probability distribution of $S_B$ for an ensemble of quantum trajectories is shown in Fig.~\ref{fig:Nonint_rands_pd}. Initially, the delta-function distribution peaked at a finite $S_B$ corresponding to a random state, broadens over time, eventually converging to an approximately stationary distribution in the long-time limit. We observe that this final stationary distribution of $S_B$ is different from the stationary distribution shown in Fig.~\ref{fig:NI_distvstime_cfill} for the initial product state $|\Psi_p\rangle$. As before, we assess the stationarity of the distribution using the TVD (Eqn. \ref{Eq:tvd}), as discussed in Appendix~\ref{sec:appendix_rs_state}.

\begin{figure}[!ht]
    \centering
\includegraphics[width=1.0\linewidth]{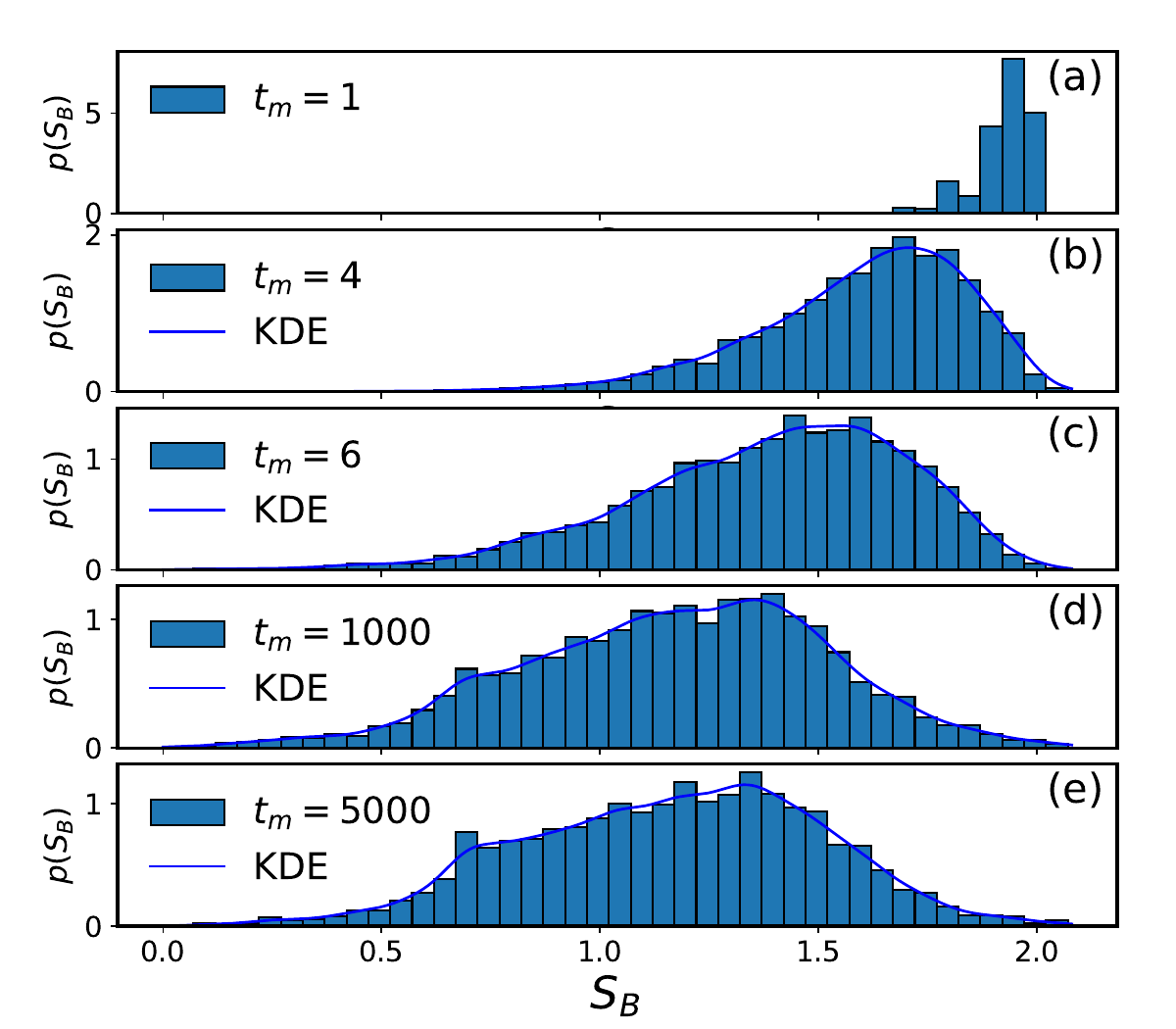}
\caption{The probability density $p(S_B)$ of $S_B$ and its smooth KDE approximation, shown at several measurement times $t_m$ for the initial random superposition state $|\Psi_{rs}\rangle$ for $\tilde{\alpha}=\pi/2$ and $L=12$. At long times, distributions converge to a stationary distribution.}
    \label{fig:Nonint_rands_pd}
\end{figure}

Additionally, we discuss results for the other initial states in Appendix~\ref{sec:appendix_rs_state}. From all this analysis, we can conclude that the distribution of $S_B$ over an ensemble of trajectories approaches stationary distributions at long times for all types of initial states considered in this work. However, the stationary distributions for different types of initial states appear to be distinct, as we analyze further below. This conclusion also holds for the entanglement entropy $S_C$ between the main and ancilla chains (corresponding to subsystem $C$, which is the union $B \cup B'$ of partitions shown in Fig.~\ref{fig:mom_cartoon}).

\subsection{Limited ergodicity in the measurement dynamics}

In the preceding section, we established that the probability density of the entanglement entropy \(S_B\) obtained over an ensemble of trajectories reaches stationary distributions starting from various initial states.
A natural question is whether this stationary distribution attained via the measurement-only dynamics follows some notion of ergodicity, e.g., whether different initial states lead to the same stationary distribution. For thermal equilibrium systems, like a system in contact with a thermal bath or an isolated system following the eigenstate thermalization hypothesis \cite{Srednicki, Deutsch}, ergodicity implies that the steady-state distribution of an observable becomes independent of the initial conditions, except for initial values of a few conserved quantities. Equivalently, the time average of an observable or its distribution over a sufficiently long time window can be replaced by its average or distribution over the static thermal ensemble average, determined only by the conserved quantities. However, for systems subjected to completely out-of-equilibrium conditions without even energy conservation, such as only repeated measurements in our case, there is no obvious notion of an equivalent static ensemble. 

As evident from Figs.~\ref{fig:NI_distvstime_cfill}(e) and \ref{fig:Nonint_rands_pd}(e) in the preceding sections, the dynamics approaches distinct stationary distributions of entanglement entropy over the ensemble of quantum trajectories for different types of initial states. Thus, only some limited form of ergodicity, if at all, can emerge through the measurement-only dynamics. Below, we look for the following limited forms of ergodicity (cf.~Ref.~\cite{Yevtushenko2024}), namely:
\begin{itemize}
\item[(1)] whether the distribution of entanglement entropy over a long-time interval along an individual trajectory for a given initial state is the same as the distribution over an ensemble of trajectories, discussed in the preceding sections;
\item[(2)] whether the long-time distributions of entanglement entropy over an ensemble of trajectories for many different initial states, all belonging to the same type, like random superposition or random product states, are the same;
\item[(3)] whether the entanglement distribution over many trajectories, each generated starting from different initial states of a given type, reaches a stationary distribution, same as the distribution over the ensemble of trajectories for one initial state of the same type.
\end{itemize}
We find that only the limited ergodicity of the form (2) above is satisfied by the measurement-only dynamics.


\subsubsection{Entanglement distribution over time}

Here, we compare the distribution of $S_B$ constructed from a long interval of time at long times for an individual trajectory with the stationary distributions for an ensemble of trajectories, shown, e.g., in Fig.~\ref{fig:Nonint_rands_pd}(e). In  Fig.~\ref{fig:Exp1_psi_rs}, the time-window distributions, along with the metric distances from the stationary distribution, are shown for several trajectories starting from an initial random superposition state. Further, we discuss the results for an initial random product state in {Appendix \ref{sup_sec:NI_prod_timewindowSB}}. Clearly, the entanglement distributions over a time interval do not match the stationary probability distribution over the ensemble of trajectories. However, the deviation of the time-window distribution from the stationary distribution is large for rare trajectories [Figs.\ref{fig:Exp1_psi_rs}(d),(e)] with low Born probabilities, whereas this deviation is much smaller for the most probable trajectory in Fig.~\ref{fig:Exp1_psi_rs}(b).

\begin{figure}[htb!]
    \centering
\includegraphics[width=0.95\linewidth]{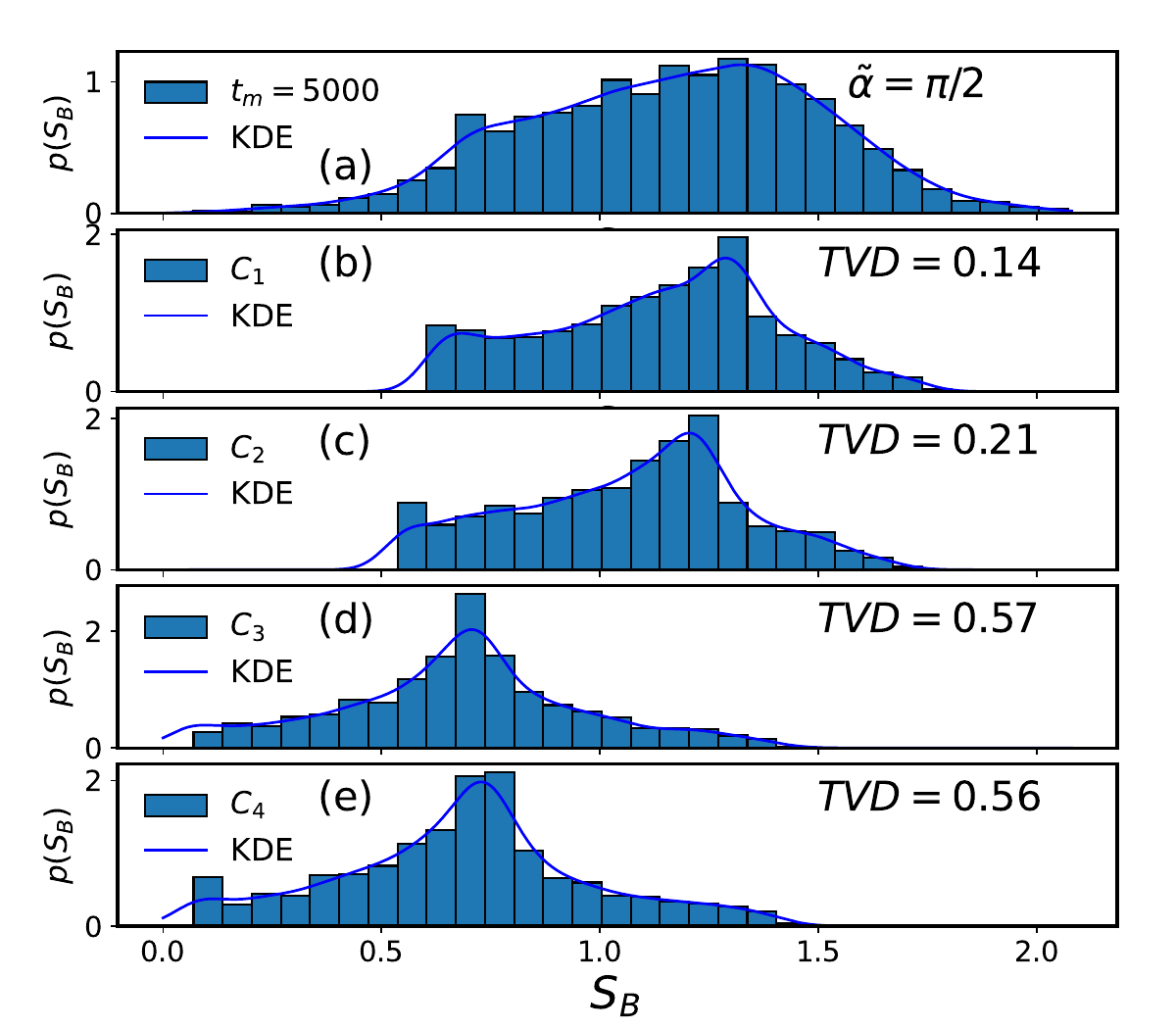}
\caption{{(a)} The stationary probability density of $S_B$ over the trajectory ensemble at large time $t_m=5000$ for random-superposition initial state and $\tilde{\alpha}=\pi/2$. Panels {(b-e)} show the distributions over a long time interval, from $t_m=1000$ to $t_m=5000$, for four trajectories. The trajectory in panel {(b)} has the maximum Born's probability among all the trajectories. We show the corresponding smoothened (KDE) approximations to the distributions in panels {(b-e)}, alongside their metric distances from the smooth approximation to the stationary distribution in panel {(a)}.}
\label{fig:Exp1_psi_rs}
\end{figure}

\subsubsection{Initial-state dependence of the stationary distribution}

As shown in Figs.~\ref{fig:NI_distvstime_cfill} and \ref{fig:Nonint_rands_pd}, different types of initial states, particularly states with very different values of the entanglement entropy---like the product and superposition states---lead to distinct stationary distributions. However, initial states of the same type, having similar entanglement entropy, lead to the same stationary distribution, as we discuss here.
Specifically, we examine whether a set of random product initial states all lead to a similar final stationary distribution. 

We first determine the long-time stationary probability density $p(S_B)$ of $S_B$ for several random product initial states. We extract the smoothened curve for each $|\Psi_{rp}\rangle$ via kernel density estimation (KDE) \cite{Kde_Węglarczyk}. In Fig.~\ref{fig:PD_diff_psi_rp_longtime}, we show the extracted curves for different $|\Psi^{(k)}_{rp}\rangle $ states ($k=1,\ldots,4$) for three values of $\x$. The probability densities for various initial random product states shown in each panel are very similar, with the metric distance (TVD) between them approaching a small value ($<0.05$). 

We perform a similar analysis for random superposition states in {Appendix~\ref{sup_sec:PsiRsInitialStateDependence}}. It is known that, in the thermodynamic limit, all random superposition states have nearly identical volume-law entanglement entropy for any subsystem cut, with small fluctuations in finite systems. Our findings confirm that all random superposition states also lead to similar stationary distributions over an ensemble of trajectories at long times. Thus, when categorized by the initial entanglement entropy $S_B$, the probability distributions for all initial states with the same $S_B$ appear to be identical at long times. Therefore, the stationary distributions follow a form of limited ergodicity where the final distribution does not depend on initial states within the same category determined by the initial entanglement entropy. 

\begin{figure}[htb!]
    \centering
    \includegraphics[width=0.95\linewidth]{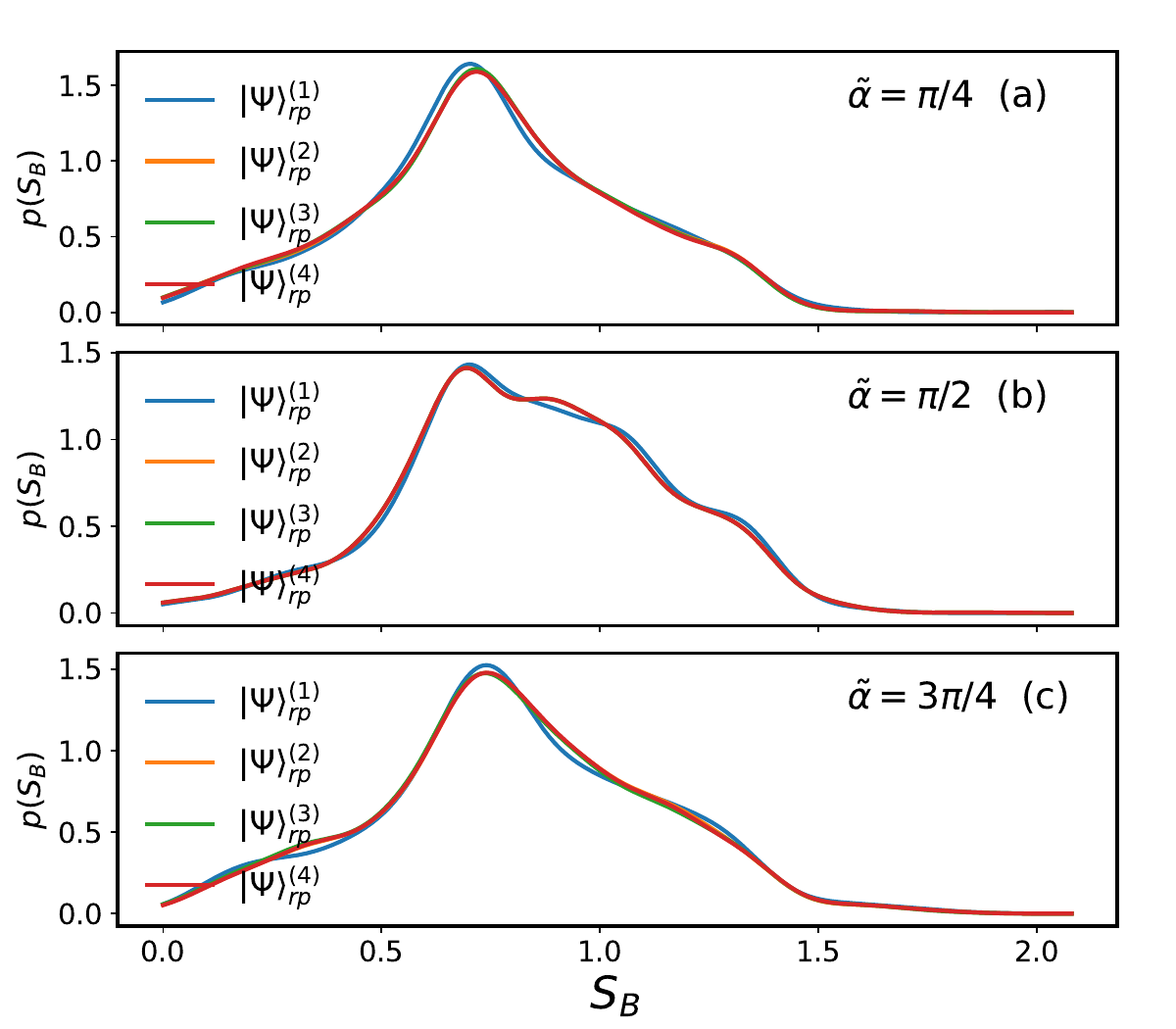}
    \caption{Stationary probability density function  $p(S_B)$ obtained from the KDE approximation for the probability-density distributions over an ensemble of trajectories for different random product initial states for three values of $\tilde{\alpha}=\pi/4,\, \pi/2,\, 3\pi/4$.}
    \label{fig:PD_diff_psi_rp_longtime}
\end{figure}

\section{Evolution of entanglement in the three-body measurement model}\label{sec:StatThreebodyMeasurement}

In this Section, we explore the measurement dynamics in the three-body measurement model of Sec.~\ref{sec:IntModel}. Again, we characterize the evolution of quantum states in terms of their entanglement properties along quantum trajectories and analyze the statistical properties of the entanglement entropy for an ensemble of trajectories.

Our results reveal a stark contrast between the three-body and one-body measurement models. Specifically, we observe that in the three-body measurement model, individual quantum trajectories evolve toward a single stationary quantum state---albeit different ones, in general, for different initial states. We establish the conditions for this stationarity and further demonstrate that the distribution of entanglement entropy of these stationary states is sharply peaked at a discrete set of values, unlike the broad and continuous distributions observed in the one-body measurement model.


\begin{figure}[ht]
    \centering
\includegraphics[width=0.98\linewidth]{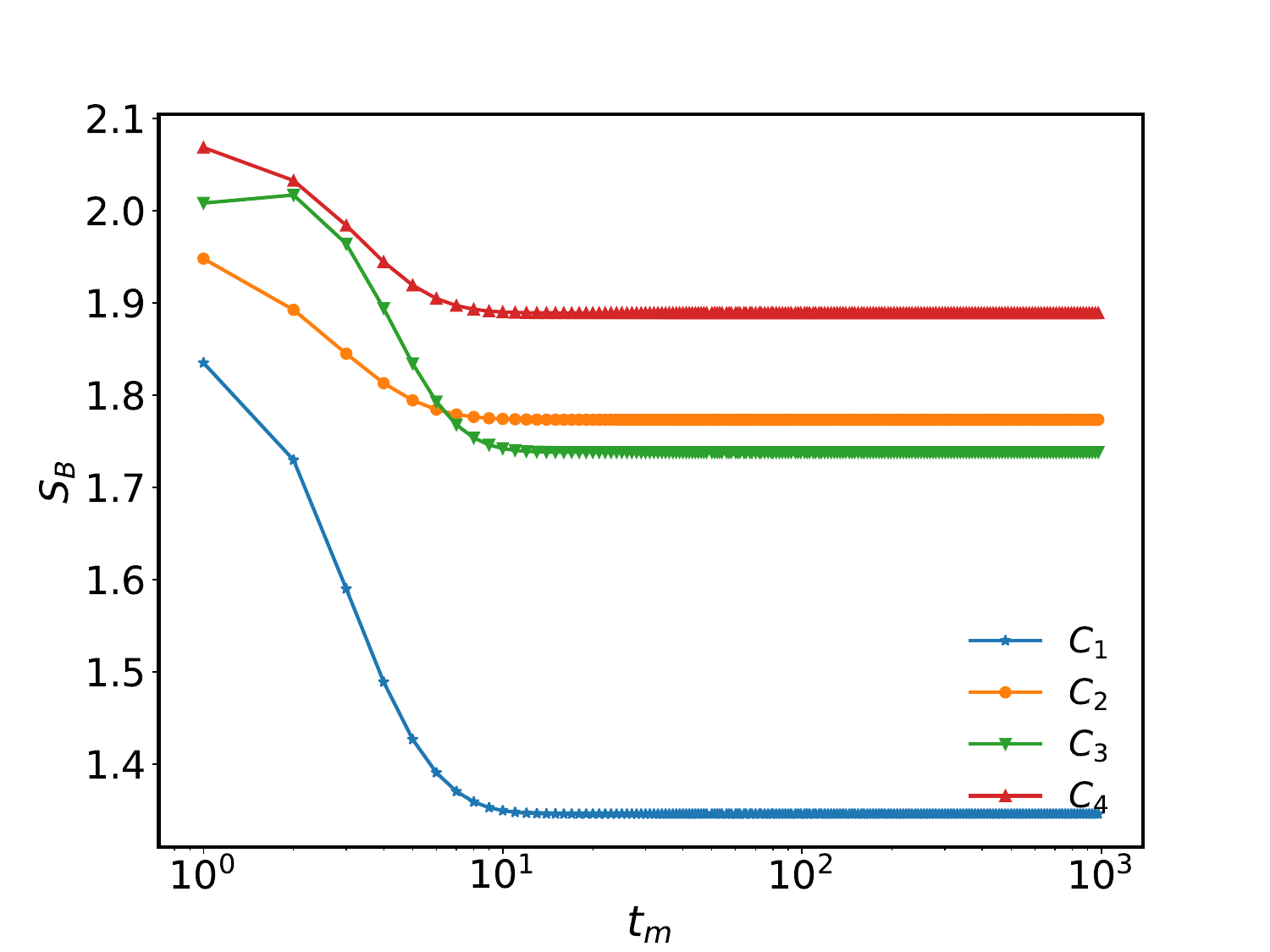}
    \caption{Entanglement entropy $S_B$ as a function of discrete measurement times $t_m$ is shown on logarithmic scale in the $x$-axis for a few quantum trajectories $\mathcal{C}_k$ ($k=1,2,3,4$) starting with a random superposition initial state $|\Psi_{rs}\rangle$  for $\tilde{\alpha}=\pi/4$. This demonstrates that each quantum trajectory reaches a stationary state in the three-body measurement model, in contrast to the one-body measurement model.}
    \label{fig:IntTraj}
\end{figure}

\subsection{Quantum trajectories}

In Fig.~\ref{fig:IntTraj}, we show the evolution of the entanglement entropy $S_B$ for several quantum trajectories $\mathcal{C}_k, \, k=1,2,3,4$, originating from the initial random superposition state $|\Psi_{rs}\rangle $. These trajectories are computed for a system size $L=12$ with $\tilde{\alpha}=\pi/4$. In contrast to the one-body measurement model, we observe that each quantum trajectory in the three-body measurement model converges to a stationary quantum state after a sufficient number of measurement steps $(t_m \gg 1)$, as evident from the lack of any fluctuations in entanglement entropy $S_B(t_m)$, and also verified from the overlap (not shown) of successive states at long times. We find that, in the three-body measurement model, stationary quantum states are consistently achieved for all trajectories, regardless of the initial states or the value of $\tilde{\alpha}$.

\subsection{Condition for stationary state}

Stationary quantum dynamics is reached under the measurement protocol of Sec.~\ref{sec:IntModel}, when a quantum state $|\Psi\rangle\otimes |\!\uparrow\rangle_i$ for the system and detector remains unchanged under the application of unitary $\mathcal{U}_i$:
\begin{align}
    \mathcal{U}_i |\Psi\rangle\otimes |\!\uparrow\rangle_i = |\Psi\rangle \otimes |\!\uparrow\rangle _i               \hspace{0.3cm} \text{for all } i.
\end{align}
This implies that the unitary operator governing the system-detector interaction acts as the identity operator. 
In the three-body measurement case, $\mathcal{U}_i$ acts non-trivially only when the occupation numbers of two neighboring sites of the main chain are either both empty or both filled, as in Eqs.~(\ref{eq:UNp1_Int}). These states are referred to as `non-trivial basis states'. In contrast, for other basis states, which we refer to as `trivial basis states', $\mathcal{U}_i = \mathbf{I}$.

Therefore, under the blockwise action of the system-detector interaction as described in Sec.~\ref{sec:IntModel}, the quantum state in the Hilbert space may evolve into a configuration that is spanned by these trivial basis states. 
A closer examination of Eq.~\ref{eqn:int_Ua} reveals that the action of $\mathcal{U}_i$ on non-trivial basis states results in a state spanned by trivial basis states within block $i$ once the detector registers a `click' (i.e., a $\downarrow$ readout). While the particle configuration in a given block can change through the action of unitary operators in neighboring blocks, over time, the system ultimately settles into a particle configuration in which each block contains only trivial states. At this stage, the unitary operator acts solely on trivial basis states, leading to the emergence of a stationary state.

Below, we list a few examples of long-time stationary states obtained when starting with a product initial state $|\Psi_p\rangle$. These stationary states are for a system size $L=8$ in the half-filling sector, obtained directly from numerical simulations:
\begin{align*}
    |\Psi(t_m \gg L)\rangle &= \frac{1}{\sqrt{2}}\bigg(|1001\rangle + |0101\rangle\bigg) \otimes |0011\rangle, \\
    |\Psi(t_m \gg L)\rangle &= -\bigg(0.99 |1010\rangle + 0.15 |0101\rangle\bigg) \otimes |1010\rangle, \\
    |\Psi(t_m \gg L)\rangle &= |0101\rangle \otimes |1100\rangle, \\
    |\Psi(t_m \gg L)\rangle &= \frac{1}{\sqrt{2}}\bigg(|1010\rangle + |1001\rangle\bigg) \otimes |1001\rangle.
\end{align*}
In all of the examples above, the basis states in any of the blocks, $|n^c_i n^c_{i+1}\rangle \otimes |n^a_i\rangle$, are spanned by the `trivial basis states'.

\subsection{Distribution of stationary states}

\subsubsection{Random superposition initial state}

We now discuss the distribution of entanglement entropy $S_B$ in the stationary states obtained starting from an entangled superposition state, such as a random superposition state $|\Psi_{rs}\rangle$, Eq.~\eqref{eq:Psi-rs}. Since the dynamics lead to a discrete set of stationary states over an ensemble of trajectories in the three-body measurement model, instead of the probability density we compute the probability {$P(S_B)$} for the distribution of $S_B$ from an ensemble of 4000 quantum trajectories.

In Fig.~\ref{fig:Fig_entB_rands}, the probability distribution at long times, after the system reaches a stationary quantum state, is shown for several values of the parameter $\x = 0.141,\, \pi/4,\, \pi/2$ for system size $L=12$. We observe that the distribution has sharp peaks at a nonzero value of $S_B$ for all values of $\x$. In {Appendix~\ref{sup_sec:IPR_Int}}, we verify that the stationary-state probability distribution $P(S_B)$ over ensemble of trajectories only contains discrete values of $S_B$ and, thus, only a discrete set of stationary states, unlike the continuum of values of $S_B$ between $0$ and $(L/4)\log{2}$ in the one-body measurement model (Sec.~\ref{sec:NonIntModel}). 
The discreteness of $P(S_B)$ can be diagnosed by computing the inverse participation ratio IPR$=\sum_{k=1}^{N_t}P^2(S_B^{(k)})$ as a function of the total number of sampled trajectories $N_t$, where $S_B^{(k)}$ is the steady-state entanglement entropy for the $k$-th trajectory. If there are only discrete values of $S_B$, the IPR saturates to a nonzero value as a function of $N_t$, as we observe for the three-body measurement model in {Appendix~\ref{sup_sec:IPR_Int}}.

\begin{figure}[htb]
    \centering
\includegraphics[width=0.95\linewidth]{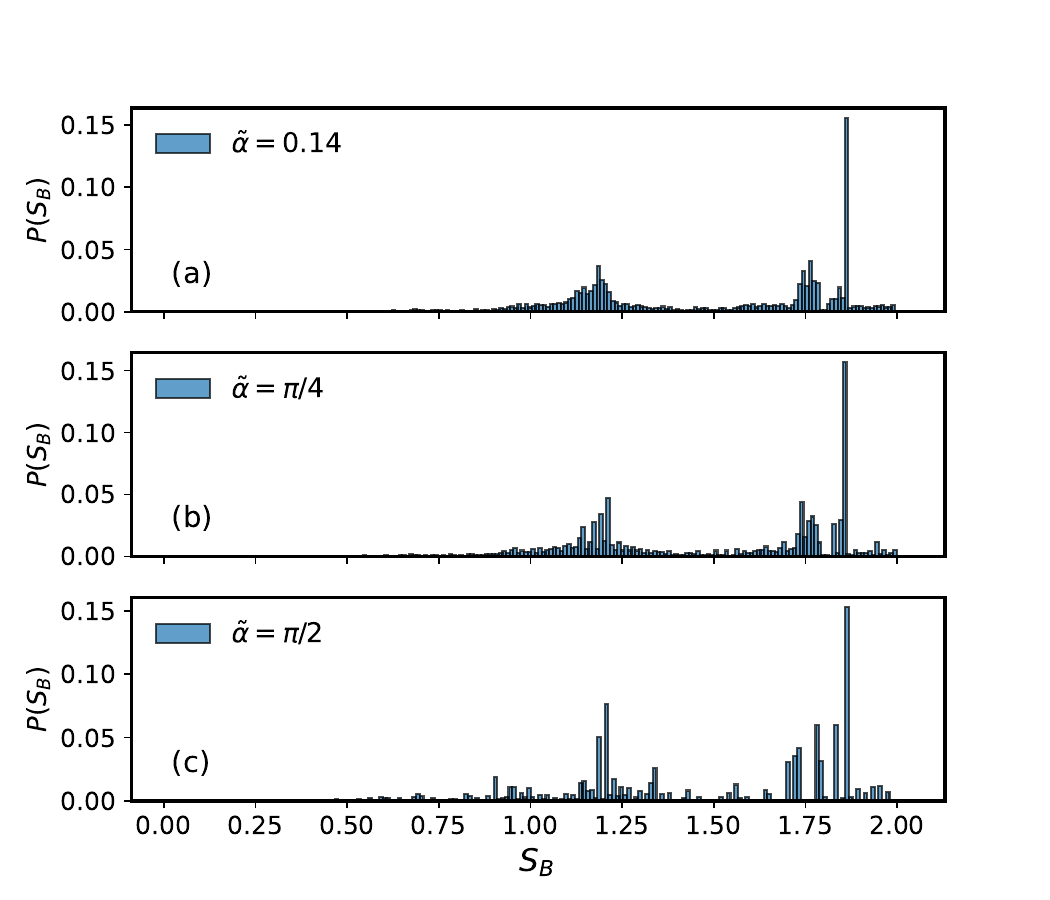}
    \caption{Probability distribution $P(S_B)$ as a function of steady-state entanglement $S_B$, shown for $\tilde{\alpha}=0.14,\, \pi/4,\, \pi/2$ for an initial random superposition state $|\Psi_{rs}\rangle$.}
    \label{fig:Fig_entB_rands}
\end{figure}

Further analysis, discussed in {Appendix \ref{sup_sec:BornProbInt}}, reveals that the peak with the maximum probability (the most probable values of $S_B$ in the probability distributions in  Fig.~\ref{fig:Fig_entB_rands}) corresponds to quantum trajectories with the maximum Born weight. These trajectories consist entirely of \emph{no-click} ($\uparrow$) outcomes for the detectors. According to Eq.~\eqref{eqn:int_Ua}, even all no-click events can lead to a non-trivial evolution of a superposition state, such as $|\Psi_{rs}\rangle$, where no-click ($\uparrow$) outcomes reduce the relative weight of states with pair(s) of occupied or empty neighboring sites by factor(s) of $|\cos{\tilde{\alpha}}|<1$ in the superposition. The most probable values of $S_B$ across different values of $\x$ are the same, as the peaks corresponds to the same quantum state derived from the same \emph{no-click} trajectory.

It is important to note that the most probable value of $S_B$ does not coincide with $S_B$ of the initial state, and it is less than $S_B(t_m=0)$ of the random initial state $|\Psi_{rs}\rangle$. In general, we find that quantum trajectories other than the most probable ones exhibit very few \emph{click} ($\downarrow$) events. Interestingly, even though the measurement-only dynamics initially allows for both click and no-click events, the stationary distribution is dominated by an almost entirely no-click trajectory. 
 We thus observe that the non-equilibrium stationary-state distribution  may reveal, within a measurement-only dynamics framework, a kind of steering towards a certain class quantum trajectories, e.g., here sequences of no-click outcomes. This phenomenon, presumably, is a manifestation of ergodicity breaking due to kinetic constrains imposed by the special (`three-body') form of the measurement operators.


\subsubsection{Product initial state}

In Fig.~\ref{fig:Fig_entB_cfill}, we show the {probability distribution $P(S_B)$}, for stationary states obtained from an ensemble of quantum trajectories, starting from the initial unentangled product state $|\Psi_p\rangle$, Eq.~\eqref{eq:Psi-p}. Unlike the previous case with the random superposition initial state $|\Psi_{rs}\rangle$, where the distribution was peaked at nonzero values, we observe that the distribution is sharply peaked at $S_B = 0$. 
This indicates that most of the quantum trajectories do not generate entanglement when starting from the zero-entanglement product state. 

\begin{figure}[htb]
    \centering
\includegraphics[width=0.95\linewidth]{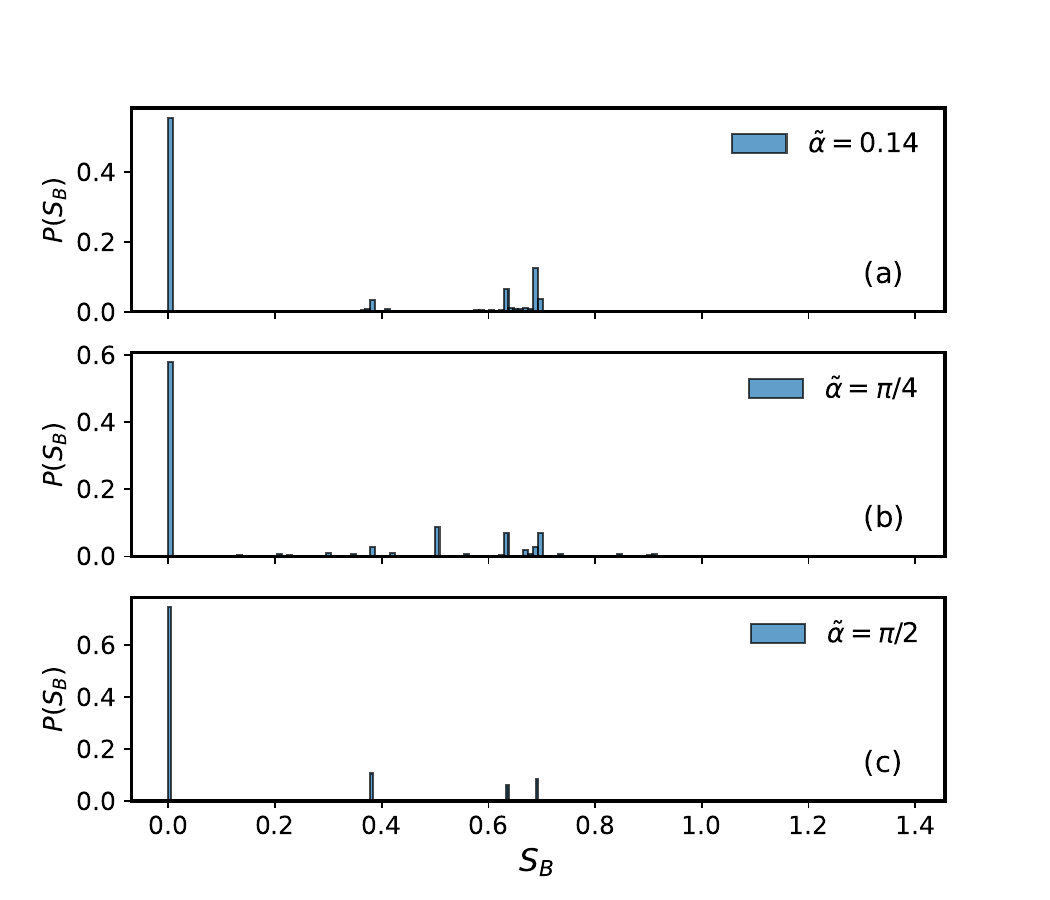}
    \caption{Probability distribution $P(S_B)$ as a function of steady state entanglement $S_B$, shown for $\tilde{\alpha}=0.14,\, \pi/4,\, \pi/2$. The initial state here is a product state $|\Psi_p\rangle$}
    \label{fig:Fig_entB_cfill}
\end{figure}

However, in addition to the dominant peak at $S_B = 0$, we find several lower peaks at $S_B \neq 0$. These additional peaks represent those quantum trajectories that do generate entanglement. In contrast to the previous case, we do not observe any trajectories consisting entirely of no-click events. We also do not find any entanglement-generating quantum trajectories with predominant Born's weight.

\section{System-size scaling of average entanglement from measurement-only dynamics} \label{sec:EntGen}

In the two previous Sections, we studied the evolution of entanglement along quantum trajectories and the approach to stationary distributions and/or stationary states over an ensemble of trajectories, starting from either unentangled or entangled initial states, under the measurement-only dynamics.
We showed that entanglement is generated in both the one-body measurement and three-body measurement models of Secs.~\ref{sec:NonIntModel} and \ref{sec:IntModel}, even when staring from initially unentangled product states. 

In this section, we examine how the trajectory-averaged entanglement entropy scales with system size at long times, and whether the entanglement generation persists as the system size increases.
As before, we consider different partitions of the system to extract distinct types of quantum correlations between the subsystem and the rest of the system.
We first discuss the average entanglement entropy for different subsystem cuts and the mutual information between two halves of the main chain in the one-body measurement model (Sec.~\ref{sec:NonIntModel}). This will be followed by a similar analysis in the three-body measurement model (Sec.~\ref{sec:IntModel}). The average entanglement entropy and mutual information results presented later in this section were computed from ensemble sizes of 6000, 4000, 2000, and 1000 trajectories for system sizes $L = 8, 12, 16, 20$, respectively.
Our discussion primarily focuses on two initial states: an unentangled product initial state and an entangled random superposition initial state. The analysis for other initial states is provided in Appendix~\ref{sec:appedix_systemsize_scaling}}, with all the results summarized in Tables \ref{table:noninteracting} and \ref{table:interacting}. 

\subsection{One-body measurement model: System-size scaling of entanglement entropy and mutual information} \label{sec:EntGenNonInt}

In the one-body measurement model, our key finding is that the long-time trajectory-averaged subsystem-$B$ entanglement entropy and mutual information between two halves of the main chain (Fig.~\ref{fig:mom_cartoon}) follow the volume-law scaling with system size. 
Moreover, we find that the auxiliary chain is completely disentangled from the main chain at each measurement step for product initial states [Eqs.~\eqref{eq:Psi-p} and \eqref{eq:Psi-rp}] and gets weakly entangled with main chain at long-times for superposition initial states [Eqs. \eqref{eq:Psi-rs}, \eqref{eq:Psi-es}]. These results demonstrate that only repeated non-commutative competing \emph{local} measurements can generate quantum correlations within the system, even in the absence of any entangling unitary evolution in the system.

 \begin{figure*}[htb!]
    \centering    \includegraphics[width=0.9\textwidth]{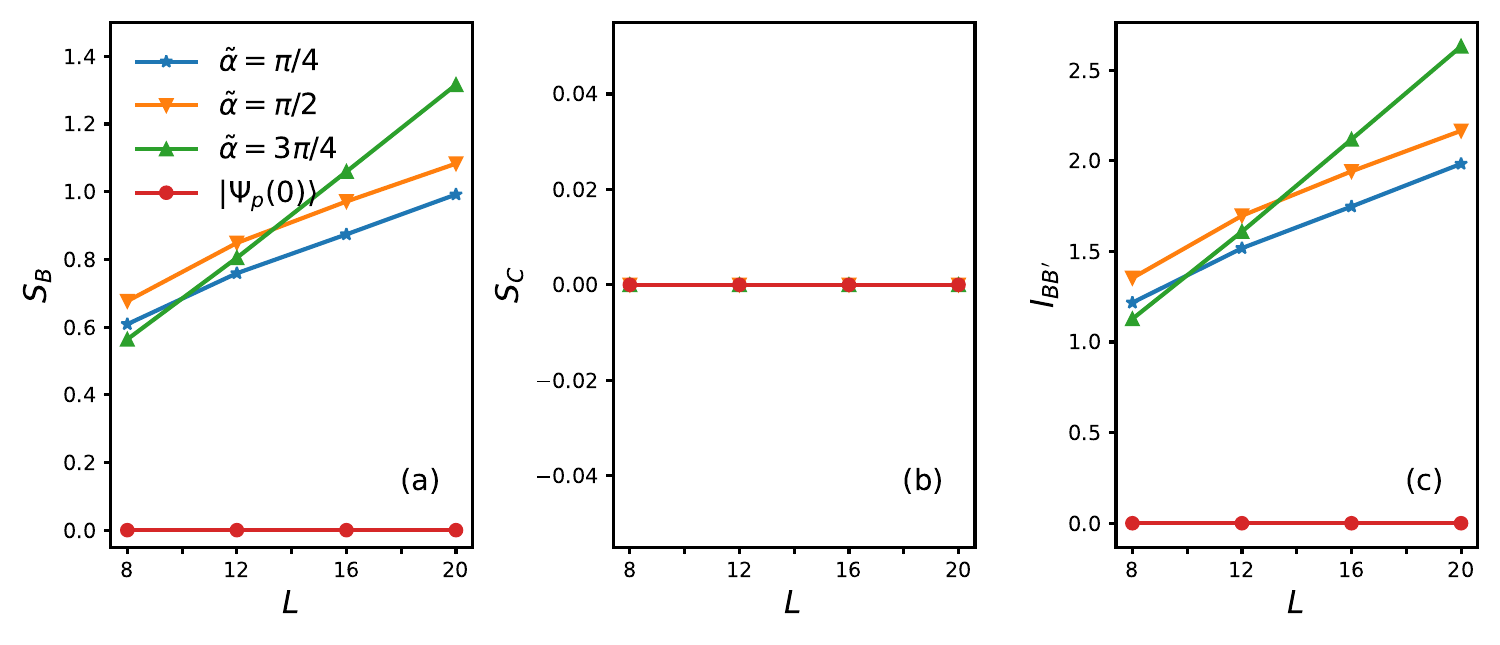}
    \caption{
    System-size scaling of long-time entanglement entropies and the mutual information for product initial state $|\Psi_p\rangle$ [Eq.~\eqref{eq:Psi-p}] in the one-body measurement model for several values of the coupling constant $\alpha$ in Eq.~\eqref{eq:HsdNonInt} ($\x = \sqrt{2}\alpha=\pi/4,\, \pi/2,\, 3\pi/4$ for blue, orange, and green data, respectively).  The values of entanglement measures for the initially unentangled state is also shown (red). (a) Entanglement entropy $S_B$ for the $B$-partition is plotted as a function of system size $L$ in the {linear scale}. (b) Entanglement entropy $S_C$ for the $C$-partition (ancillary chain traced out) as a function of $L$ remains zero [equal to $S_C(t=0)$] at long times. (c) Mutual information between two halves of the main chain $I_{BB'}=2S_B$ (exactly) as expected follows a volume-law scaling with $L$.}
    \label{fig:scaling_cfill_NI}
\end{figure*}

\subsubsection{Product Initial State} 

 In Fig.~\ref{fig:scaling_cfill_NI}(a), we plot the average entanglement entropy for the half-subsystem of the main chain, $S_B$, as a function of the system size $L$ for different values of the parameter $\x$ in the stationary state at long times. Here, the initial state is the product state $|\Psi_p\rangle$, Eq.~\eqref{eq:Psi-p}. We observe the volume-law scaling, $S_B\propto L$, up to the maximum system size ($L=20$) accessible in our numerics. The value of $S_B$ varies with $\x$, however, the overall scaling behavior remains consistent.

In Fig.~\ref{fig:scaling_cfill_NI}(b), we show the long-time average entanglement entropy $S_C$ for the $C$-partition in Fig.~\ref{fig:mom_cartoon}, which characterizes entanglement between the main chain and the auxiliary chain. The numerical finding $S_C=0$ is expected for product initial state as discussed in Sec.~\ref{sec:Role-aux-chain}.  
The complete disentangling between the main chain and the auxiliary chain implies that the ancilla chain facilitates entanglement generation without acting as a bath, unlike what is typically expected for unitary Hamiltonian dynamics. 
Consequently, the entanglement generated in the main chain is genuine quantum correlation, as evidenced by the volume-law scaling of the mutual information $I_{BB'}$ between two halves of the main chain, $B$ and $B'$, shown in Fig.~\ref{fig:scaling_cfill_NI}(c). As expected in this case, the mutual information and the entanglement entropy are related to each other: 
$$I_{BB'}=2 S_B.$$

{As discussed in Appendix \ref{app:TMI}, to probe the nature of the long-time volume-law states further, we compute the tripartite mutual information (TMI) \cite{Cerf1998,Kitaev2006,Levin2006,Iyoda2018,Caceffo2023} for the states. The TMI, $I_3$, is obtained by dividing the system into three subsystems, consisting of sites of the main chain, and the rest of the main and auxiliary chains. A negative $I_3$ implies genuine multi-partite quantum correlations, while $I_3 = 0$ and $I_3>0$ indicate solely bi-partite entanglement and classical correlations, respectively. We find that the long-time distribution of $I_3$ over quantum trajectories is asymmetrical around $I_3=0$, exhibiting much longer tail for $I_3<0$ than that for $I_3>0$. As a result, the average value of $I_3$ is negative. These indicate genuine multi-partite entanglement, beyond just bi-partite quantum correlations, for the long-time volume-law states generated by the one-body-measurement-only dynamics.}

\subsubsection{Random superposition initial state}

\begin{figure*}[htb!]
    \centering    \includegraphics[width=0.9\textwidth]{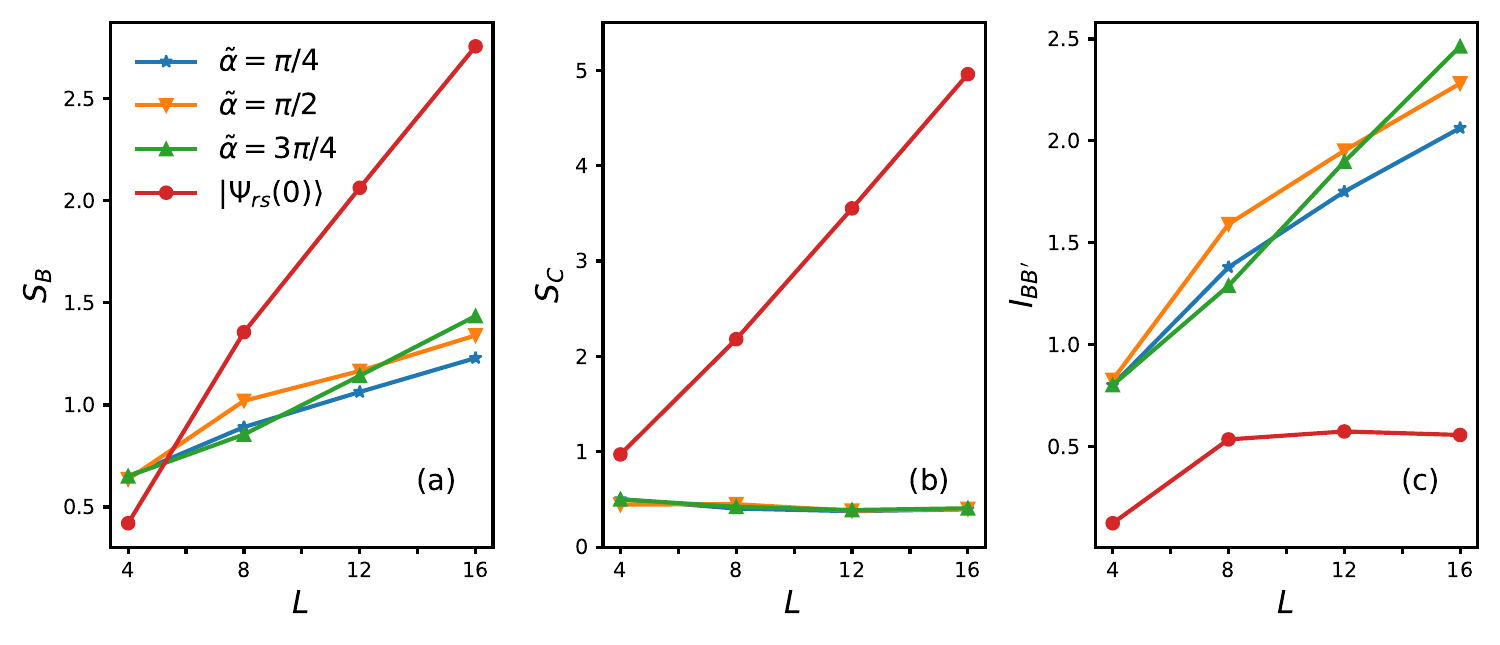}
    \caption{System-size scaling of long-time entanglement entropies and the mutual information for random superposition initial state $|\Psi_{rs}\rangle$ [Eq.~\eqref{eq:Psi-rs}] in the one-body measurement model. (a) Volume-law scaling of entanglement entropy $S_B$ for $B$-partition (Fig.~\ref{fig:mom_cartoon})  is compared with the volume-law scaling of the initial random state. (b)  Entanglement entropy $S_C$ for $C$-partition exhibits the area-law scaling, in contrast to the volume-law scaling of $S_C$ for the initial random state. (c) Long-time average mutual information between two halves of the main chain $I_{BB'}$ shows volume-law scaling, as opposed to the area-law scaling in the initial random state.
    }
    \label{fig:scaling_rands_NI}
\end{figure*}

We now consider how the measurement-only dynamics in the one-body measurement model affect an initially entangled state, such as a random superposition state, which follows a maximal (Page value \cite{Page}) volume-law scaling of entanglement entropy. In Fig.~\ref{fig:scaling_rands_NI}(a), the average entanglement entropy $S_B$ is shown at long times as a function of $L$, demonstrating a linear behavior with $L$. Although the average entanglement entropy $S_B$ decreases relative to its initial maximum value in the random superposition state, the volume-law scaling remains robust under the measurement dynamics. 

In Appendix \ref{sup_sec:average_SC}, we show the trajectory-averaged entanglement entropy $S_C$, characterizing entanglement between the main and ancillary chains, as a function of time for a random superposition initial state. We find that the ancilla chain is initially strongly entangled with the main chain, but over time, this entanglement gradually decreases, as evident by very small entanglement values at long times. In Fig.~\ref{fig:scaling_rands_NI}(b), we plot this small long-time entanglement $S_C$, as a function of $L$. This entropy does not grow with $L$, indicating area-law scaling. Long-time disentanglement between the main and auxiliary chain, suggests that the auxiliary chain eventually ceases to act like a bath to the main chain under the measurement-only dynamics. As a result, we again observe the volume-law scaling of the mutual information $I_{BB'}$ between two parts of the main chain, as shown in Fig.~\ref{fig:scaling_rands_NI}(c).

Typically, in quantum circuits with random unitaries or generic interacting fermionic chains, repeated local projective measurements reduce entanglement, eventually resulting in an area-law phase above a critical measurement strength or rate. Interestingly, we find that our measurement-only quantum dynamics induces (or preserves) the volume-law scaling in the final non-equilibrium stationary state, starting from either a disentangled product state or entangled states. We discuss similar calculations for the atypical entangled equal-superposition initial state $|\Psi_s\rangle$ in Appendix~\ref{sup_sec:scaling_NI}. In this case, however, we find that the system-size scaling of the average mutual information at long times is more consistent with a logarithmic scaling, i.e., $I_{BB'}\sim \log{L}$, modulo, of course, the limitation of small system sizes that we access here.

\subsection{Three-body measurement model: System-size scaling of entanglement entropy and mutual information}

Turning now to the three-body measurement model of Sec.~\ref{sec:IntModel}, we find that the direct interaction between neighboring main chain sites in the system-detector coupling leads to less-entangled long-time stationary states compared to the one-body measurement model. 
This is in contrast to the typical behavior observed in Hamiltonian systems, where interactions generally increase entanglement. This reduction occurs because the particular form of the interaction in Eq.~\eqref{eqn:HsdInt} imposes kinetic constraints on the dynamics, as already discussed. Moreover, we observe that the auxiliary chain gets strongly entangled with the main chain for entangled initial states, unlike in the one-body measurement case. Consequently, the mutual information within the main chain follows the area-law scaling with system size, indicating reduced entanglement generation in the three-body measurement model.

\subsubsection{Random superposition initial state}

\begin{figure*}[htb]
    \centering
    \includegraphics[width=0.95\textwidth]{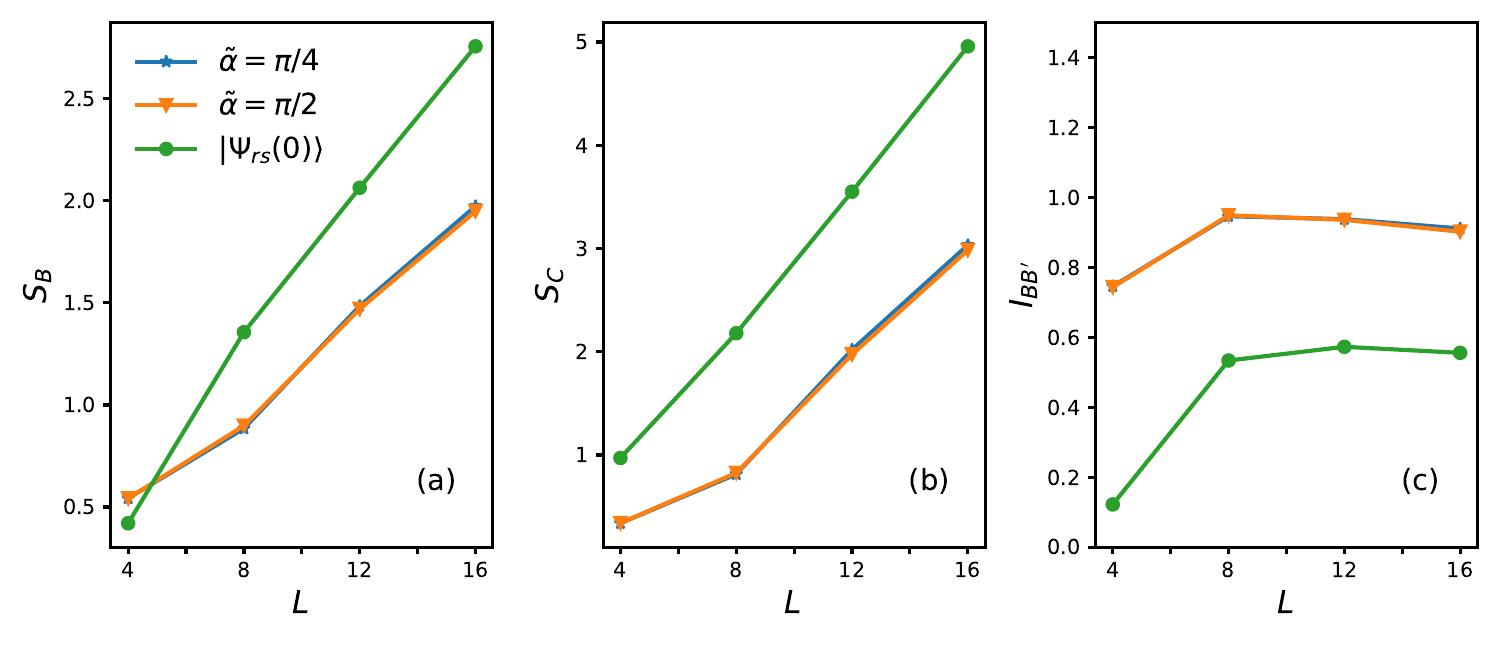}
    \caption{System-size scaling of average stationary-state entanglement entropies and the mutual information for random superposition initial state $|\Psi_{rs}\rangle$, Eq.~\eqref{eq:Psi-rs}, in the three-body measurement model defined in  Sec.~\ref{sec:IntModel}.
(a) The average entanglement entropy $S_B$ in the stationary state is plotted against the system size $L$ and compared with the maximum entanglement entropy of the initial random state.
(b) The entanglement entropy $S_C$, characterizing entanglement between the main chain and the ancilla chain, follows the volume-law scaling, like $S_C$ in the initial random superposition state.
(c) The mutual information $I_{BB'}$ within the main chain follows an area-law scaling similar to the initial random state. The numerically obtained curves for the coupling constants $\x=\pi/4$ and $\pi/2$ (blue and orange) are almost indistinguishable in all three panels. 
}
    \label{fig:Fig_ABCI_int}
\end{figure*}

In Fig.~\ref{fig:Fig_ABCI_int}(a), we present the steady-state trajectory-averaged entanglement entropy $S_B$ for random superposition initial states as a function of system size $L$. The results show a linear dependence of $S_B$ on $L$ within the range of system sizes accessible in our numerics. This indicates that the entanglement entropy between half of the main chain and the rest of the system follows a volume-law scaling, although the average $S_B$ is lower than the initial maximal entanglement entropy of $|\Psi_{rs}\rangle$. We note that the long-time entanglement entropy $S_B$ in the three-body measurement model is larger than that in the one-body measurement model. The results for the values of the coupling parameter $\x = \pi/4, \pi/2$ show very little variation with respect to $\x$.

Next, in Fig.~\ref{fig:Fig_ABCI_int}(b), we plot the average stationary-state entanglement entropy $S_C$, as a function of system size $L$. Here, $S_C$ also follows a volume-law scaling, implying strong entanglement between the main and auxiliary chains at long time in the three-body measurement model, in contrast to the one-body measurement case. We conclude that, in the three-body measurement dynamics, the auxiliary chain acts as a bath for the main chain. 

The stationary value of the average mutual information $I_{BB'}$ between two halves of the main chain is plotted as a function of system size $L$ in Fig.~\ref{fig:Fig_ABCI_int}(c). The results show that the long-time mutual information is higher than its initial value but remains constant as $L$ increases, following an area-law scaling. This finding is in stark contrast to the one-body measurement case, where the ancilla chain effectively disentangles from the main chain, which is accompanied by establishing a volume-law scaling of the mutual information in the main chain.


\begin{figure*}[htb]
    \centering
    \includegraphics[width=0.95\textwidth]{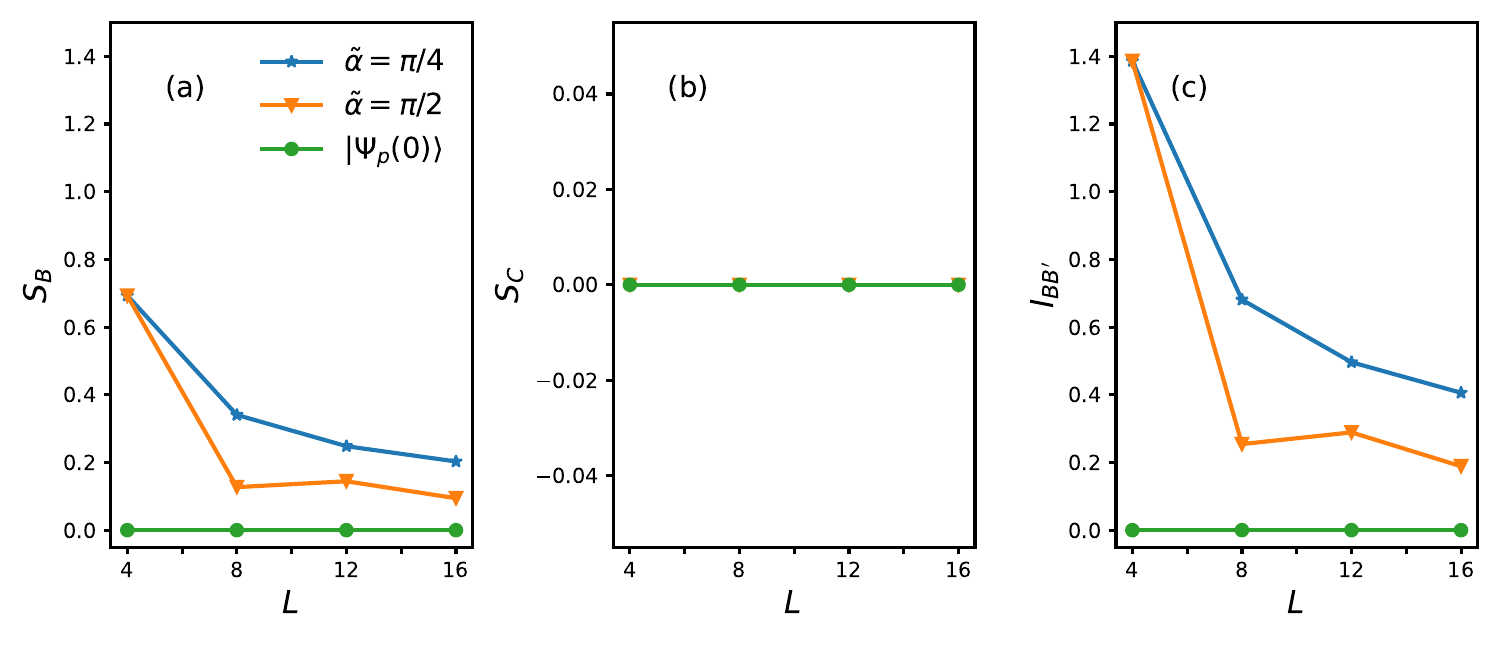}
    \caption{System-size scaling of average stationary state entanglement entropies and mutual information for product initial state $|\Psi_{p}\rangle$, Eq.~\eqref{eq:Psi-p}, in the three-body measurement model.
(a) The average entanglement entropy $S_B$ in the stationary state decreases with system size $L$.
(b) The entanglement entropy $S_C$ vanishes as in the initial product state.
(c) The mutual information $I_{BB'}=2S_B$ (exactly) within the main chain decreases with $L$.
}
    \label{fig:Fig_ABCI_int_prod}
\end{figure*}

\subsubsection{Product Initial State}

Finally, in Fig.~\ref{fig:Fig_ABCI_int_prod}(a), the long-time average entanglement entropy $S_B$ is shown as a function of system size $L$ for an unentangled initial state. As discussed in Sec.~\ref{sec:IntModel}, the majority of quantum trajectories in this case did not generate entanglement when starting from a product state. We find that the stationary average entropy $S_B$ does not increase with system size, obeying an area-law scaling.

The entanglement entropy $S_C$ is shown in Fig.~\ref{fig:Fig_ABCI_int_prod}(b). We see that $S_C$ is exactly zero, as expected under the measurement dynamics for the product initial state discussed in Sec.~\ref{sec:Role-aux-chain}. The mutual information between two halves of the main chain, $I_{BB'}$ also follows an area-law behavior, as shown in Fig.~\ref{fig:Fig_ABCI_int_prod}(c).
Thus, unlike the one-body measurement-only dynamics with volume law-mutual information generation in the main chain, only area-law entanglement generation is observed in the three-body measurement model. As discussed above, this can be traced back to the effect of kinetic constraints that effectively block entangling dynamics.

\section{Discussion and conclusion}\label{sec:Discussion}

In this work, we have constructed models of generalized local measurement-only dynamics capable of producing highly entangled volume-law states at long times. Our models invoke explicit local non-random system-detector Hamiltonian and consequent non-commuting measurements involving a main and an ancilla fermionic chains, as well as a set of detector qubits. These measurement setups differ in terms of locality of measurements from those employed in several recent works \cite{Ippoliti_Khemani_PRX,  Nicolai_PRB,  Qian_2024, Cheng_2024, Vu_PRL2024, Kuno_PRB2023} on the generation of entanglement via only random multi-site measurements. We illustrate rich statistical properties of entanglement in ensembles of quantum trajectories generated through the local-measurement-only dynamics starting from different types of unentangled and entangled initial states, as well as the resultant approach to stationary distributions or stationary states. 

Curiously, we demonstrate that volume-law entangled states can be generated via simple one-body measurements. This result is in striking variance with many recent studies of non-interacting fermions, evolving under both entangling unitary dynamics and local measurements, where one can typically achieve area-law entanglement in the long-time steady state, or at best, entanglement varying logarithmically with system size. 

In Ref.~\cite{Lumia2023}, the authors demonstrate a measurement-induced phase transition by weakly breaking Gaussianity through current measurements within an effective single-particle description based on natural orbitals of the one-body reduced density matrix. In the measurement-only limit (very high measurement rate), Ref.~\cite{Lumia2023} observed the area-law phase.   
In our work, by breaking Gaussianity via measurement operators in a measurement-only dynamics setting, we show that volume-law entanglement generation is indeed possible for one-body Gaussianity-breaking measurement operators in the absence of the intrinsic dynamics in the system.

Further, by inducing direct density-density interaction within the main chain through the system-detector Hamiltonian, we illustrate how the generation of entanglement can be manipulated and reduced by introducing kinetic constraints on the dynamics. Intriguingly, we show that the kinetic constraints can self-tune the system to an entirely \emph{no-click} trajectory with non-trivial dynamics for certain types of initial states, thus producing a purely non-stochastic dynamics describable by a non-Hermitian model without the need for post-selection.

In future, it will be worthwhile to study the continuous-time limit of dynamics, which can be described by quantum-state diffusion, by a stochastic Schr{\"o}dinger equation \cite{Ghirardi1990, CaoLuca,  AlbertonDiehl}, or a sigma-model-type effective field theory \cite{Nahum, Jian2023,  Igor2, Poboiko2024, Chahine2024, Starchl2024, Poboiko2025, guo2024,  Poboiko2025Levy}.
It will be interesting to see whether the volume-law entangled states are generated in this continuum limit for the one-body measurement model.
Finally, by randomly employing the three-body measurement and one-body measurement operators, one can expect to drive the measurement-induced transitions within this measurement-only framework.

{Finally, we emphasize that, beyond the specific models considered here, the main objective of our work is to demonstrate that strongly entangled quantum states can be generated only through repeated quasi-local non-random generalized measurements. The main ingredient for entanglement generation here is the non-commutativity of the measurement operators. Our work elucidates this broad principle, going beyond earlier work on measurement-only models with repeated random projective measurements \cite{Ippoliti_Khemani_PRX}. The specific generalized measurement models with two fermionic chains, the main and the auxiliary chains, coupled to qubit detectors, has been taken for the purpose of concreteness.}

{When the main and auxiliary chains are viewed as parts of the same system, our explicit separation into two chains is merely a specific constraint on the measurement operators that now distinguish between the sites of the entire system: only every second site is directly coupled to the detector. This emphasizes that local discrete symmetry can be important for measurement-induced phenomena, and tuning the nature microscopic couplings between the system and detectors may strongly affect the resulting phases of a monitored system. Other complementary examples of such a microscopically-structured knob are presented in Refs.~\cite{Lumia2023} and \cite{Poboiko2025Levy}, where measurements involving two sites (e.g., monitoring local currents in Ref.~\cite{Lumia2023}) were shown to \textit{enhance} entanglement. Here, we have shown that structuring the quasi-local measurement operators is a promising way to  \textit{generate the volume-law entanglement} in the system without any underlying unitary dynamics.}

{In fact, the one-body measurement-only model, which leads to the volume-law entanglement generation, has fairly generic fermionic hopping terms coupled to qubit detectors or spins, and thus is realizable in more general quantum many-body systems. The analogous terms with similar physics can be easily written for bosons or qubits (spins). 
As a result, our one-body measurement model is realizable in superconducting qubit-based systems \cite{Makhlin2001,Kjaergaard2020}, which are well-suited for system-detector-based generalized measurements setups \cite{Averin2000,Weber2014,Vijay2012,Murch2013,Koh2022,Lin2025}, including mid-circuit measurements.}


\section{Acknowledgments}
We thank I. Poboiko for useful discussions. Y.G. and I.V.G. were supported by the Deutsche Forschungsgemeinschaft (DFG,
German Research Foundation) through grants SH 81/8-1 (Y.G.) and GO 1405/7-1 (I.V.G. and Y.G.). Y.G. also acknowledges support by the National Science Foundation (NSF)--Binational Science Foundation (BSF) through grant 2023666. S.Ba. acknowledges support from CRG, SERB (ANRF), DST, India (File No. CRG/2022/001062), and STARS, MoE, Govt. of India (File. No. MoE-STARS/STARS-2/2023-0716).


\appendix

\section{Action of system-detector Hamiltonian on basis states}\label{sup_sec:Uact}

In this Appendix, we derive the action of the unitary operator $$\mathcal{U}_i = \exp\left[{-\ci H^{(i,i+1)}_{sd}}\right],$$ 
given by Eq.~(\ref{eq:UnitaryOp}) of the main text, on the computational basis states of the system-detector setup. 
The computational-basis states for the system are spanned by the occupation numbers of the $L$ fermionic sites with a fixed particle number $L/2$:  
$$
|n\rangle = |n^c_1 \ldots n^c_i \ldots n^c_{L/2}\rangle |n^a_1 \ldots n^a_i \ldots n^a_{L/2}\rangle,
$$
where $n^c$ and $n^a$ denote the site occupations of the main chain and the ancilla chain, respectively.

As mentioned in the main text, in our measurement dynamics, the system-detector Hamiltonian $H^{(i,i+1)}_{sd}$ is switched on for an instantaneous time in a blockwise manner. Consequently, the Hamiltonian $H^{(i,i+1)}_{sd}$ acts non-trivially, depending on the occupation numbers of the fermionic sites in the $i$-th block. Below, we analyze the action of $\mathcal{U}_i$ on the effective part of the basis states spanned by  
$$
|n\rangle_i = |n^c_in^c_{i+1}\rangle |n^a_i\rangle,
$$
where $|n\rangle_i$ represents the effective part of the computational basis $|n\rangle$ for the $i$-th block. The $i$-th block consists of two main chain sites and one ancilla site, giving a total of three sites. We can determine the action of $\mathcal{U}_i$ on each particle sector $N_p = 0, 1, 2, 3$ separately, as the particle number is a good quantum number in our model. 

For $N_p=1$ sector, the basis states are given by:
 \begin{subequations}
    \begin{align}
 |10\rangle|0\rangle &=c^{\dagger}_{i}|\text{vac}\rangle, 
 \\
|00\rangle|1\rangle&=a^{\dagger}_i|\text{vac}\rangle, 
\\
  |01\rangle|0\rangle &= c^{\dagger}_{i+1}|\text{vac}\rangle.
\end{align} 
 \end{subequations}
The basis states for $N_p=2$ sector are as follows:
\begin{subequations}
   \begin{align}
    |10\rangle|1\rangle &=c^{\dagger}_i a^{\dagger}_i|\text{vac}\rangle,
    \\
    |01\rangle|1\rangle &= a^{\dagger}_ic^{\dagger}_{i+1}|\text{vac}\rangle,
    \\
    |11\rangle|0\rangle &=c^{\dagger}_{i}c^{\dagger}_{i+1}|\text{vac}\rangle.
\end{align} 
\end{subequations}
For $N_p=0$ sector, the basis state is $|00\rangle|0\rangle$, and for $N_p=3$ sector, the basis state is $$|11\rangle|1\rangle=c^{\dagger}_i a^{\dagger}_i c^{\dagger}_{i+1}|{\rm vac}\rangle.$$  Below, we find the action of the Hamiltonian and the evolution operator $\mathcal{U}_i$  on these basis states for both the one-body measurement and three-body measurement models.

\subsection{One-body measurement model}\label{sup_sec:nonintUact}
For better readability, we write the system-detector Hamiltonian, Eq.~(\ref{eq:HsdNonInt}) for one-body measurement model here:
$$
    H^{(i, i+1)}_{sd} = \alpha\,(c^{\dagger}_i + c^{\dagger}_{i+1})\,a_i\, \sigma^{x}_{i} + \text{H.c.} $$
The Hamiltonian (\ref{eq:HsdNonInt}) acts on the system-detector basis states $|n\rangle_i |\sigma\rangle$ (where $|\sigma\rangle$ represents the spin state ${\sigma=\uparrow,\downarrow}$) as follows:
\begin{subequations}\label{sup_eq:actionH_Np1}
    \begin{align}
        H^{(i,i+1)}_{sd} |00\rangle|1\rangle|\sigma\rangle &= \alpha ( |10\rangle|0\rangle +  |01\rangle|0\rangle)|\overline{\sigma}\rangle,
        \\
        H^{(i,i+1)}_{sd}  |01\rangle|0\rangle|\sigma\rangle &= \alpha |00\rangle|1\rangle|\overline{\sigma}\rangle, 
        \\
        H^{(i,i+1)}_{sd} |10\rangle|0\rangle|\sigma\rangle &= \alpha |00\rangle|1\rangle|\overline{\sigma}\rangle.
    \end{align}
\end{subequations}
Here and below, $|\overline{\sigma}\rangle$ represents the spin state opposite to $|\sigma\rangle$. 
The action of the unitary operator $\mathcal{U}_i$ can be determined by expanding $e^{-iH}$ in a Taylor series as follows:
\begin{align}\label{sup_eqn:Taylor}
    \exp(-iH) &= 1 - iH + \frac{(-i)^2}{2!}H^2 + \frac{(-i)^3}{3!}H^3 + \dots
\end{align}

As mentioned in the main text, the detector qubit is initialized in the spin-up state ${|\!\uparrow\rangle}$ before switching on $H_{sd}$. Therefore, we consider the action of $\mathcal{U}_i$ in the spin-up sector. By regrouping the terms for the spin-up and spin-down amplitudes in Eq.~(\ref{sup_eqn:Taylor}), we immediately obtain Eqs.~\eqref{eqn:nonint_Ua}, \eqref{eqn:nonint_Ub}, and \eqref{eqn:nonint_Uc} of the main text.

Similarly, we repeat the above steps to determine the action of  $\mathcal{U}_i$ on the basis states in the $N_p=2$ sector. The Hamiltonian (\ref{eq:HsdNonInt}) acts on the system-detector basis states in the $N_p=2$ sector as follows:
\begin{subequations}
    \begin{align}
        H^{(i,i+1)}_{sd}  |10\rangle|1\rangle|\sigma\rangle &= \alpha |11\rangle|0\rangle|\overline{\sigma}\rangle,
        \\
        H^{(i,i+1)}_{sd} |01\rangle|1\rangle|\sigma\rangle &= \alpha |11\rangle|0\rangle|\overline{\sigma}\rangle,
        \\
        H^{(i,i+1)}_{sd} |11\rangle|0\rangle|\sigma\rangle &= \alpha ( |01\rangle|1\rangle +  |10\rangle|1\rangle)|\overline{\sigma}\rangle. 
    \end{align}
\end{subequations}
As before, by regrouping the terms for the up-spin and down-spin amplitudes in Equation \ref{sup_eqn:Taylor}, we obtain the following set of equations: 
\begin{subequations}\label{sup_eq:UNp2}
\begin{align}
\mathcal{U}_i |10\rangle|1\rangle|\!\uparrow\rangle &= \bigg({\frac{\cos \x+1}{2}|10\rangle} + \frac{\cos \x-1}{2}|01\rangle\bigg)|1 \rangle |\!\uparrow \rangle \label{sup_eqn:nonint_UaNp2}
\notag \\ &
- \frac{i\sin{\x}}{\sqrt{2}} |11\rangle|0\rangle |\!\downarrow\rangle \\
\mathcal{U}_i |01\rangle|1\rangle|\!\uparrow\rangle &= \bigg({\frac{\cos \x+1}{2}|01\rangle} + \frac{\cos \x-1}{2}|10\rangle\bigg)|1 \rangle |\!\uparrow \rangle  \label{sup_eqn:nonint_UbNp2}
\notag \\ &
- \frac{i\sin{\x}}{\sqrt{2}} |11\rangle|0\rangle |\!\downarrow\rangle \\
\mathcal{U}_i|11\rangle |0\rangle  |\!\uparrow\rangle &= \cos{\x}|11\rangle|0\rangle |\!\uparrow \rangle  \notag\\
&-\frac{i\sin \x}{\sqrt{2}} \big(|10\rangle + |01\rangle  \big) |1 \rangle|\!\downarrow\rangle  \label{sup_eqn:nonint_UcNp2}
\end{align}
\end{subequations}
Notice that these results can also be obtained by simply interchanging $0$ and $1$ in Eqs.~\eqref{eqn:nonint_Ua}, \eqref{eqn:nonint_Ub}, and \eqref{eqn:nonint_Uc} derived for the $N_p = 1$ sector. This is because of the particle-hole symmetry of the Hamiltonian. 

Finally, in the $N_p = 0$ and $N_p = 3$ sectors, the action of the unitary operator is trivially the identity operation. Indeed, the Hamiltonian acting on the basis states $|00\rangle|0\rangle$ and  $|11\rangle|1\rangle$ in these sectors has a zero eigenvalue.

\subsection{Three-body measurement model}\label{sup_sec:intUact}

Again, for the reader's convenience, we present  the system-detector Hamiltonian for the three-body measurement model, given by Eq.~\eqref{eqn:HsdInt}, here:
\begin{align*}
 \tilde{H}^{(i, i+1)}_{\rm s-d}\! &= \alpha\Big(\big[a^{\dagger}_i\,(c_i+c_{i+1})\,n_i n_{i+1} \big]\otimes \sigma^{-}  \notag
 \\   +&\big[(c^{\dagger}_i+c^{\dagger}_{i+1})\,a_i\,(1-n_i) (1-n_{i+1}) \big]\otimes \sigma^{-}\! + {\rm H.c.} \Big). 
\end{align*}
This Hamiltonian acts on the basis states in the $N_p=1$ particle sector and $|\sigma=\uparrow\rangle$ spin sector as follows:
 \begin{subequations}
 \begin{align}
\tilde{H}^{(i,i+1)}_{sd} |00\rangle|1\rangle|\!\uparrow\rangle &= \alpha (|01\rangle|0\rangle+|10\rangle|0\rangle)|\!\downarrow\rangle, \\  
\tilde{H}^{(i,i+1)}_{sd} |01\rangle|0\rangle|\!\uparrow\rangle &= 0,\\
\tilde{H}^{(i,i+1)}_{sd} |10\rangle|0\rangle|\!\uparrow\rangle &= 0.
\end{align}    
 \end{subequations}
For the $N_p=1$ particle sector and $|\sigma=\downarrow\rangle$ spin sector, we get:
\begin{subequations}
  \begin{align}
     \tilde{H}^{(i,i+1)}_{sd} |00\rangle|1\rangle|\!\downarrow\rangle &= 0, \\  
     \tilde{H}^{(i,i+1)}_{sd} |01\rangle|0\rangle|\!\downarrow\rangle &= \alpha |00\rangle|1\rangle|\!\uparrow\rangle,\\
    \tilde{H}^{(i,i+1)}_{sd} |10\rangle|0\rangle|\!\downarrow\rangle &=\alpha |00\rangle|1\rangle|\!\uparrow\rangle.
\end{align}  
\end{subequations}

By regrouping the spin-up and spin-down amplitudes in the expansion of the unitary operator, Eq.~\eqref{sup_eqn:Taylor}, we obtain  Eqs.~\eqref{eqn:int_Ua}, \eqref{eqn:int_Ub}, and \eqref{eqn:int_Uc} of the main text.
Similarly, we obtain the action of the unitary operator in the $N_p=2$ sector by knowing the action of the Hamiltonian on basis states and then using Eq.~\eqref{sup_eqn:Taylor} as in the one-body measurement model. Since the three-body measurement Hamiltonian also has particle-hole symmetry, we can obtain this by interchanging $0$ and $1$ in Eqs.~\eqref{eqn:int_Ua}, \eqref{eqn:int_Ub}, and \eqref{eqn:int_Uc} derived for the $N_p=1$ sector.
The action of the unitary operator in the $N_p=0$ or $N_p=3$ sectors is again trivially the identity operation.

\begin{figure}[H]
    \centering
\includegraphics[width=1.0\linewidth]{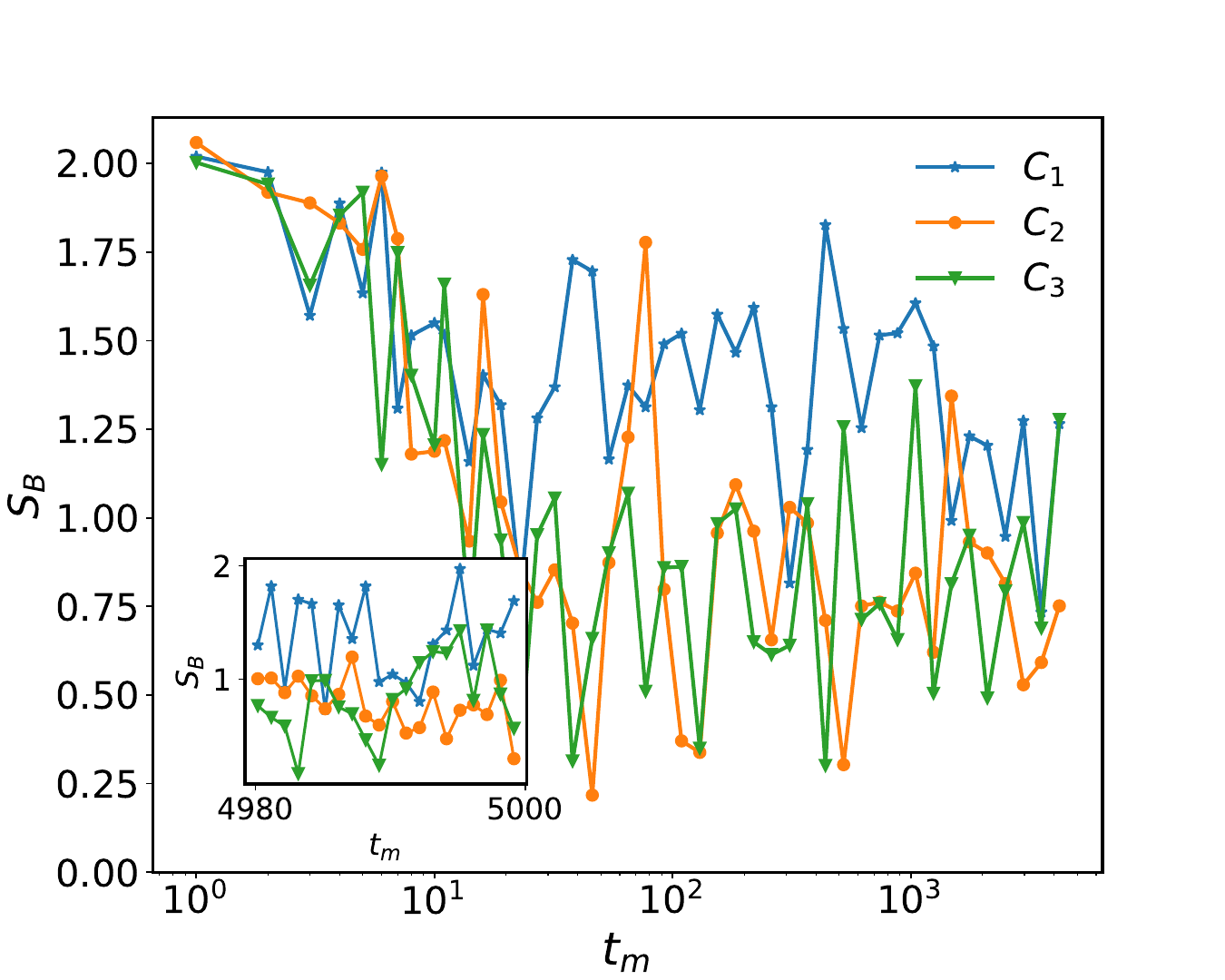}
\caption{$S_B$ as a function of discrete time $t_m$ is shown on a log-normal scale for three representative trajectories ($C_i$). The inset shows $S_B$ as a function of $t_m$ for the last 20 measurement sweeps on a normal scale. The initial state is $|\Psi_{rs}\rangle$ with $\tilde{\alpha} = \pi/4$ computed from system size $L=12$.
}
\label{fig:Nonint_rands_traj}
\end{figure}
\begin{figure}[htb]
    \centering
\includegraphics[width=1.0\linewidth]{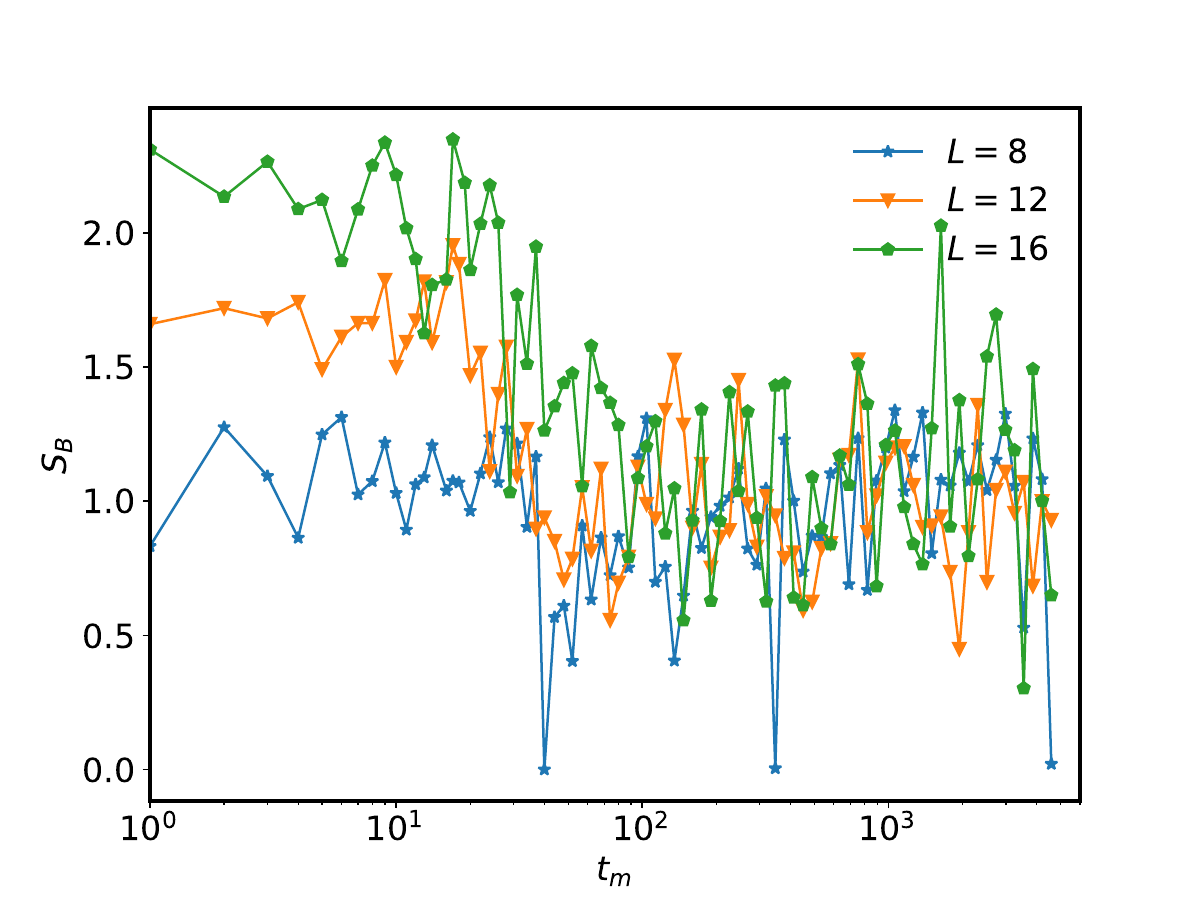}
\caption{$S_B$ as a function of discrete time $t_m$ is shown on a log-normal scale for  arbitrary representative trajectories from the system sizes $L=8, 12, 16$.  The initial state is $|\Psi_{rs}\rangle$ with $\tilde{\alpha} = \pi/4$. 
}
\label{fig:Nonint_rands_traj_vs_N}
\end{figure}

\section{Quantum trajectories and long-time states  in one-body measurement model}\label{sec:appendix_rs_state}

In this Appendix, we present an analysis of quantum trajectories and the stationarity check for a random superposition state $|\Psi_{rs}\rangle$, Eq.~(\ref{eq:Psi-rs}), for the one-body measurement model. We find a very similar behavior for the atypical equal superposition state $|\Psi_s\rangle$,  Eq.~(\ref{eq:Psi-es}), which we do not show here.

\subsection{Entanglement along quantum trajectories}\label{sup_sec:QuantumTraj_PsiRs}
In Fig.~\ref{fig:Nonint_rands_traj}, we show the $B$-subsystem entanglement entropy $S_B$ as a function of discrete measurement time $t_m$ on a log-normal scale for three representative quantum trajectories ($C_i$, $i=1, 2, 3$) under the measurement-only dynamics for a random superposition initial state in the one-body measurement model. Similar to the product initial state, as discussed in the main text, we find that individual quantum trajectories do not reach a stationary state or exhibit stationary values of $S_B$, instead, they fluctuate. Therefore, the individual trajectories do not reach stationary states.

In Fig.~\ref{fig:Nonint_rands_traj_vs_N}, we compare the trajectories generated from different system sizes $L=8, 12, 16$. This shows that the fluctuation of $S_B(t)$ does not decrease with increasing system sizes.

\subsection{Stationarity check for long-time distribution}
\begin{figure}[htb]
    \centering
\includegraphics[width=1.0\linewidth]{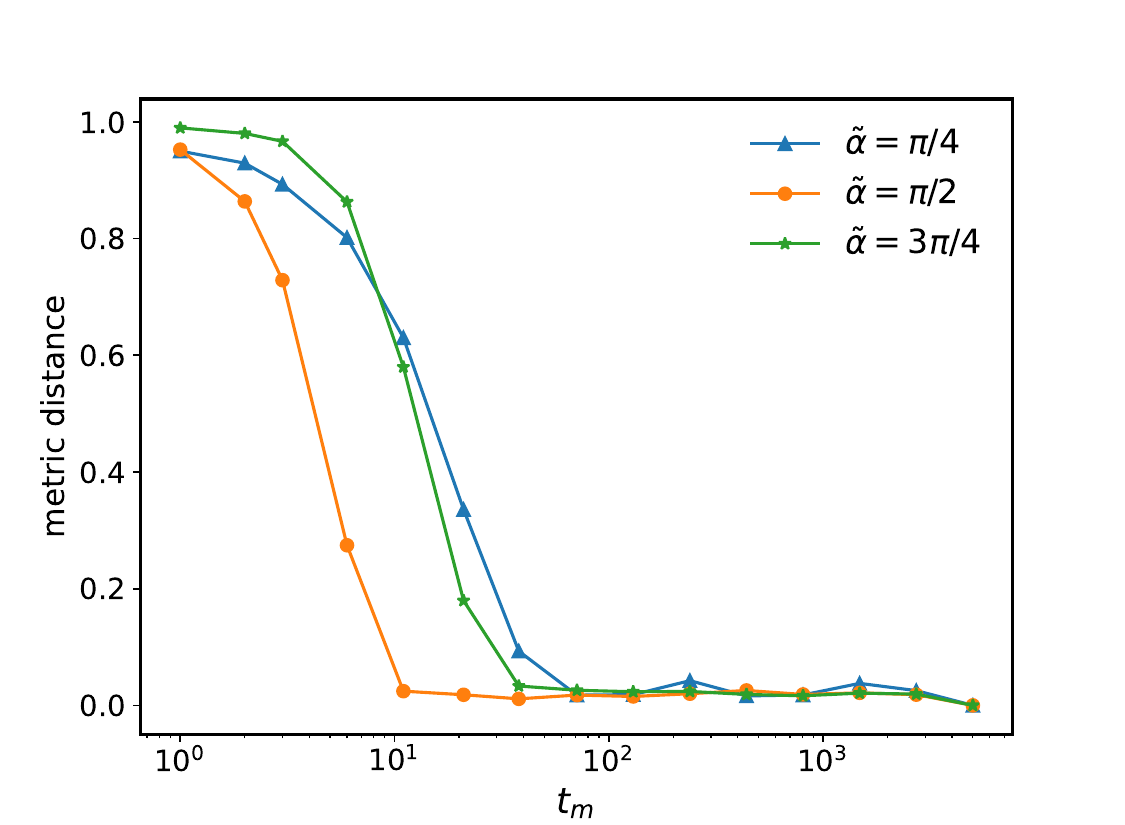}
\caption{The total variation distance (TVD) between the distribution at time \( t_m \) and the long-time distribution at {$t_m=5000$} is plotted as a function of \( t_m \) for different values of \( \tilde{\alpha} \). The results show a rapid decrease in TVD, indicating convergence to a stationary distribution. This is shown for $|\Psi_{rs}\rangle$ from $L=12$.}
\label{fig:Nonint_rands_TVD}
\end{figure}

We examine whether the long-time distribution for the initial state $|\Psi_{rs}\rangle$ reaches a stationary distribution, similar to the product initial state discussed in the main text. To this end, we compute the metric distance (TVD) introduced earlier. Using KDE, we smooth the distribution of $S_B$ at various time instances and compute the TVD between the distribution at time $t_m$ and the large-time distribution at $t_m = 5000$. In Fig.~\ref{fig:Nonint_rands_TVD}, we plot the metric distance as a function of $t_m$. The results show that the metric distance quickly approaches very small values, indicating the stationarity of the distribution.

\section{Ergodicity check of long-time distribution in one-body measurement model}

\subsection{Entanglement distribution over a time-window} \label{sup_sec:NI_prod_timewindowSB}
In the main text (Fig.~\ref{fig:Exp1_psi_rs}), we presented the distribution of entanglement entropy $S_B$ over a time window $t_m \in [1000, 5000]$ for a few trajectories, comparing it to the stationary distribution at $t_m = 5000$ for an initial random superposition state.

\begin{figure}[htb]
    \centering
\includegraphics[width=1.0\linewidth]{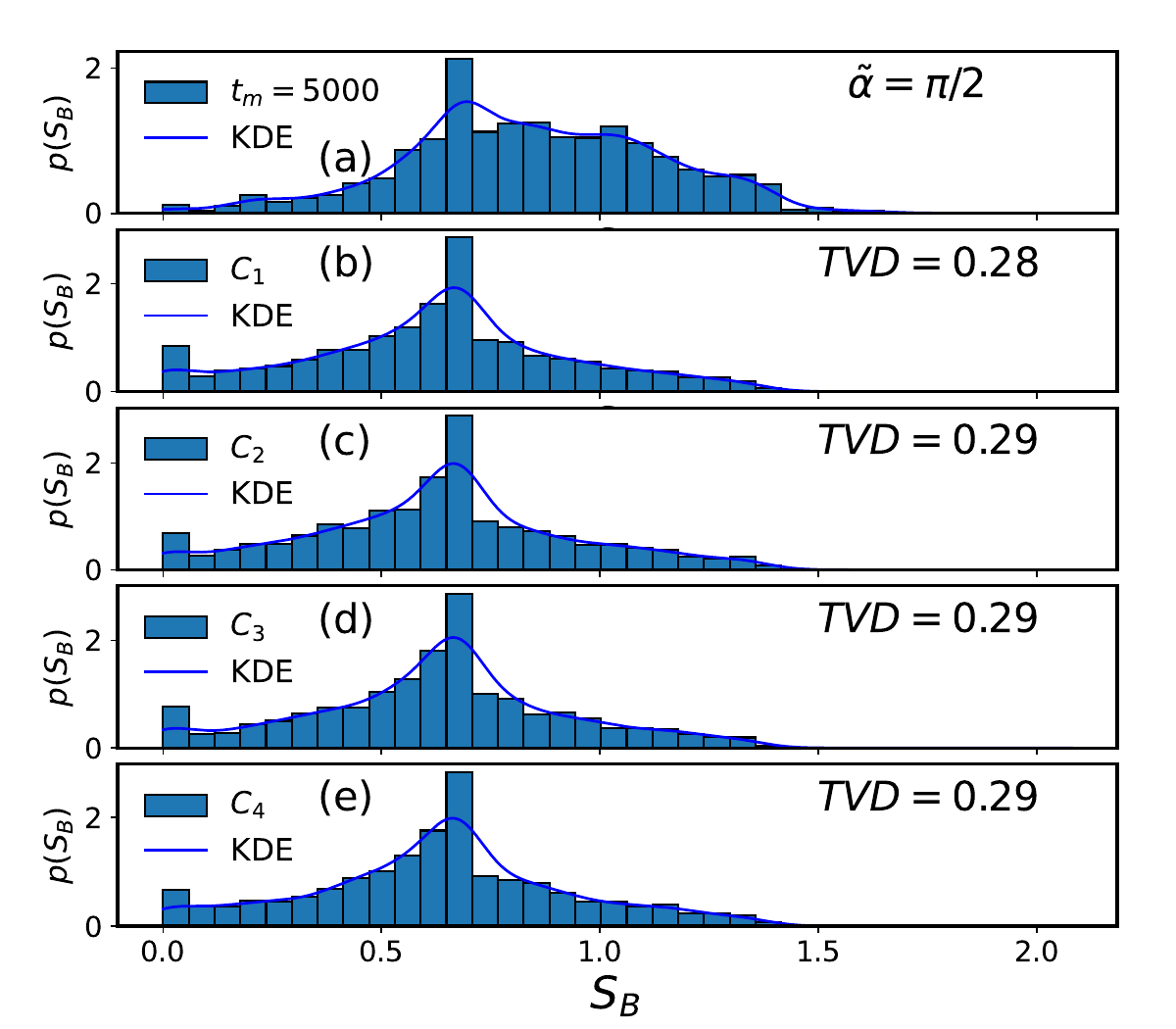}
\caption{{(a)} The stationary probability density of $S_B$ over the trajectory ensemble at large time $t_m=5000$ for an initial product state and $\tilde{\alpha}=\pi/2$. {(b-e)} show the distributions over a long time interval, from $t_m=1000$ to $t_m=5000$, for four trajectories. The trajectory in {(b)} has the maximum Born probability among all the trajectories. We show the corresponding smooth (KDE) approximations to the distributions in {(b-e)} along with their metric distances from the smooth approximation to the stationary distribution in {(a)}.}
\label{fig:Exp1_psi_rp}
\end{figure}

Here, we perform a similar analysis for a product initial state under one-body, measurement-only dynamics. In Fig.~\ref{fig:Exp1_psi_rp}(a), we show the stationary probability density of $S_B$ at $t_m = 5000$. Figures \ref{fig:Exp1_psi_rp}(b-e) present the probability density of $S_B$ over the time window $t_m \in [1000, 5000]$ for representative trajectories labeled $\mathcal{C}_k = 1, 2, 3, 4$. The distributions are smoothed using kernel density estimation (KDE), and the total variation distance (TVD) between the time-window distributions of individual trajectories and the stationary distribution is computed, as indicated in Figs.\ref{fig:Exp1_psi_rp}(b-e). Similar to  the results for the random superposition initial state discussed in the main text, we find that the time-window distribution of a quantum trajectory starting from product initial does not converge to the stationary  distribution of the trajectory ensemble. This demonstrates that the time-averaged entanglement entropy is not equivalent to the avearge over stationary distribution.

\subsection{Initial state dependence of the stationary distribution}\label{sup_sec:PsiRsInitialStateDependence}

\begin{figure}[htb]
    \centering
    \includegraphics[width=0.95\linewidth]{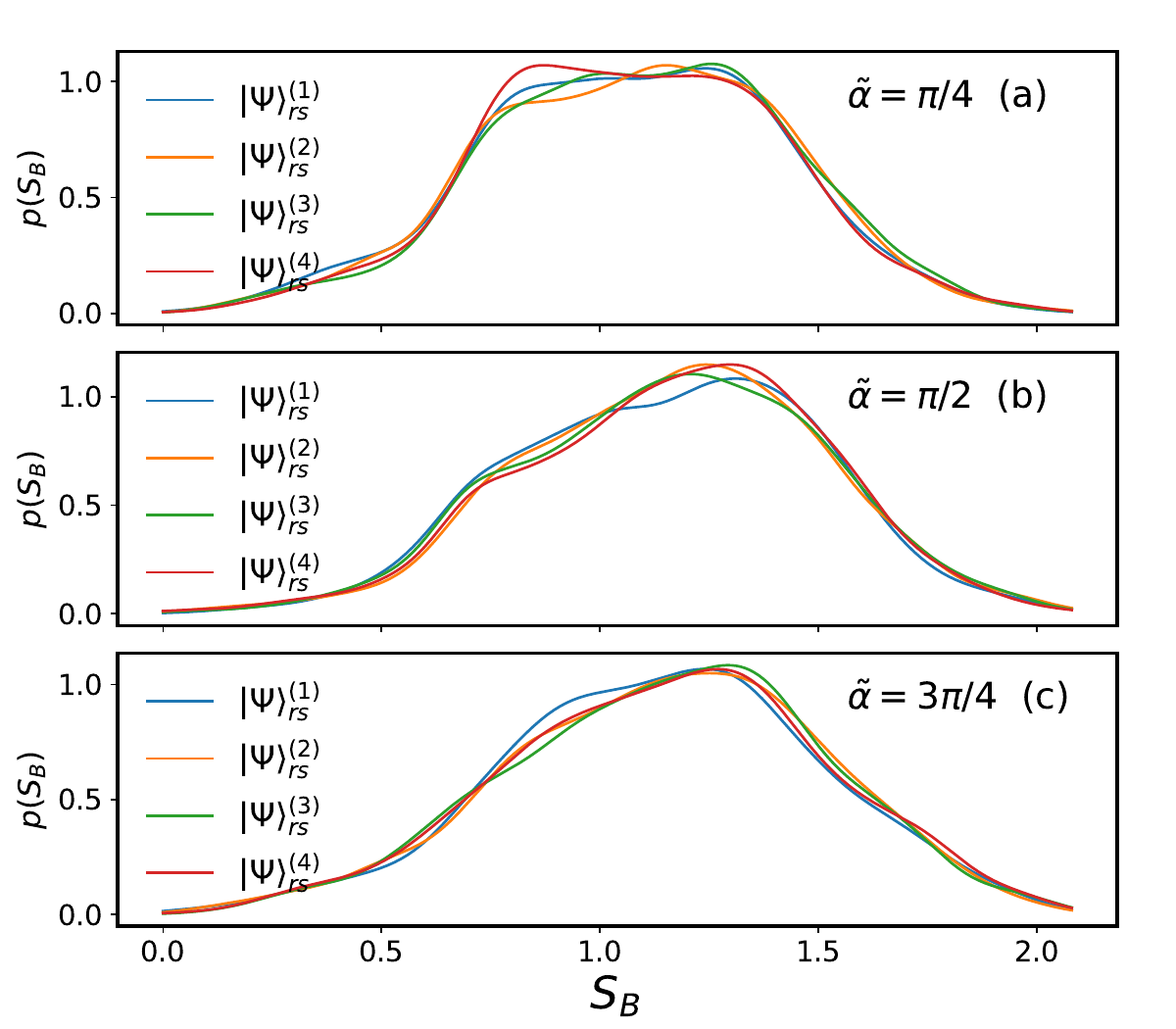}
    \caption{Stationary probability density function  $p(S_B)$ obtained from KDE approximation of the probability density distributions over ensemble of trajectories for different random superposition initial states for three values of $\tilde{\alpha}=\pi/4, \pi/2, 3\pi/4$.}
    \label{fig:PD_diff_psi_rs_longtime}
\end{figure}

In the main text (Fig.~\ref{fig:PD_diff_psi_rp_longtime}), we discussed that the stationary distributions for an ensemble of trajectories originating from a set of random product states lead to very similar stationary distributions. Here, we perform a similar analysis for a set of random superposition states.

It is well known that random superposition states exhibit the same entanglement entropy for any subsystem cut in the thermodynamic limit. For finite systems, the entanglement entropy for different $|\Psi_{rs}\rangle$ fluctuates around an average value, following a volume law scaling with the system size. We generate the stationary distribution of entanglement for each random superposition state $|\Psi^{(i)}_{rs}\rangle$ ($i = 1, 2, 3, 4$) and extract the smoothed probability density function $p(S_B)$. In Fig.~\ref{fig:PD_diff_psi_rs_longtime}, we show $p(S_B)$ for different $|\Psi^{(i)}_{rs}\rangle$ across various values of the parameter $\tilde{\alpha}$. The curves in each panel are visually similar, and the total variation distance (TVD) between $p(S_B)$ in each case approaches a very small value ($< 0.05$). This demonstrates that the stationary distributions are similar for a set of initial states of the same type, categorized by their initial state entanglement, similar to the results for random product states (Fig.~\ref{fig:PD_diff_psi_rp_longtime}).

\subsection{Entanglement distribution from  quantum trajectories originating from different initial states of a given type }

Here, we consider the distribution of entanglement entropy $S_B$ for an ensemble of trajectories, where each trajectory originates from a different initial state of a given type, such as a set of random superposition states. We investigate whether this distribution reaches a stationary distribution, and if so, whether it reaches the stationary distribution constructed from a single initial state.

\begin{figure}[htb]
    \centering
    \includegraphics[width=0.99\linewidth]{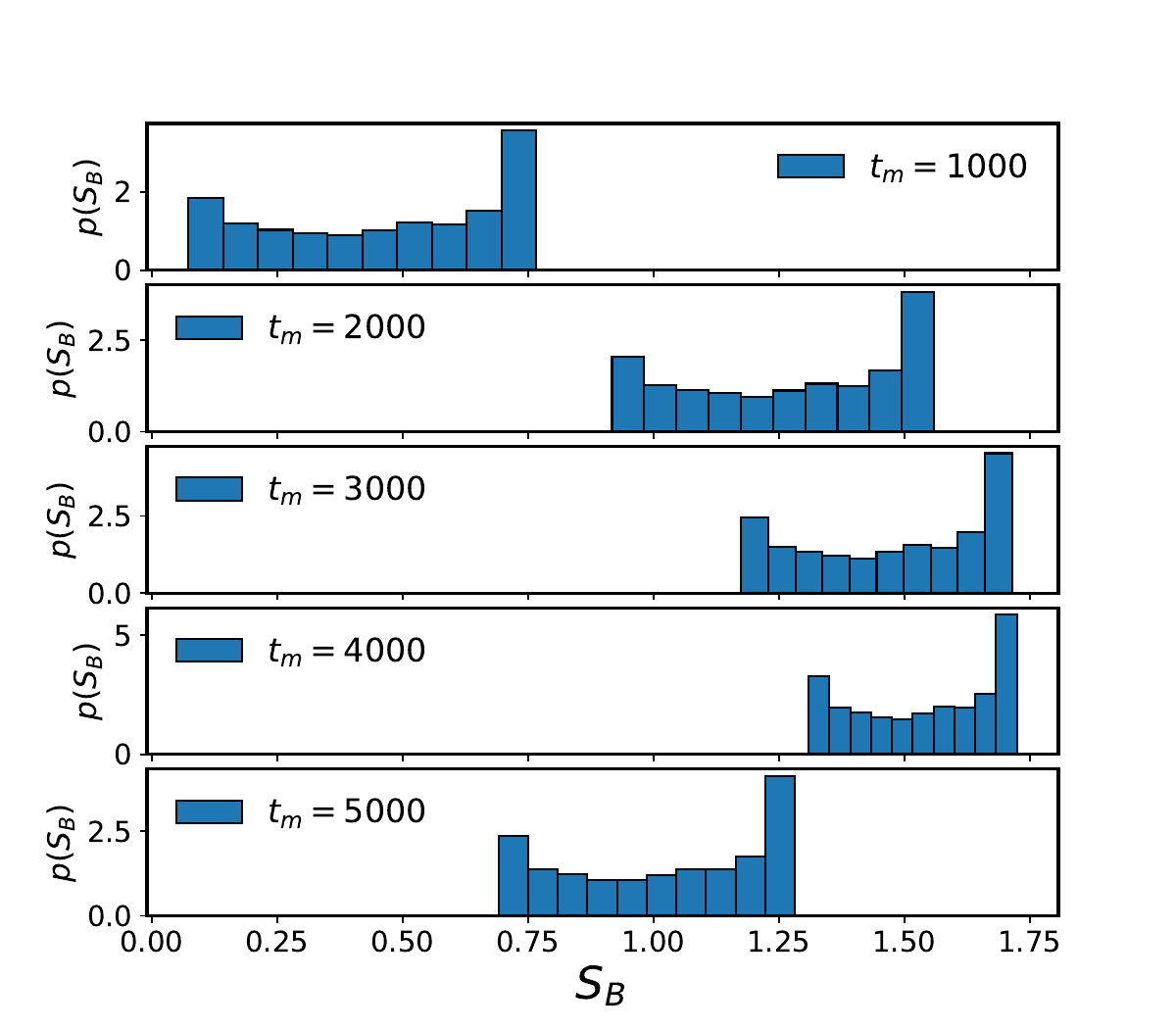}
    \caption{The probability density function $p(S_B)$ constructed from ensemble of $|\Psi_{rs}\rangle$ are shown at different time $t_m$ as indicated in  the legend in different panel. This is computed for $\x=\pi/4$ and $L=12$}
    \label{fig:exp3_Psi_rs}
\end{figure}

To this end, we construct the probability density function $p(S_B)$ at time $t_m$ for the ensemble of $|\Psi_{rs}\rangle$, where each initial state contributes a single representative trajectory. In Fig.~\ref{fig:exp3_Psi_rs}, we show $p(S_B)$ at different values of $t_m$, as indicated in the legend of each panel. Clearly, the $p(S_B)$ at different $t_m$ are not identical. This demonstrates that the $p(S_B)$ constructed from a single representative trajectory of an ensemble of initial states does not reach a stationary distribution. We expect a similar phenomenon to occur for a set of random product states as well.

\section{Time evolution of average  $S_C$ in one-body measurement model for $|\Psi_{\rm rs}\rangle$}\label{sup_sec:average_SC}



\begin{figure}[htb]
    \centering
    \includegraphics[width=0.99\linewidth]{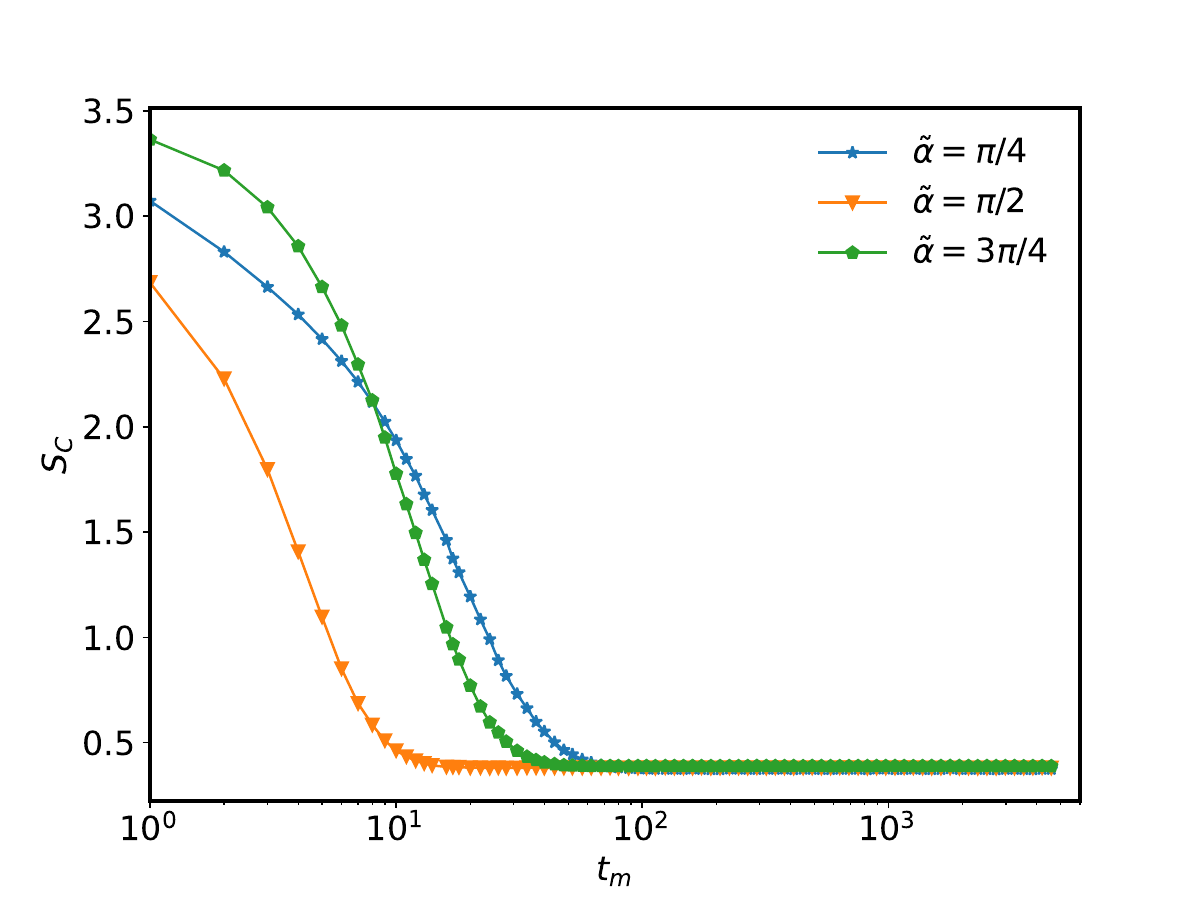}
    \caption{The time ($t_m$) evolution of trajectories averaged $S_C$ is shown for random superposition initial state under one-body measurement  only dynamics.}
    \label{fig:SC_avg_tm_rands}
\end{figure}

In Fig.~\ref{fig:SC_avg_tm_rands}, we show the trajectory-averaged $S_C$ for an initial random superposition state. In this case, we observe that the main chain and the auxiliary chain remain entangled over a time interval starting from an initial nonzero value of entanglement corresponding to the random superposition state. However, in the long-time limit, the auxiliary chain effectively disentangles from the main chain, leading to a very small entanglement $S_C$. 

\section{Click outcome in one-body measurement model}\label{sup_sec:BornProbClick_NI}
Here, we analyze the distribution  click outcomes for different quantum trajectories in the one-body measurement-only dynamics. 



\begin{figure}[htb]
    \centering
    \includegraphics[width=0.99\linewidth]{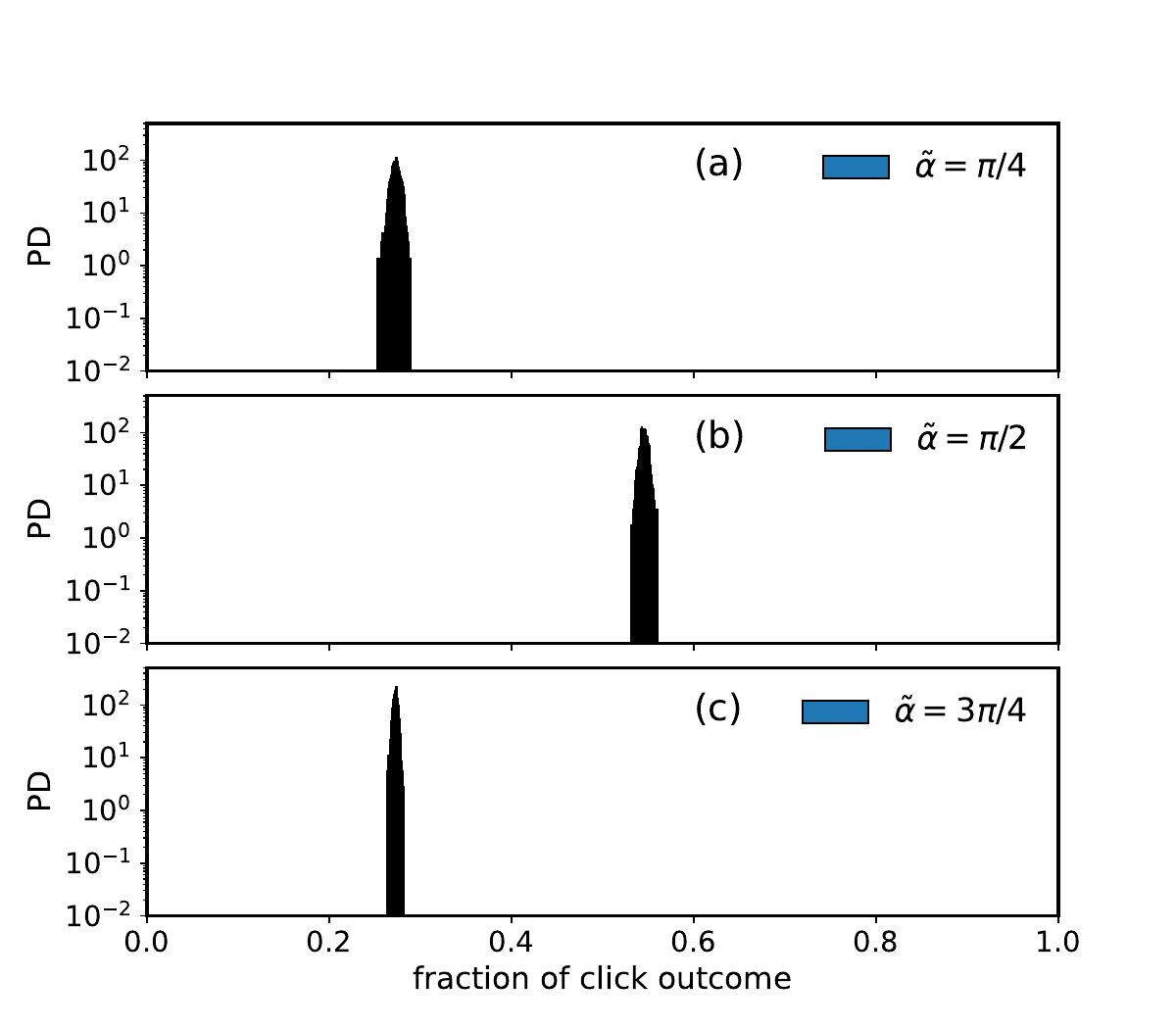}
    \caption{The probability density (PD) for the distributions of the fraction click outcomes along quantum trajectories generated by one-body measurement evolution are shown for different $\tilde{\alpha}$. This is computed from $L=12$.}
    \label{fig:clickOutcome_NI}
\end{figure}

In Fig.~\ref{fig:clickOutcome_NI}, we show the distribution of the fraction of click outcomes for an initial product state. The fraction of click outcomes is computed as the total number of click outcomes in each quantum trajectory divided by the total number of {measurements}. Irrespective  of the initial state, we find that individual quantum trajectories exhibit a finite, nonzero fraction of click outcomes.

\section{Additional results on stationary-state distribution for three-body measurement model}

\subsection{ Stationary distribution for $|\Psi_s\rangle$}

\begin{figure}[htb]
    \centering
    \includegraphics[width=0.95\linewidth]{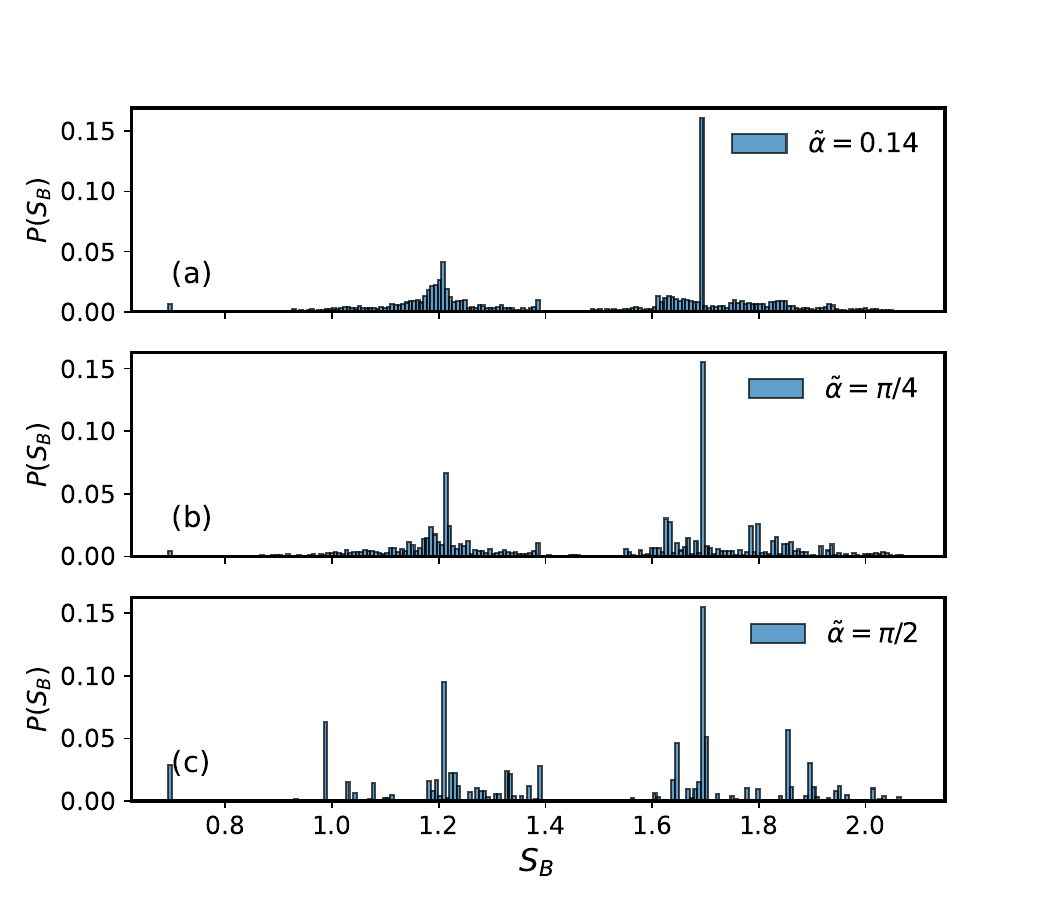}
     \caption{Probability distribution $P(S_B)$ as a function of steady-state entanglement $S_B$ are shown for $\tilde{\alpha}=0.14, \pi/4, \pi/2$ for an initial equal superposition state $|\Psi_{s}\rangle$.}
    \label{fig:Prob_Psisuper}
\end{figure}

In Fig.~\ref{fig:Prob_Psisuper}, we show the probability distribution $P(S_B)$ for different values of the parameter $\tilde{\alpha} = 0.14, \pi/4, \pi/2$ for equal amplitude superposition state $|\Psi_s\rangle$. The distribution is very similar to that for random superposition states, as shown in Fig.~\ref{fig:Fig_entB_rands} in the main text. For each $\tilde{\alpha}$, the distribution exhibits a discrete nature with a prominent peak corresponding to the `no-click' outcomes as we discuss it later. 

\subsection{Born weights and `click' outcome distribution }\label{sup_sec:BornProbInt}
The \emph{unnormalized} Born weight of a quantum trajectory $\mathcal{C}_k$ is determined by the product of local Born probabilities at each measurement step $t_{i,m}$, as given by:  
\begin{align}
    P[\mathcal{C}_k] = \prod_{i,m} P_{\sigma_{i,m}}(t_{i,m}),
\end{align}
where $P_{\sigma_{i,m}}(t_{i,m}) = \||\Psi_{\sigma_{i,m}}(t_{i,m})\rangle \|^2$ is the local Born probability for measurement on $i$-th block at the measurement step $m$ with the outcome $\sigma_{i,m}$. {Since we already sampled the trajectories according to their Born weights, we cannot normalize the above expression by summing the weights over all the sampled trajectories. For the latter, we need to sample the trajectories uniformly, i.e., consider forced measurement trajectories. Thus, in our case, we look at the Born weight of a trajectory relative to the trajectory with maximum Born weight, i.e.,
\begin{align}
    \tilde{P}[\mathcal{C}_k] = \frac{P[\mathcal{C}_k]}{\mathrm{max}_{\{\mathcal{C}_k\}}(P[\mathcal{C}_k])}.
\end{align}}

\begin{figure}[htb]
    \centering
\includegraphics[width=0.99\linewidth]{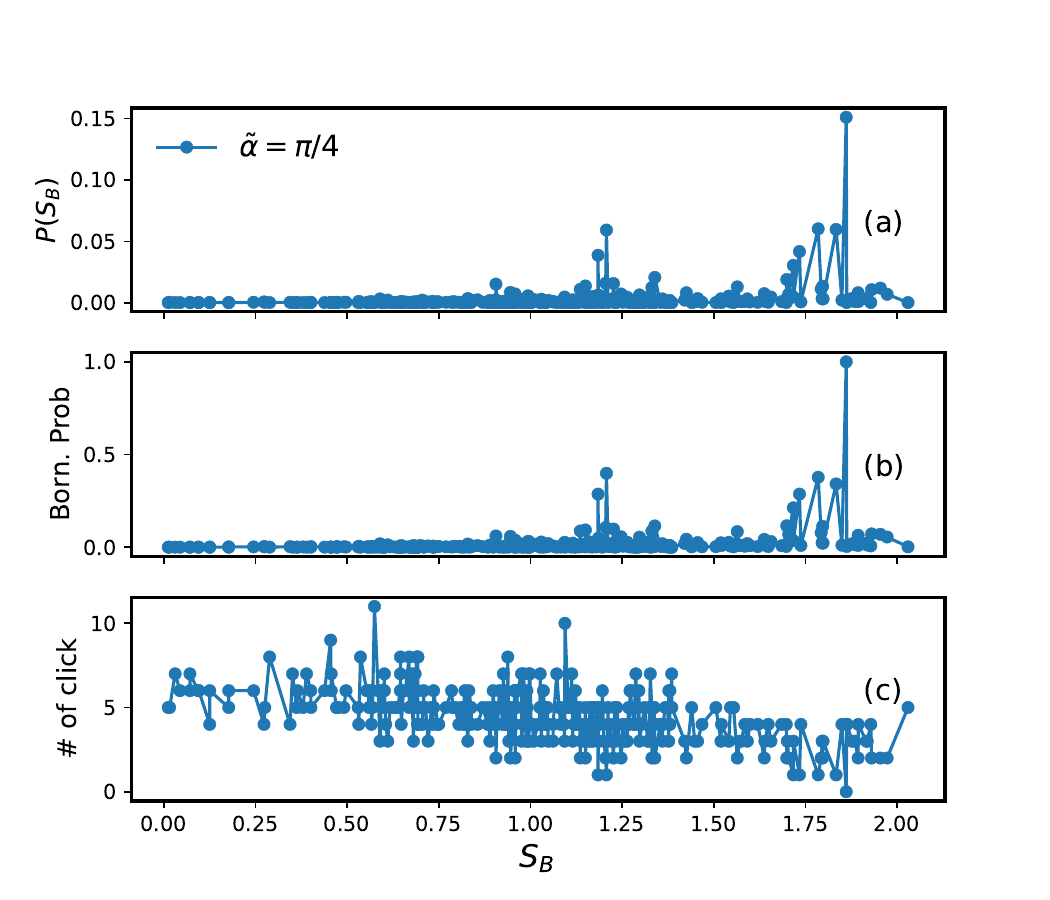}
    \caption{(a) Stationary probability distribution $P(S_B)$ as a function of long-time entanglement entropy $S_B$ for the three-body measurement model with random superposition initial state is shown [same as Fig.~\ref{fig:Fig_entB_rands}(b)].  
{(b) The Born weights of quantum trajectories relative to the trajectory with maximum Born weight are presented as a function of long-time entanglement entropy $S_B$.  
(c) The total number of click outcomes as a function of long-time entanglement entropy $S_B$ is shown. This is computed for a system size of $L=12$ for the random superposition state $|\Psi_{rs}\rangle$.}  
}
    \label{fig:Bornprob}
\end{figure}

In Fig.~\ref{fig:Bornprob}(a), we plot the stationary probability distribution $P(S_B)$ (also shown as a histogram in Fig.~\ref{fig:Fig_entB_rands} in the main text) as a function of $S_B$ for random superposition initial state. In Fig.~\ref{fig:Bornprob}(b), we plot the relative Born weights $\tilde{P}[\mathcal{C}_k]$ of trajectories $\mathcal{C}_k$, characterized by their long-time entanglement $S_B$, as a function of $S_B$. We observe that the most probable value of $S_B$ in Fig.~\ref{fig:Bornprob}(a) coincides with the peak in Born weight in Fig.~\ref{fig:Bornprob}(b), indicating that the most probable value of entanglement entropy corresponds to the trajectory with the maximum Born weight.  

Furthermore, we compute the number of click outcomes in each trajectory $\mathcal{C}_k$ and plot this quantity as a function of the long-time entanglement $S_B$ for the trajectories. We find that the trajectory with most probable $S_B$ and maximum Born weight does not exhibit any click outcomes. This result demonstrates that the most probable trajectory originates entirely from `no-click' outcomes. We observe that other trajectories also have very few click outcomes $\mathcal{O}(1)$ compared to total number of measurement sweeps $t_m=1000$.  

We also find that measurement dynamics for the equal superposition initial state, whose stationary distribution $P(S_B)$ is shown in Fig.~\ref{fig:Prob_Psisuper}, exhibits a Born weight and `click' outcome distribution for quantum trajectories similar to Fig.~\ref{fig:Bornprob}(b, c) for random superposition states. The most probable value of $S_B$ corresponds to `no-click' outcomes and has the highest Born weight, similar to  random superposition states.

\begin{figure}[!hb]
    \centering 
\includegraphics[width=0.99\linewidth]{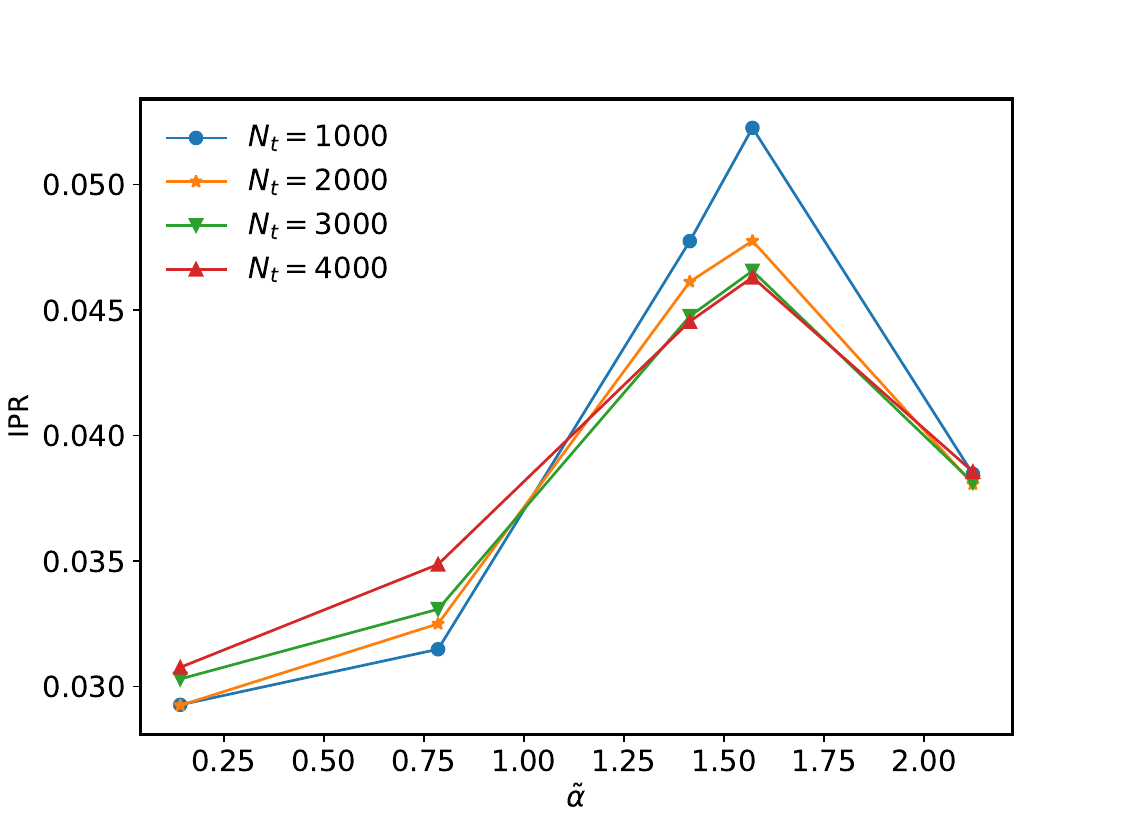}
    \caption{The inverse participation ratio (IPR) for the stationary probability distribution of the random superposition state is shown as a function of the ensemble size $N_t$ of quantum trajectories  for different $\tilde{\alpha}$. This is computed for a system size of $L=12$.  
.}
    \label{fig:IPR_psirs}
\end{figure}

\subsection{IPR of stationary distribution}\label{sup_sec:IPR_Int}

The stationary state distribution for different initial states in the three-body measurement model appears discrete compared to the continuous stationary distribution observed in the one-body measurement model. The distinction between a discrete (localized) and continuous (delocalized) distribution can be quantified using the inverse participation ratio (IPR). The IPR is widely used to characterize the localized or delocalized nature of wave functions in Hilbert space, particularly in the context of many-body localization \cite{PhysRevB.83.184206}. It captures the extent to which the wave function amplitudes are distributed across the Fock space basis. For a localized distribution, the IPR saturates to a finite value, whereas for a delocalized system, it approaches zero as the Hilbert space dimension increases.

We compute the IPR of the distribution using the following expression:  
\begin{align}
    \text{IPR} = \sum^{N_t}_{k=1} P^2(S^{(k)}_B),
\end{align}
where $P(S^{(k)}_B)$ is the steady-state probability of the entanglement entropy for the $k$-th trajectory, and $N_t$ denotes the total number of trajectories.  

\begin{figure}[!ht]
    \centering
\includegraphics[width=0.99\linewidth]{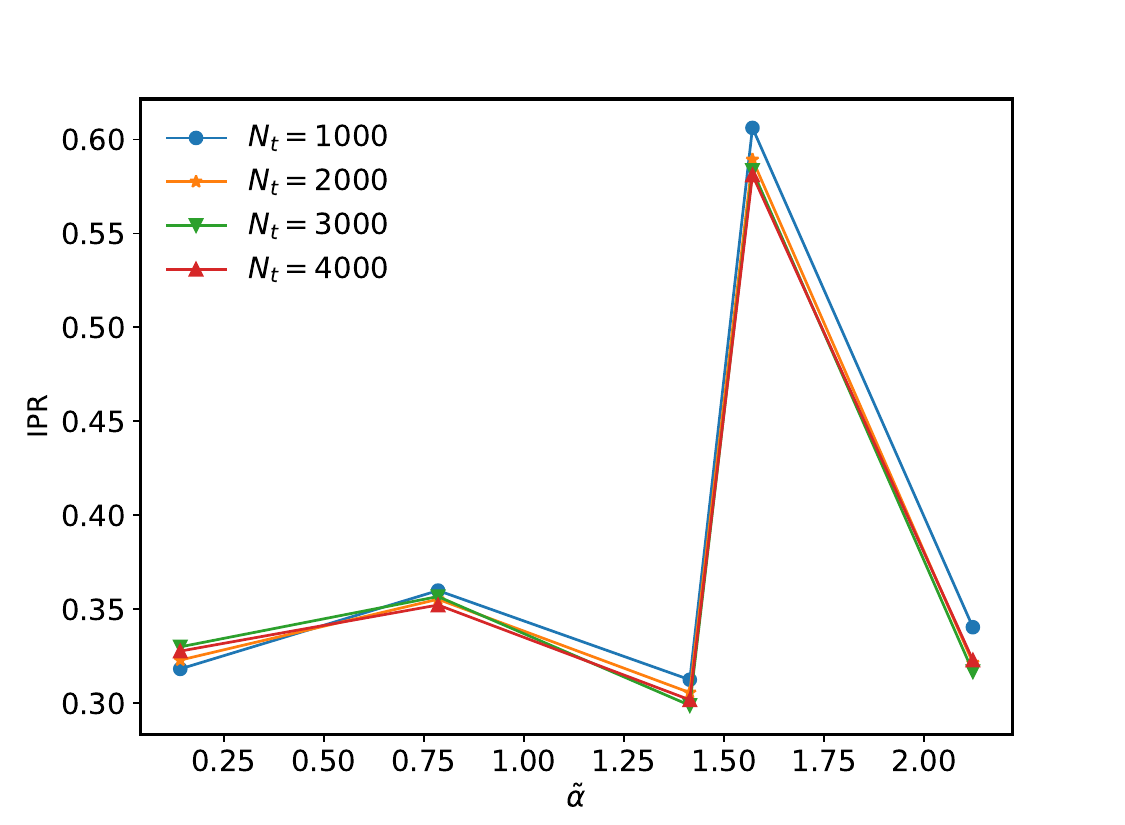}
    \caption{The inverse participation ratio (IPR) for the stationary probability distribution of product initial state is shown as a function of the ensemble size $N_t$ of quantum trajectories for different $\tilde{\alpha}$. This is computed for a system size of $L=12$.  
.}
    \label{fig:IPR_psiprod}
\end{figure}

In Fig.~\ref{fig:IPR_psirs}, we plot the IPR of the stationary probability distribution for random superposition states as a function of the ensemble size $N_t$ of quantum trajectories. We observe that the IPR does not vary significantly with the total number of trajectories $N_t$ for different values of the parameter $\tilde{\alpha}$. This confirms that the distribution remains discrete as the ensemble size $N_t$ increases. Similarly, for equal superposition state, the stationary state distribution being similar to that of random superposition state, it's IPR looks qualitatively very similar to \ref{fig:IPR_psirs}, confirming discreteness of the distribution.

The stationary state distribution for the product initial state, as discussed in the main text, appears significantly more discrete compared to that of the random superposition or equal superposition states. In Fig.~\ref{fig:IPR_psiprod}, we show the IPR of the stationary state distribution for the product initial state as a function of the ensemble size $N_t$ for different $\tilde{\alpha}$. We find that the IPR does not vary with $N_t$, confirming that the stationary state distribution remains discrete as the ensemble size increases.  Furthermore, we observe that the saturation value of the IPR for the product state is larger than that for the superposition states. This indicates that the stationary state distribution for the product initial state exhibits a more localized nature compared to the stationary state distributions for the superposition states.

\begin{figure*}[htb]
    \centering
\includegraphics[width=0.95\textwidth]{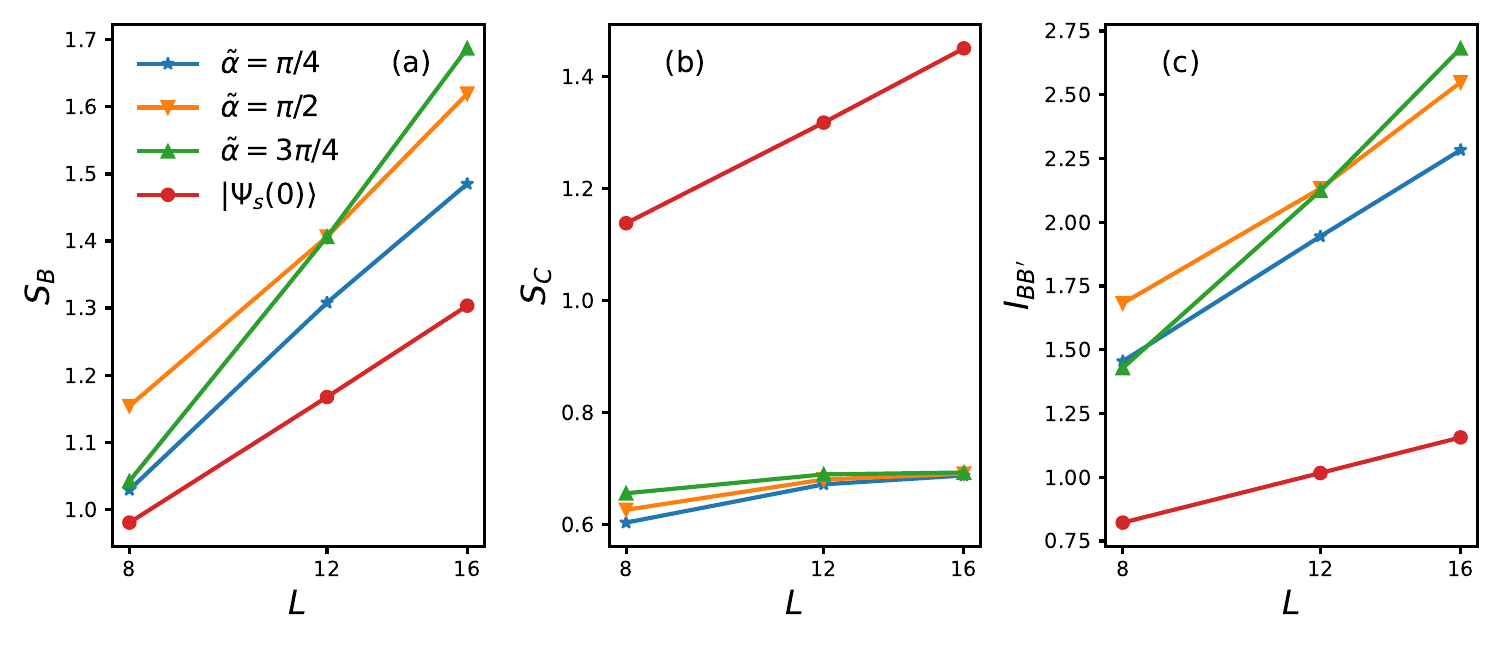}
    \caption{The system-size scaling of long-time entanglement entropies and mutual information for the equal superposition initial state $|\Psi_{s}\rangle$ in the one-body measurement model is shown in log-normal scale. 
(a) The scaling of $B$-subsystem (Fig.~\ref{fig:mom_cartoon}) entanglement entropy $S_B$ is similar to the $\log L$ scaling of the initial state.  
(b) The $C$-subsystem (Fig.~\ref{fig:mom_cartoon}) entanglement entropy $S_C$ exhibits area-law scaling, in contrast to the $\log L$ scaling of $S_C$ for the initial state.  
(c) The long-time average mutual information between the two halves of the main chain, $I_{BB'}$, shows $\log L$ scaling, similar to the initial state.
  }
    \label{fig:NI_entscaling_super}
\end{figure*}

\begin{figure*}[htb]
    \centering
\includegraphics[width=0.95\textwidth]{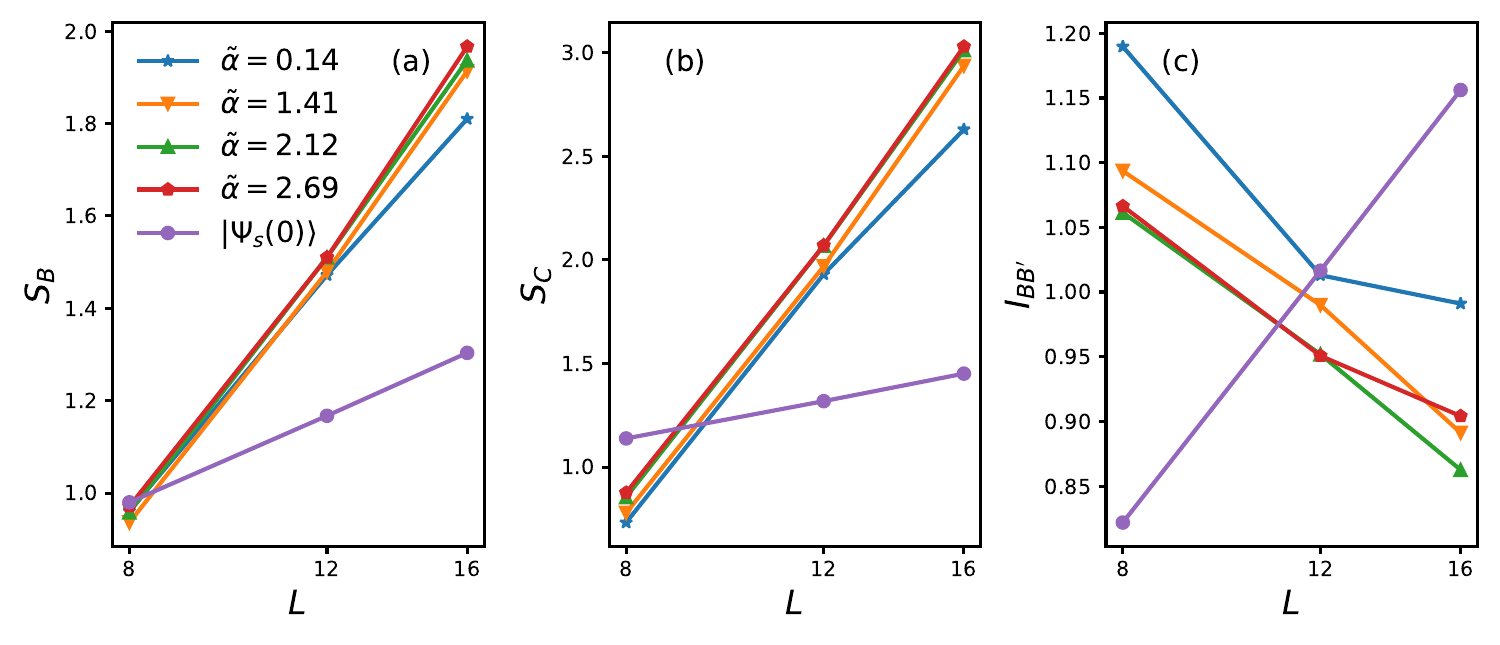}
    \caption{The system-size scaling of stationary-state entanglement entropies and mutual information for the equal superposition initial state $|\Psi_{s}\rangle$ in the three-body measurement model is presented in a log-normal scale. Both $S_B$ and $S_C$ in (a) and (b), respectively, exhibit $\log L$ scaling, similar to the scaling behavior of the initial state. However, the mutual information in (c) decreases with $L$ unlike the initial-state mutual information which increases as $\log L$.}
    \label{fig:Int_entscaling_super}
\end{figure*}

\section{System-size scaling of entanglement for equal superposition states}\label{sec:appedix_systemsize_scaling}

\subsection{One-body measurement model}\label{sup_sec:scaling_NI}
In Fig.~\ref{fig:NI_entscaling_super}, we present the system-size scaling of the average entanglement entropy for the stationary distribution under measurement-only dynamics in the one-body measurement model, for the equal superposition state as the initial state. 

Figure \ref{fig:NI_entscaling_super}(a) shows the $B$-subsystem entanglement entropy $S_B$ as a function of system size $L$ on a semi-logarithmic scale ($\log x$). The results show that the average entanglement entropy of the stationary distribution scales logarithmically with $L$. The entanglement values are larger than the $\log L$ scaling observed for the initial state $|\Psi(0)\rangle$.  For the $C$-subsystem, the average entanglement entropy of the stationary distribution follows an area-law scaling, as shown in Fig.~\ref{fig:NI_entscaling_super}(b), which is also presented in semi-logarithmic scale ($\log x$). Finally, in Fig.~\ref{fig:NI_entscaling_super}(c), we show the mutual information $I_{BB'}$ between two equal halves of the main chain as a function of $L$, also on a semi-logarithmic scale. The mutual information exhibits a $\log L$ scaling, and its value is larger than the mutual information of the initial state's $I_{BB'}$, showing true entanglement generation in the stationary distribution.

\subsection{Three-body measurement model}\label{sup_sec:scaling_Int}
In Fig.~\ref{fig:Int_entscaling_super}, we present the system-size scaling of the average entanglement entropy for the stationary state distribution under three-body measurement model measurement-only dynamics, for the equal superposition state as the initial state. All plots in Fig.~\ref{fig:Int_entscaling_super}(a, b, c) are shown on a semi-logarithmic scale.

\begin{figure*}[htb]
    \centering   \includegraphics[width=0.495\linewidth]{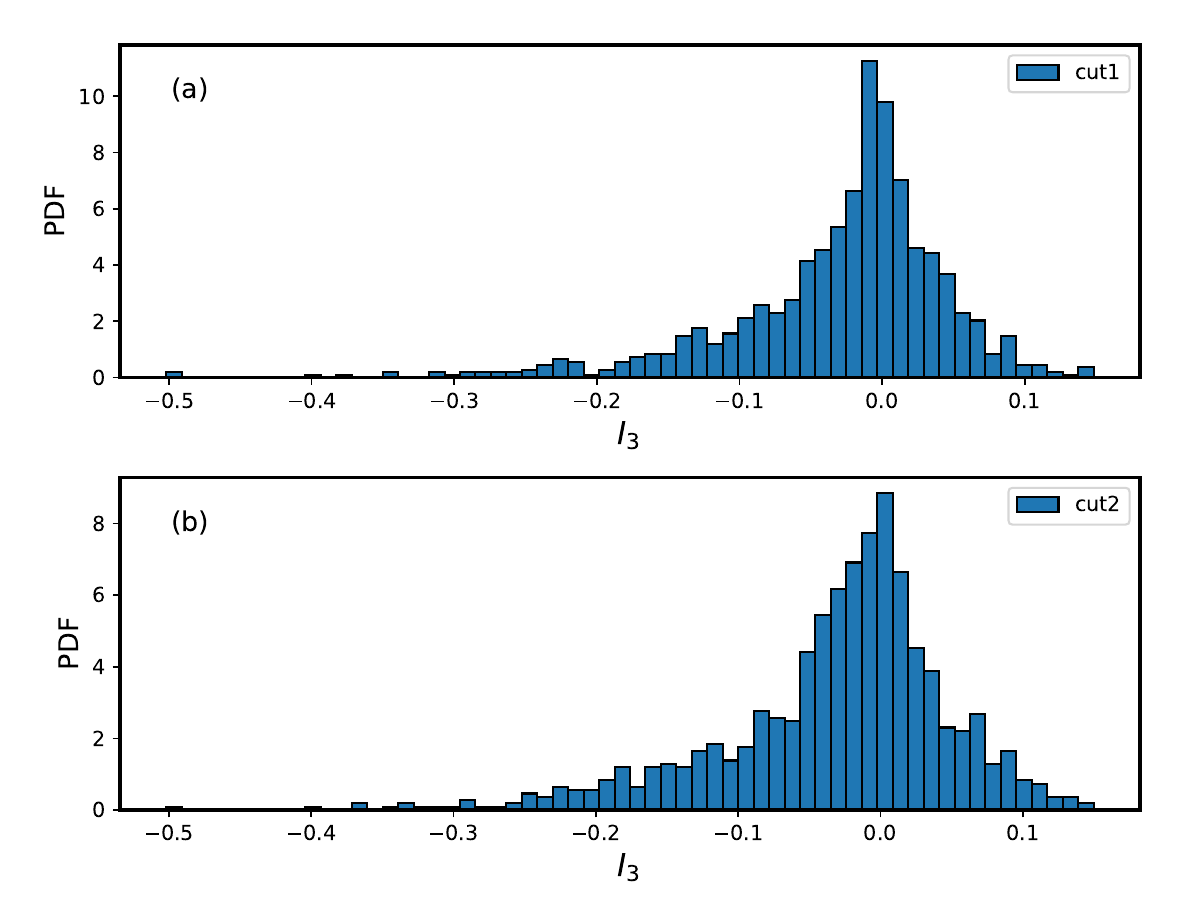}
\includegraphics[width=0.495\linewidth]{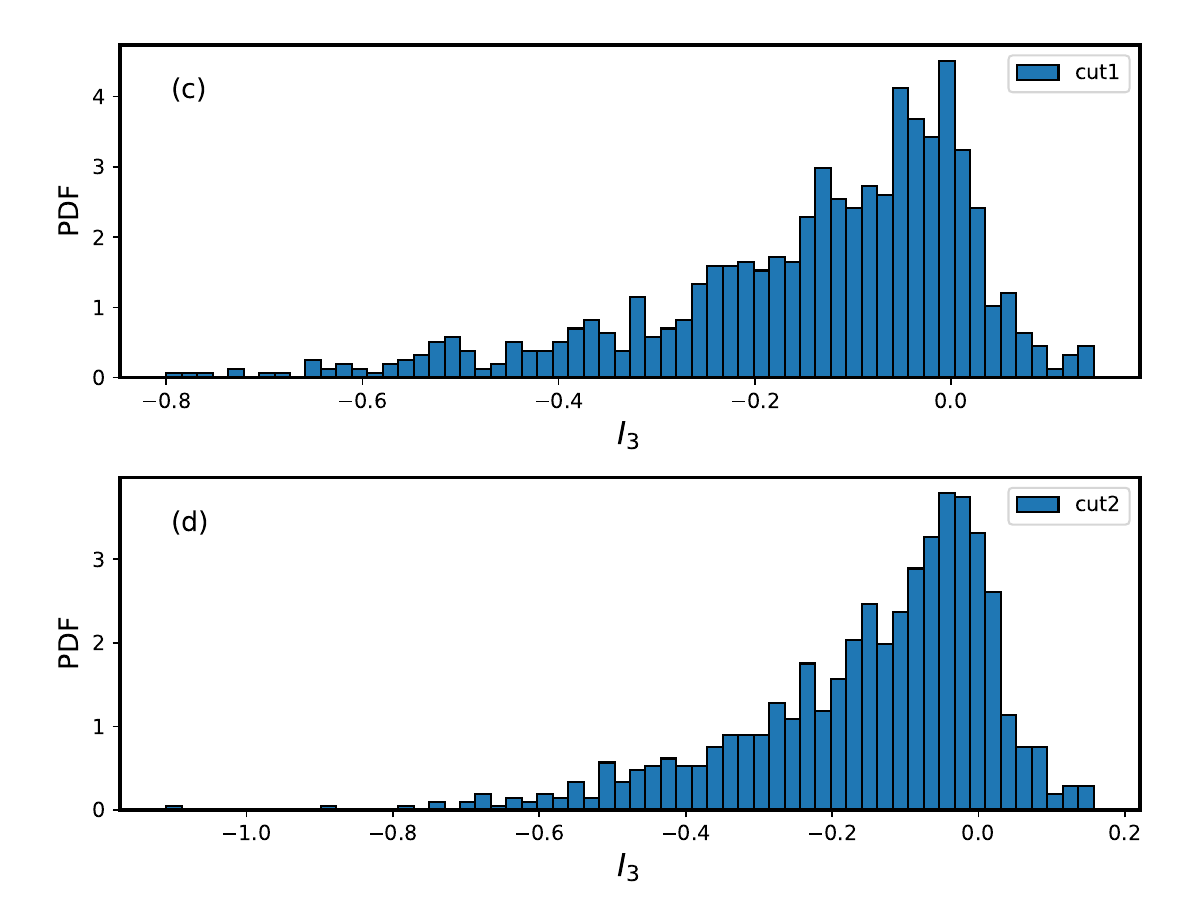}
    \caption{The long-time probability distribution of tripartite mutual information (TMI) over quantum trajectories for the main chain is shown for two distinct subsystem choices, referred to as ``cut1'' and ``cut2'' in the legends of the upper (a, c) and lower (b, d) panels, respectively. Here the results are obtained for one-body measurement dynamics [Sec.\ref{sec:NonIntModel}] starting with a product state. The left panels (a, b) correspond to $\tilde{\alpha}=\pi/4$, and the right panel (c, d) corresponds to $\tilde{\alpha}=3\pi/4$. The results are computed for a system size of $L=16$.}
    \label{fig:TMI_pi_4}
\end{figure*}

From Figs.~\ref{fig:Int_entscaling_super}(a) and \ref{fig:Int_entscaling_super}(b), we observe that both the $B$-subsystem and $C$-subsystem entanglement entropies of the stationary distribution follow a $\log L$ scaling, similar to the initial state. However, unlike the one-body measurement model, the mutual information $I_{BB'}$ decreases with increasing system size, as shown in Fig.~\ref{fig:Int_entscaling_super}(c). This decrease in mutual information show that, due to the strong entanglement between the main chain and the auxiliary chain (captured by the $C$-subsystem), there is no true entanglement generation within the main chain itself.

\section{{Study of tri-partite mutual information under one-body measurement model}} \label{app:TMI}

{
As discussed in Sec.\ref{sec:EntGenNonInt}, in our one-body measurement model (Sec.\ref{sec:NonIntModel}), we find that the stationary distribution exhibits volume-law entanglement scaling with system size, as well as volume-law mutual information between the two halves of the main chain. This suggests that the one-body-measurement-only dynamics generates highly entangled states. 

In this Appendix, to probe the nature of the long-time volume-law states further, we present the tripartite mutual information (TMI) \cite{Cerf1998,Kitaev2006,Levin2006,Iyoda2018,Caceffo2023} for the states. TMI is the most computationally straightforward way to analyze multipartite entanglement in our system, but there are other measures like ``entanglement contour''\cite{Chen_2014, PhysRevB.101.195134} for Gaussian and non-Gaussian states.

The TMI is defined for three disjoint subsystems $A, B, C$ as follows:
\begin{align}
    I_3(A: B: C) &= S_A + S_B + S_C - S_{AB} - S_{BC} - S_{AC}\notag \\
    &+ S_{ABC}.
\end{align}
$I_3 < 0$ indicates genuine multi-partite quantum correlations among different parts of the system, while $I_3 = 0$ and $I_3>0$ imply purely bi-partite entanglement and classical correlations, respectively. The TMI is often used to characterize multipartite entanglement, particularly in mixed states.

In our one-body measurement model, the quantum state of the main chain remains disentangled from the auxiliary chain, and thus always stays pure, for initial product states of the main and the auxiliary chain [Sec.\ref{sec:Role-aux-chain}]. In Fig.~\ref{fig:TMI_pi_4}, we present the distribution of TMI over quantum trajectories at long times, starting from a product state for a total system size $L=16$ for $\tilde{\alpha}=\pi/4,  3\pi/4$. The length of the main chain is then $N=L/2=8$. 

To compute the TMI, we consider different types of divisions of the main chain into subsystem, where the sites are labeled as $i=1,1,\dots,8$. For example, one choice, denoted as ``cut1", is three contiguous subsystems, namely
$A=\{1,2\}$, $B=\{3,4\}$, and $C=\{5,6\}$, while keeping other main chain sites and the auxiliary chain as the complement. Alternatively, we choose spatially separated subsystems,
$A=\{1,2\}$, $B=\{4,5\}$, and $C=\{7,8\}$, denoted as ``cut2".

As shown in Figs.~\ref{fig:TMI_pi_4}, we find that the long-time distribution of $I_3$ for both the subsystem choices, is asymmetric around $I_3=0$ and exhibits a long tail for $I_3<0$. Consequently, the average value of $I_3$ is negative. These indicate genuine multi-partite quantum correlations, beyond just bi-partite entanglement, for the long-time volume-law states for the one-body-measurement-only model.
}



\clearpage

\bibliography{ref}
\end{document}